\documentclass[11pt]{article}
\usepackage[utf8]{inputenc}
\usepackage[T1]{fontenc}
\usepackage[DIV12]{typearea}
\usepackage{amsmath,amsfonts,amssymb,amscd}
\usepackage[all]{xy}
\usepackage{graphicx}
\usepackage{cite}
\usepackage{mathrsfs}
\usepackage{bbm}
\usepackage{bbold}
\usepackage{braket}
\usepackage[dvipsnames]{xcolor}
\usepackage[normalem]{ulem}
\usepackage{tensor}
\usepackage{mathtools}
\usepackage[bookmarks=false]{hyperref}
\usepackage{srcltx}

\newcommand{\be}{\begin{equation}}
\newcommand{\ee}{\end{equation}}
\newcommand{\bea}{\begin{eqnarray}}
\newcommand{\eea}{\end{eqnarray}}

\newcommand{\g}{\gamma}

\newcommand{\B}{\mathcal{B}}

\newcommand{\n}{\mathbf{n}}
\newcommand{\m}{\mathbf{m}}

\DeclareMathOperator{\Tr}{Tr}

\def\nn{\nonumber}
\newcommand{\ketbra}{\rangle \langle}
\def\ket{\rangle}
\def\bra{\langle}

\def\r{\rho}
\def\a{\alpha}
\def\b{\beta}
\def\g{\gamma}

\def\d{\delta}

\def\e{\epsilon}

\def\m{\mu}
\def\n{\nu}

\def\lam{\lambda}

\def\s{\sigma}

\def\conc{{\cal C}}
\def\bell{{\cal B}}
\def\BM{{\cal B}_{\rm M}}
\def\BS{{\cal B}_{\rm S}}
\def\B4{{\cal B}_{442}}

\def\eBM{\bra {\cal B}_{\rm M} \ket}
\def\eBS{\bra {\cal B}_{\rm S} \ket} 
\def\eB4{\bra {\cal B}^{\rm sym}_{442} \ket}

\def\eBB4{ \bra {\cal B}_{442} \ket }

\definecolor{dg}{rgb}{0,.5,0}

\usepackage{soul}

\def\mytitle{Three-Body Non-Locality in Particle Decays}

\title{\mytitle}

\begin{document}

\begin{titlepage}
\setlength{\topmargin}{0.0 true in}
\thispagestyle{empty}

\vspace{-10mm}

\begin{flushright}
IPPP/25/12 
\end{flushright}

\vspace{10mm}

\begin{center}
{\LARGE\textbf{\mytitle}}

\renewcommand*{\thefootnote}{\fnsymbol{footnote}}

\vspace{1.1cm}
Pawe{\l} Horodecki${}^{1,2}$\footnote[1]{
\href{pawhorod@pg.edu.pl}{pawhorod@pg.edu.pl}
},
Kazuki Sakurai${}^{3,}$\footnote[2]{
\href{kazuki.sakurai@fuw.edu.pl}{kazuki.sakurai@fuw.edu.pl}
},
Abhyoudai S. Shaleena ${}^{1}$\footnote[3]{
\href{abhyoudai.sajeevkumarshaleena@phdstud.ug.edu.pl}{abhyoudai.sajeevkumarshaleena@phdstud.ug.edu.pl}
},
Michael Spannowsky${}^{4,}$\footnote[4]{
\href{michael.spannowsky@durham.ac.uk}{michael.spannowsky@durham.ac.uk}
}
\\[5mm]

\normalsize

\textit{
${}^1$
International Centre for Theory of Quantum Technologies (ICTQT), University of Gda{\'n}sk, Jana Ba{\.z}ynskiego 8, 80-309 Gda{\'n}sk, Poland
} \\[2mm]
\textit{
${}^2$
Faculty of Applied Physics and Mathematics, Gda{\'n}sk University of Technology,
Gabriela Narutowicza 11/12, 80-233 Gda{\'n}sk, Poland
} \\[2mm]
\textit{
${}^3$Institute of Theoretical Physics, Faculty of Physics,\\
University of Warsaw, Pasteura 5, 02-093, Warsaw, Poland
} \\[2mm]
\textit{
${}^4$Institute for Particle Physics Phenomenology, Department of Physics, \\ Durham University, Durham DH1 3LE, U.K.%
}

\end{center}
\vspace*{5mm}

\begin{abstract}\noindent

The exploration of entanglement and Bell non-locality among multi-particle quantum systems offers a profound avenue for testing and understanding the limits of quantum mechanics and local real hidden variable theories.
In this work, we examine non-local correlations among three massless spin-1/2 particles generated from the three-body decay of a massive particle, utilizing a framework based on general four-fermion interactions.
By analyzing several inequalities, we address the detection of deviations from quantum mechanics as well as violations of two key hidden variable theories: fully local-real and bipartite local-real theories. 
Our approach encompasses the standard Mermin inequality and the tight $4 \times 4 \times 2$ inequality, providing a comprehensive framework for probing three-partite non-local correlations. 
Our findings provide deeper insights into the boundaries of classical and quantum theories in three-particle systems, advancing the understanding of non-locality in particle decays and its relevance to particle physics and quantum foundations.

\end{abstract}
\thispagestyle{empty}
\clearpage

\end{titlepage}

\section{Introduction}

Quantum mechanics stands apart from classical physics due to its inherent probabilistic nature and the baffling phenomenon of entanglement \cite{Schrödinger_1935, Einstein:1935rr}.  Entanglement describes a situation where two or more particles become intrinsically linked, sharing a combined quantum state that cannot be factored into independent states for each  
particle.  This interconnectedness, regardless of spatial separation, leads to correlations that profoundly challenge our classical intuitions and are fundamentally incompatible with local realism — the idea that physical systems have definite properties independent of measurement and that influences cannot propagate faster than the speed of light.

The seminal work of John Bell \cite{PhysicsPhysiqueFizika} provided a rigorous framework for experimentally testing the predictions of quantum mechanics against the constraints imposed by local realism.  
Bell-type inequalities, mathematical expressions derived from the assumption of local realism, predict upper bounds on the correlations measurable in experiments. 
The violation of the CHSH inequality \cite{Clauser:1969ny}, one of the Bell-type inequalities for the two-qubit system, in numerous experiments \cite{Freedman:1972zza,Aspect:1981nv,Weihs:1998gy,Hensen:2015ccp,Giustina:2015yza,Stevens:2015awv,Handsteiner:2016ulx,Storz:2023jjx}, has decisively confirmed the non-locality of quantum mechanics, demonstrating the existence of correlations that cannot be explained by any local realistic theory.
The correlation measurements and corresponding observables involving Bell-type inequalities also provide a powerful framework for testing quantum mechanics itself. 
Quantum mechanics often gives upper bounds on these observables \cite{Tsirelson}; if a violation of these upper bounds is experimentally confirmed, quantum mechanics would be falsified.

While the violation of CHSH inequality and its experimental verification have firmly established non-locality in two-particle systems, the extension to multipartite systems presents significantly greater complexity and remains an active area of research.  
The intricacies of multipartite entanglement, involving three or more particles, introduce a rich landscape of correlations far exceeding the complexity of two-particle scenarios.  

Understanding these multipartite correlations is crucial not only for advancing fundamental physics but also for developing emerging quantum technologies, such as quantum computing and quantum communication, where multipartite entanglement plays a central role.
Following early works \cite{PhysRevA.63.062112, Hiesmayr2012},
there has recently been a surge of interest in exploring quantum information-theoretic properties within the context of high-energy particle physics \cite{
Afik:2020onf, 
Ashby-Pickering:2022umy, 
Barr:2021zcp, 
Barr:2022wyq, 
Afik:2022kwm, 
Aguilar-Saavedra:2022uye, 
Aoude:2022imd, 
Severi:2022qjy, 
Fabbrichesi:2022ovb, 
Altakach:2022ywa, 
Acin:2000cs,
Sakurai:2023nsc, 
Morales:2024jhj, 
Subba:2024mnl,
Aoude:2023hxv, 
Bernal:2023ruk, 
Han:2023fci, 
Dong:2023xiw, 
Fabbrichesi:2023jep, 
Maltoni:2024tul, 
Afik:2024uif, 
Maltoni:2024csn, 
Aguilar-Saavedra:2024fig, 
Aguilar-Saavedra:2024hwd, 
Aguilar-Saavedra:2024vpd, 
Aguilar-Saavedra:2024whi,
Grabarczyk:2024wnk,
Carena:2023vjc,Kowalska:2024kbs,Chang:2024wrx,Thaler:2024anb,
Low:2024mrk,Low:2024hvn,
Altomonte:2024upf,
Han:2024ugl,
Han:2025ewp}
(see also \cite{Barr:2024djo} for the recent review). 
Particle scattering and the decay of unstable particles provide a natural laboratory where multipartite entanglement and nonlocal correlations emerge spontaneously (see, however, \cite{Abel:1992kz}).\footnote{There are issues to conduct a loophole-free test of local realism at conventional collider experiments.
High-energy collider experiments have imperfect acceptance and detection efficiency, resulting in the detection loophole. 
Additionally, the spin correlation between particles A and B is only inferred indirectly from the momentum distribution of decay products of A and B, assuming Standard Model predictions or those decays behave the same way as they do in an independent experiment.  
Even if we accept those assumptions, we do not have the freedom to choose the spin measurement axes, as the spin measurement is constituted indirectly and statistically by analysing the momentum distributions. 
This makes the local realism test prone to the so-called freedom-of-choice loophole.
Finally, the momenta of observed particles are essentially commuting observables.
Therefore, there is always some hidden variable theory that can explain the observed momentum data.  
The interest of this paper is not a high-energy test of local realism.
Our goal is to study what type of spin correlation emerges in different phase-space regions of three-body decays in different underlying particle interactions within the quantum field theory framework.  
We also discuss the spin correlation beyond quantum theories.}
In this context, the ATLAS and CMS collaborations have recently measured spin correlations in the $t\bar{t}$ system produced in $pp$ collisions at the LHC, observing entanglement within a specific phase-space region at a 5-$\sigma$ significance level \cite{ATLAS:2023fsd, CMS:2024pts, CMS:2024zkc}.
Precise measurements of entanglement and non-locality observables have been recognized as valuable tools for probing non-perturbative effects within the Standard Model \cite{Maltoni:2024tul, Aguilar-Saavedra:2024mnm}, as well as exploring physics beyond the Standard Model \cite{Aoude:2022imd, Severi:2022qjy, Fabbrichesi:2022ovb, Altakach:2022ywa}. 

Another motivation lies in the potential of collider measurements of quantum properties to directly test quantum mechanics at high energy scales.
It is plausible that quantum mechanics undergoes modifications at some short distance scales to achieve compatibility with gravity.
Such modifications could, in principle, be detected by measuring Bell-type observables or through quantum process tomography, as outlined in \cite{Altomonte:2024upf}.

Finally, studying particle physics through the lens of quantum information theory offers the potential to uncover new insights into quantum field theory.
For instance, recent research has explored the interplay between a theory’s internal (emergent) symmetries and the entanglement generated by specific scattering processes \cite{Carena:2023vjc, Kowalska:2024kbs, Chang:2024wrx, Thaler:2024anb}.
Additionally, efforts have been made to establish connections between scattering cross sections and the entanglement entropy produced during these processes \cite{Low:2024mrk, Low:2024hvn, Seki:2014cgq, Peschanski:2016hgk, Peschanski:2019yah}.

While the present literature has focused on quantum properties in two-particle systems, systematic investigations of three-particle entanglement and non-locality in particle physics are significantly underexplored \cite{Acin:2000cs, Sakurai:2023nsc, Morales:2024jhj, Subba:2024mnl}.

This paper addresses this gap by investigating the three-body decay of a massive fermion into three massless spin-1/2 particles, a scenario that offers a tractable yet non-trivial setting for studying three-partite entanglement and non-locality. 
This specific choice allows for a detailed theoretical analysis while maintaining relevance to experimentally accessible scenarios in particle physics.  
Utilizing a theoretical framework with general four-fermion interactions, we systematically analyze the correlations emerging from this decay process using a variety of entanglement measures and Bell-type inequalities. 
While we have analyzed the entanglement generated in the aforementioned decay process in our earlier work \cite{Sakurai:2023nsc}, the present study shifts focus to the non-locality arising from the same decay process.
This study aims to quantify the degree of non-local correlations and identify the conditions under which deviations from local realistic predictions are maximized in the decay kinematic phase space.  
The study strives to provide a deeper and more comprehensive understanding of three-body entanglement and non-locality, bridging the gap between fundamental physics and the rapidly advancing field of quantum information science.

The rest of the paper is organised as follows. 
In the next section, we provide a short review of entanglement and non-locality, supplementing the necessary background and introducing some observables and formulae used in the later sections. 
Readers familiar with multiparticle entanglement and non-locality may wish to skip this section.
In section \ref{sec:kin}, we describe the three-body kinematics and the resulting spin states. 
The theoretical and numerical analyses of entanglement and non-locality in the particle decays are presented in section \ref{sec:analysis}.
We explore these properties in three different types of four fermion interactions.
Section \ref{sec:concl} is devoted to the conclusion.

\section{A Review of Basic Features of Non-locality and Entanglement}  

This section provides a concise overview of entanglement and nonlocality in bipartite and tripartite quantum systems.
The content summarized here can be found in several review articles, such as \cite{Horodecki:2009zz, Guhne:2008qic,Friis_2018}.

\subsection{Entanglement}

Entanglement is known as {\it the characteristic trait} of quantum mechanics, distinguishing it from classical mechanics \cite{Schrödinger_1935}.
When two subsystems, A and B, are entangled, the prediction of measurement outcomes for A cannot be fully described independently of the state of B, and vice versa, no matter how far they are spatially separated.
In contrast, quantum states without entanglement, called {\it separable states}, can be expressed as a simple tensor product of the local states of A and B:
\be
| \psi_{AB}^{\rm sep} \rangle = | \psi \rangle_A \otimes | \psi \rangle_B.
\label{sep}
\ee
The concept of separability extends naturally to mixed states. A mixed state is considered separable if its density operator can be written as a convex combination of tensor products of local density operators \cite{PhysRevA.40.4277}
\be
\rho_{AB}^{\rm sep} = \sum_k p_k  \, \rho_{A,k} \otimes \rho_{B,k},
\ee
where $p_k \geq 0$ are probabilities satisfying $\sum_k p_k = 1$, and  $\rho_{A,k}$ and $\rho_{B,k}$ are density operators for the subsystems A and B, respectively.\footnote{We denote the Hilbert spaces of the subsystems A and B as ${\cal H}_A$ and ${\cal H}_B$, respectively.
The Hilbert space of the combined system is given by ${\cal H}_{AB} = {\cal H}_{A} \otimes {\cal H}_{B}$.
The set of all density operators (positive semi-definite Hermitian operators with unit trace) of the Hilbert space ${\cal H}$ is denoted as ${\cal S}({\cal H})$.
In this notation, $| \psi_A \ket \in {\cal H}_A$, $| \psi_B \ket \in {\cal H}_B$, $| \psi_{AB} \ket \in {\cal H}_{AB}$,
$\rho_A \in {\cal S}({\cal H}_A)$, $\rho_B \in {\cal S}({\cal H}_B)$ and $\rho_{AB} \in {\cal S}({\cal H}_{AB})$.
}

If the system is inseparable, the degree of entanglement can be measured by a class of functions, $E(\rho)$, called {\it entanglement measures (or entanglement monotones)} \cite{Vidal_2000}.
Those functions must vanish for all separable states and positive for entangled states. 
Additionally, entanglement measures must be invariant under local unitary transformations, reflecting changes in the local basis.
Another important property involves Local Operations and Classical Communication (LOCC) \cite{PhysRevA.54.3824}. 
In this class of operations, two experimenters, Alice and Bob, each control subsystems A and B, respectively. 
They are allowed to manipulate their particles by local unitary operations and making local measurements while also communicating classically.
Motivated by the fact that separable states cannot be transformed into entangled states under LOCC, entanglement measures are required to be non-increasing under LOCC.\footnote{This property contains the invariance under local unitaries as unitary operations are invertible.}
The above consideration also introduces the notion of {\it maximally entangled states}.
Maximally entangled states cannot be reached from non-invertible LOCC from any other states. 

Among several entanglement measures proposed in the literature, in this work, we use a particular one called {\it concurrence}, which is defined for the pure state, $| \psi_{AB} \ket \in {\cal H}_{AB}$, as   
\be
{\cal C}( | \psi \ket_{AB} ) \equiv \sqrt{ 2 ( 1 - \Tr{\varrho_A^2 ) }},
\label{conc_pure}
\ee
where $\varrho_A = \Tr_B \rho_{AB}$ is the reduced density operator of subsystem A, obtained by tracing out the degrees of freedom of subsystem B.
The same result can be obtained by exchanging the role of subsystems A and B: ${\cal C}( | \psi \ket_{AB} ) =  \sqrt{ 2 ( 1 - \Tr \varrho_B^2 ) }$.
With this definition, the concurrence takes the maximum value, $\conc = 1$, for maximally entangled two-qubit states.

For a mixed state $\rho$, the concurrence is defined as the convex roof \cite{Uhlmann1998} of the pure state concurrence as
\be
{\cal C}(\rho) \,\equiv\, \inf_{p_k, |\psi_k\ket} \sum_k p_k {\cal C}( | \psi_k \ket ),
\label{conc_mix}
\ee
where the infimum is taken over all possible decompositions of $\rho$ into sets of $\{ p_k \}$ and $\{ | \psi_k \ket \}$ with $\rho = \sum_k p_k | \psi_k \ketbra \psi_k |$. 

The advantage of the concurrence is that the analytical formula of ${\cal C}(\rho)$ is known for two-qubit systems \cite{PhysRevLett.80.2245}: 
\be
{\cal C}(\rho) \,=\,
\max(0,\, \eta_1 - \eta_2 - \eta_3 - \eta_4 ) \,\in\, [0, 1],
\ee
where $\eta_i$ are the decreasingly ordered eigenvalues of $\sqrt{ \sqrt{\rho} (\sigma_y \otimes \sigma_y) \rho^*  (\sigma_y \otimes \sigma_y) \sqrt{\rho} }$, obtained in the computational basis.

\subsection{Three-particle entanglement}

The spin state of a multiparticle final state produced from the decay of a massive particle is generally pure, provided it is evaluated at a specific point in momentum phase space.\footnote{Spin and momentum degrees of freedom are generally entangled. The spin state becomes mixed if the momentum degrees of freedom are traced out (i.e., averaged over).}
Accordingly, we focus our analysis on the general pure three-qubit state, $| \Psi \ket \in {\cal H}_{ABC} = {\cal H}_A \otimes {\cal H}_B \otimes {\cal H}_C$.
In the three-qubit system, there are two types of separable states. 
One is called {\it fully separable} state, which can be written as
\be
|\psi\ket_{\rm fs} = | \alpha \ket_A \otimes | \beta \ket_B \otimes | \gamma \ket_C \,.
\label{fs}
\ee
Fully separable states contain no entanglement of any kind.
The second is called {\it bi-separable} states, which are of the types 
\be
| \alpha \ket_A \otimes | \delta \ket_{BC},
~~
| \beta \ket_B \otimes | \delta \ket_{AC},
~~
| \gamma \ket_C \otimes | \delta \ket_{AB},
\ee
where $| \delta \ket_{IJ}$ may not be factorised as $| \delta_1 \ket_I \otimes | \delta_2 \ket_J$.

If the state is neither fully separable nor bi-separable, it is called {\it genuine tripartite entangled} (GTE).
There are two important GTE states: GHZ \cite{PhysRevD.35.3066, GHZ} and W \cite{Dur:2000zz} states. 
The former is defined by
\be
| GHZ \ket = \frac{1}{\sqrt{2}} \left( |000 \ket + | 111 \ket \right),
\label{GHZ}
\ee
while the latter is given as
\be
| W \ket = \frac{1}{\sqrt{3}} \left( |100 \ket + | 010 \ket + | 001\ket \right)\,.
\label{W}
\ee
The importance of these states is appreciated by the fact that any GTE state can be transformed by local invertible operation, $| \Psi \ket \to ({\cal O}_A \otimes {\cal O}_B \otimes {\cal O}_C) | \Psi \ket$, either into $| GHZ \ket$ or $| W \ket$, respectively.  
Since these operations are invertible, they naturally introduce an equivalence relation.  
Namely, all GTE states can be classified into either the GHZ or the W class. 

The generalisation of the Schmidt decomposition to three-qubit systems implies that any three-qubit pure state can be transformed by local unitaries into the canonical form \cite{PhysRevLett.85.1560}
\be
\xi_0 | 0 0 0 \ket + e^{i \varphi} \xi_1 | 100 \ket 
+ \xi_2 | 101 \ket + \xi_3 | 110 \ket + \xi_4 | 111 \ket 
\label{schmidt}
\ee
with $\xi_i \geq 0$, $\sum_i \xi_i^2 = 1$ and $0 \leq \varphi \leq \pi$.  
This means that six real parameters essentially characterise the entanglement and non-local properties of three-qubit pure states. 
Also, it has been shown that $\xi_4 = \varphi = 0$ is required for the W class states, implying that W class states are much rare compared to the GHZ class states.

For three-qubit systems, one can consider three types of entanglement. 
One is an entanglement between two individual particles, say between A and B.
This {\it one-to-one} entanglement can be computed by first tracing out subsystem C and then using formula \eqref{conc_mix}:
\be
{\cal C}_{AB} = {\cal C}(\varrho_{AB}),~~~~
\varrho_{AB} = {\rm Tr}_C ( | \Psi \ket \bra \Psi | )
\label{CAB}~.
\ee
The one-to-one entanglement between A-C (${\cal C}_{AC}$) and B-C (${\cal C}_{BC}$) can also be computed in the similar way.  
These entanglement measures are nonvanishing even if the state is bi-separable. 
For example, ${\cal C}_{AB}$ can be nonzero for $| \gamma \ket_C \otimes | \delta \ket_{AB}$, while it vanishes for $| \alpha \ket_A \otimes | \delta \ket_{BC}$ and $| \beta \ket_B \otimes | \delta \ket_{AC}$.

Another type is an entanglement between one particle and the rest of the system,
known as {\it one-to-other} bipartite entanglement. 
The concurrence between A and the composite subsystem BC is computed as follows.
First, we write the general pure state as
\be
| \Psi \ket 
\,=\,
\sum_{ijk} c_{ijk} | i j k \ket
\,=\,
c_{0 0} | 0 \ket_A \otimes | 0 \ket_{BC} 
+
c_{1 1} | 1 \ket_A \otimes | 1 \ket_{BC} 
\ee
with $| 0 \ket_{BC} = \sum_{jk} c_{0jk} | j k \ket_{BC}$
and $| 1 \ket_{BC} = \sum_{jk} c_{1jk} | j k \ket_{BC}$,
and pretend that it is a two-qubit pure state. 
The concurrence formula for two-qubit pure states \eqref{conc_pure} may then be used:
\be
\conc_{A(BC)} = \sqrt{2 (1 - {\rm Tr} \varrho^2_{BC})},
~~~~
\varrho_{BC} = {\rm Tr}_A ( | \Psi \ket \bra \Psi | )~.
\label{Ci(jk)}
\ee
The one-to-other entanglement for other combinations, ${\cal C}_{B(AC)}$ and ${\cal C}_{C(AB)}$, can be computed in the similar manner. 
The one-to-other entanglement measures may be nonvanishing for bi-separable states. 
For instance, $\conc_{A(BC)}$ is nonvanishing for $| \beta \ket_B \otimes | \delta \ket_{AC}$ and $| \gamma \ket_C \otimes | \delta \ket_{AB}$, but vanishing for $| \alpha \ket_A \otimes | \delta \ket_{BC}$.

The one-to-one and one-to-other entanglement are related by the monogamy relations \cite{Coffman:1999jd,Osborne2006}: 
\bea
{\cal C}^2_{A(BC)} &=& {\cal C}^2_{AB} + {\cal C}^2_{AC} + \tau \,,
\nonumber \\
{\cal C}^2_{B(AC)} &=& {\cal C}^2_{AB} + {\cal C}^2_{BC} + \tau \,,
\nonumber \\
{\cal C}^2_{C(AB)} &=& {\cal C}^2_{AC} + {\cal C}^2_{BC} + \tau \,,
\label{monogamy}
\eea
where $\tau$ is called three-tangle and given by
\be
\tau = 4 \xi_0^2 \xi_4^2 \,,
\ee
where $\xi_i$ are the coefficients in the Schmidt decomposition \eqref{schmidt}.
The first monogamy relation implies that there is a trade-off between A's entanglements with B and C.
For example, when A is maximally entangled with B, A cannot be entangled with C; $\conc_{AB} = 1 \Rightarrow \conc_{AC} = 0$.  
More generally, the sum of the squared one-to-one concurrence measures is bounded from above by the corresponding one-to-other concurrence square: e.g.\ $\conc_{AB}^2 + \conc_{AC}^2 \leq \conc_{A(BC)}^2$.   

For the GHZ state, Eq.\ \eqref{GHZ}, $\tau = 1$.
We therefore have ${\cal C}_{A(BC)} = {\cal C}_{B(AC)} = {\cal C}_{C(AB)} = 1$, while 
${\cal C}_{AB} = {\cal C}_{BC} = {\cal C}_{AC} = 0$.
In other words, any one particle is maximally entangled with the rest of the system, although all particle pairs are individually separable.  
On the other hand, $\tau = 0$ for the W state in Eq.\ \eqref{W}.
One-to-one and one-to-other concurrence values are found to be
${\cal C}_{AB} = {\cal C}_{BC} = {\cal C}_{AC} = \frac{2}{3}$
and 
${\cal C}_{A(BC)} = {\cal C}_{B(AC)} = {\cal C}_{C(AB)} = \frac{2\sqrt{2}}{3}$.

The third type of entanglement is entanglement among genuinely three particles. 
A good genuine tripartite entanglement (GTE) measure should satisfy the following conditions:
(1) invariant under local basis changes,
(2) vanishes for all fully-separable and bi-separable states,
(3) positive for all GTE states,
and (4) non-increasing under LOCC.
A GTE measure satisfying all these criteria has recently been found  \cite{Jin2023} (see \cite{Guo_2022} for alternative construction).
The measure is motivated by the observation that the sum of any two monogamy relations is larger than the remaining one, which can be seen in Eq.\ \eqref{monogamy}.
Since all terms in Eq.\ \eqref{monogamy} are positive, one can consider the square root of each term and obtain 
\be
\conc_{A(BC)} + \conc_{B(AC)} > \conc_{C(AB)} \,,
\ee
and the similar inequality for other combinations. 
These inequalities guarantee that one can always draw a triangle whose three sides are given by the three one-to-other concurrence measures ({\it concurrence triangle}).
\begin{figure}[t!]
\centering
\includegraphics[scale=.3]{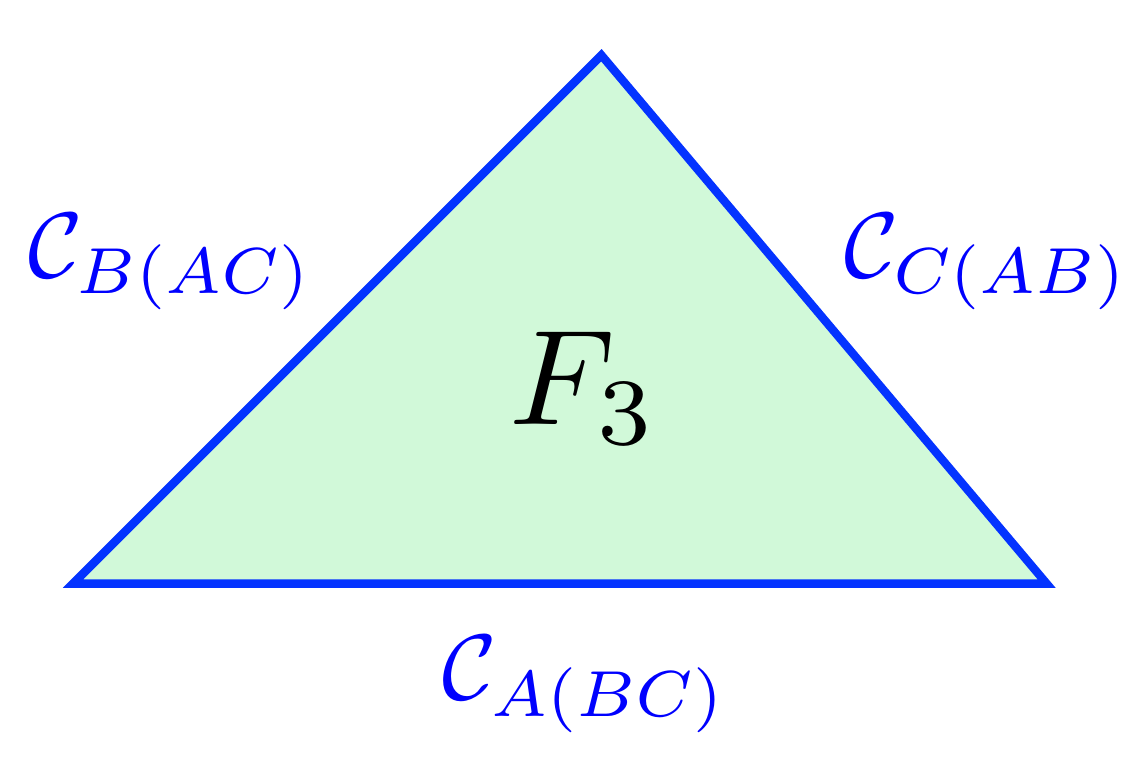}
\caption{\small The concurrence triangle and 
a genuine tripartite  entanglement measure $F_3$.
}
\label{triangle}
\end{figure}
The area of the concurrence triangle gives a GTE measure satisfying all criteria \cite{Jin2023} (see Fig.\ \ref{triangle}).
With Heron's formula, we define the GTE measure 
\be
F_3 \,\equiv\, \frac{4}{\sqrt{3}}
\cdot \sqrt{
Q ( Q - {\cal C}_{A(BC)}) ( Q - {\cal C}_{B(AC)}) ( Q - {\cal C}_{C(AB)}) 
}
\,,
\label{F3}
\ee
with $Q = \frac{1}{2}[{\cal C}_{A(BC)} + {\cal C}_{B(AC)} + {\cal C}_{C(AB)}]$.
With the normalisation factor $\frac{4}{\sqrt{3}}$, $F_3$ takes values between 0 and 1.
For the symmetric case, ${\cal C}_{A(BC)} = {\cal C}_{B(AC)} = {\cal C}_{C(AB)} = {\cal C}$, we have $F_3 = {\cal C}^2$.
One can see that the GHZ state is maximally GTE, $F_3( | GHZ \ket ) = 1$.
On the other hand, for the W state, 
$F_3( | W \ket ) = \frac{8}{9}$; the state is highly entangled but not maximally GTE.   

We stress that all the analyses here concern pure states only. 
The analogous quantities for mixed states are usually very hard to compute (see e.g.\ \cite{PhysRevLett.97.260502, Osterloh:2025wwr}).

\subsection{Bell nonlocality}

Bell-type inequalities distinguish between correlations that can be achieved by local real hidden variable (LRHV) theories and those that cannot.
Typically, the following experiment is considered. 
A pair of particles, A and B, which locally interacted with each other in the past, are spatially separated far apart. 
Two observers, Alice and Bob, who are also separated far apart, have access to each particle. 
Alice can measure one of the two spin components (axes) $A_1$ or $A_2$ of particle A, and Bob can measure one of the two spin components (axes) $B_1$ or $B_2$ of particle B. 
The corresponding outcomes of those spin component measurements are denoted by $a$ and $b$ for Alice and Bob, respectively. 
We assume the outcomes are either $+1$ or $-1$; $a,b \in \{+1, -1\}$.
The experiment is repeated many times. 
Each time, Alice and Bob can choose which spin components they measure.
After collecting a large number of measurement results, we compute the correlations $\bra A_i B_j \ket$.

If the theory is local, Alice’s measurement outcome cannot be influenced by Bob’s choice of measurement axes and vice versa.
If the theory is real, their measurement outcomes are predetermined before the act of measurement.
Under the assumptions of locality and realism, their measurement outcomes may be described by a set of common hidden variables $\lambda$ with the probability distribution $P(\lambda)$.
In particular, the joint conditional probability that Alice and Bob measure $A_i$ and $B_j$ then obtain the outcomes $a$ and $b$ may be expressed in terms of $\lambda$ as
\be
P(a,b|A_i,B_j) \,=\, \sum_\lambda P_\lambda
{\cal A}_\lambda(a | A_i)
{\cal B}_\lambda(b | B_j)\,.
~~~~\cdots~~({\rm LRHV})
\label{prob_HV}
\ee
The function ${\cal A}_\lambda(a | A_i)$ can be interpreted as Alice's conditional probability of finding her outcome $a$ for given $\lambda$ and $A_i$.
The analogous interpretation follows for Bob's function ${\cal B}_\lambda(b | B_i)$.
The form of Eq.\ $\eqref{prob_HV}$ encompasses the deterministic, i.e., realism picture. 
One can always assume that Alice's probability of finding outcome $a$, for given $\lambda$ and $A_i$, is either 0 or 1, invoking a probability function $\overline {\cal A}_\lambda(a| A_i) \in \{ 0, 1 \}$. 
The corresponding deterministic probability for Bob is denoted by $\overline {\cal B}_\lambda(b| B_j) \in \{ 0, 1 \}$.
The same conditional probability \eqref{prob_HV} can be described as $P(a,b|A_i,B_j) \,=\, \sum_\lambda \overline P_\lambda
\overline {\cal A}_\lambda(a | A_i)
\overline {\cal B}_\lambda(b | B_j)$, 
with the relations 
$\overline P_\lambda = P_\lambda Q_A(\lambda) Q_B(\lambda)$,
${\cal A}_\lambda = Q_A(\lambda) \overline {\cal A}_\lambda$
and 
${\cal B}_\lambda = Q_B(\lambda) \overline {\cal B}_\lambda$ with appropriate functions $Q_A(\lambda)$ and $Q_B(\lambda)$.

In  general, Bell-type inequalities have a form
\be
| \bra \bell \ket_{\rm LRHV} | \, \leq \, \bell_{\rm LRHV}^{\rm bound}\,,
\ee
where $\bra \bell \ket_{\rm LRHV}$ is the expectation value of some observable ${\cal B}$ in a given LRHV theory, while
$\bell_{\rm LRHV}^{\rm bound}$ is the upper bound on $\bra \bell \ket_{\rm LRHV}$ within this theory. 
One of the well-known Bell-type observables is Clauser-Horne-Shimony-Holt (CHSH) \cite{Clauser:1969ny} observable 
\be
{\cal B}_{\rm CHSH} = A_1 B_1 + A_1 B_2 + A_2 B_1 - A_2 B_2\,. 
\label{CHSH}
\ee
The bound by LRHV theories can be estimated by invoking locality, writing ${\cal B}_{\rm CHSH} = A_1 (B_1 + B_2) + A_2( B_1 - B_2)$. By the assumption of realism,
when the outcomes of $B_1$ and $B_2$ being predetermined, either $(B_1+ B_2)$ or $(B_1-B_2)$  has to vanish. 
The largest value for the other term is 2.  
Therefore, one can find the bound of Bell-type inequality by LRHV theory to be,
\be
| \bra {\cal B}_{\rm CHSH} \ket_{\rm LRHV} | \leq 2 \,.
\ee

This bound is violated by quantum mechanics.
For example, by taking $A_1$ and $A_2$ to be $\sigma_z$ and $\sigma_x$ and $B_1$ and $B_2$ be $(\sigma_z \pm \sigma_x)/\sqrt{2}$, one finds 
\be
\bra {\cal B}_{\rm CHSH} \ket_{|\psi \ket_{00}} = 2 \sqrt{2},
\ee
with $|\psi \ket_{00}$ being the spin-0 singlet state, $|\psi \ket_{00} = (|01\ket - |10\ket)/\sqrt{2}$.
The quantum states which violate Bell-type inequality are called Bell-nonlocal.

In fact, $2 \sqrt{2}$ is the upper bound in the quantum theory \cite{Tsirelson}.  
This can be seen by noting 
\be
({\cal B}_{\rm CHSH})^2 = 4 - [A_1, A_2] [B_1, B_2]\,.
\ee
With $[\sigma_i, \sigma_j] = 2 \epsilon_{ijk} \sigma_k$, we see that the maximum of the right-hand-side is 8, which leads to the quantum bound $2 \sqrt{2}$.
The presence of the quantum bound implies that the Bell-type observables are also important in testing quantum mechanics experimentally. 
For example, {\it if a violation of quantum bound is experimentally confirmed, then quantum mechanics is falsified.}
In fact, it has been known that a local probability distribution $P(a,b|A_i, B_j)$ leads to a violation of the quantum bound \cite{PopescuRohrlich94}. 
This class of local distributions is called No-signalling distributions. 
Within such distributions, the algebraic maximum 4 of \eqref{CHSH} can be reached.\footnote{Since $a,b \in \{ +1, -1 \}$, the expectation value of each term of Eq.\ \eqref{CHSH} is bounded by 1.
Therefore, $| \bra {\cal B}_{\rm CHSH} \ket|$ has the algebraic upper bound 4. 
We have $\bra \bell_{\rm CHSH} \ket_{\rm NS} \leq 4$
for general No-signalling distributions.}
We recall the No-signalling statistics that saturate this bound in Appendix \ref{sec:NS}.

Violation of Bell-type inequality is related to the entanglement. 
By definition, separable states have a factorised form \eqref{sep}. 
For this state, the joint conditional probability is factorised as
\bea
P(a,b|A_i, B_j) &=& \Tr \left[ \rho^{\rm sep} 
 \left( M(a| A_i) \otimes M(b| B_j) \right) \right]
 \nonumber \\
 &=&
 \sum_k p_k \Tr \left[ \rho^A_k 
 M(a| A_i) \right]
 \Tr \left[ \rho^B_k 
 M(b| B_j) \right]\,,
\eea
where $M(a| A_i)$ and $M(b| B_i)$ are the measurement projection operators for Alice and Bob, respectively. 
The latter expression coincides with the defining probability distribution of the LRHV theories. 
This implies that separable quantum states cannot violate the LRHV bound, meaning that the set of all Bell-nonlocal states is a subset of all entangled states, i.e.,
Bell-nonlocal $\subset$ Entangled.  
On the other hand, however, for bipartite pure states, all entangled states violate the CHSH inequality \cite{Capasso:1973wt, GISIN1991201,GISIN199215}.

\subsection{Three-particle non-locality}

In a three-particle system, the Bell-type experiment is naturally extended by introducing a third particle, C, and its corresponding observer, Charlie.
Charlie performs spin measurements along two axes, $C_1$ and $C_2$, with the measurement outcome denoted by $c \in \{ +1, -1 \}$.
The correlations $\langle A_i B_j C_k \rangle$ are computed by repeating the experiment multiple times.

Two types of local-real theories exist in the three-particle system \cite{Cereceda_2002}. 
One is {\it fully} local-real (FLR) theories, where no nonlocal correlation exists among any pair of particles and between any particle and the rest of the system.
In such theories, the joint probability distribution fully factorises as
\be
P(a,b,c|A_i,B_j,C_k) = \sum_\lambda P_\lambda 
{\cal A}_\lambda(a|A_i)
{\cal B}_\lambda(b|B_j)
{\cal C}_\lambda(c|C_k) ~~~\cdots~~({\rm FLR})
\label{PFLR}
\ee

One can test this theory with the Mermin observable \cite{PhysRevLett.65.1838}
\be
{\cal B}_{\rm M} = A_1 B_1 C_2 + A_1 B_2 C_1  + A_2 B_1 C_1 - A_2 B_2 C_2\,.
\label{Marmin}
\ee
The observable is symmetric under the permutation of the three qubits.
Notice that the product of all four terms results in $(A_1 B_1 C_1)^2(A_2 B_2 C_2)^2$.  
In the fully local-real theory, where all outcomes are predetermined before the measurements, this product is 1.
Therefore, if the first three terms have the same sign, the last term (without the minus sign) must also have the same sign.  
This means at least one pair of terms must be cancelled, leading to the Mermin inequality \cite{PhysRevLett.65.1838}
\be
| \bra {\cal B}_{\rm M} \ket_{\rm FLR} | \,\leq\, 2.
\label{MLRbound}
\ee
Since fully-separable quantum states always lead to the probability distribution of the form \eqref{PFLR}, the quantum state must be entangled, i.e.\ ${\cal C}_{i(jk)} > 0$ for some $i,j,k \in \{ A, B,C\}$, to violate this bound.

It has been shown that quantum mechanics can violate the FLR limit.
In fact, quantum mechanics can saturate the algebraic maximum 
\be
| \bra {\cal B}_{\rm M} \ket_{\rm QM} | \,\leq\, 4.
\label{MQMbound}
\ee
In particular, the GHZ state can saturate the quantum bound \cite{PhysRevLett.65.1838}.  
The saturation of the algebraic maximum within quantum mechanics implies that the Mermin observable cannot be used to test quantum mechanics like the CHSH observable.

The other local-real theory is a hybrid type, where two of the particles are nonlocally correlated, but that subsystem is separated from the third particle \cite{PhysRevD.35.3066, PhysRevLett.88.170405, PhysRevA.70.060101}. 
The most general joint probability distributions in such {\it bipartite} local-real (BLR) theories are written as
\bea
&& P(a,b,c|A_i,B_j,C_k) \,=\, p_1 
\sum_\lambda P_\lambda 
{\cal A}_\lambda(a|A_i)
\overline {\cal A}_\lambda(b,c|B_j, C_k)
\label{PBLR}
\\
&&~~
+\,
p_2 \sum_\lambda P_\lambda 
{\cal B}_\lambda(b|B_j)
\overline {\cal B}_\lambda(a,c|A_i, C_k)
+
p_3 \sum_\lambda P_\lambda 
{\cal C}_\lambda(c|C_k)
\overline {\cal C}_\lambda(a,b|A_i, B_j)
\,,
~~~\cdots~~({\rm BLR})
\nonumber 
\eea
with $p_i \geq 0$ and $\sum_i p_i = 1$.
This probability distribution is more general than that in Eq.\ \eqref{PFLR}.
If the probability distribution of this type cannot explain the results of correlation measurements, we say that there is genuine three-partite non-locality in the system.

The bipartite local-real theory can be tested using the Svetlichny observable \cite{PhysRevD.35.3066, PhysRevA.70.060101}
\be
{\cal B}_{\rm S}
= A_1 B_1 C_1 + A_1 B_1 C_2 + A_1 B_2 C_1 + A_2 B_1 C_1
- A_2 B_2 C_2 - A_2 B_2 C_1 - A_2 B_1 C_2 - A_1 B_2 C_2
\label{Svetlichny}
\ee
with the bound 
\be
| \bra {\cal B}_{\rm S} \ket_{\rm BLR} | \,\leq \, 4\,.
\label{SLRbound}
\ee
This bound can be understood as follows.
In the calculation of the expectation value, the probability distribution appears linearly; $\bra A_i B_j C_k \ket_{\rm BLR} = \sum_{a,b,c} P(a,b,c| A_i, B_j, C_k) A_i B_j C_k$.
The three terms in Eq.\ \eqref{PBLR} can, therefore, be considered separately and then added together.  
In the first term, we can interpret that particle A is described independently from the rest.
For this contribution, we rewrite the Svetlichny observable as 
\be
{\cal B}_{\rm S} \,=\, (A_1 + A_2) B_1 C_1 
+ (A_1 - A_2) B_1 C_2
+ (A_1 - A_2) B_2 C_1
- (A_1 + A_2) B_2 C_2
\ee
Within this contribution, we can assume the measurement outcomes of $A_1$ and $A_2$ are predetermined before the act of Alice's measurement and take values of either $+1$ or $-1$. 
Two of the above terms must, therefore, vanish. 
Since the largest value of each term is 2, we understand the maximum contribution to ${\cal B}_{\rm S}$ from this part of the probability distribution is $4 p_1$.
The analogous argument follows for the other two contributions in Eq.\ \eqref{PBLR}.
Adding all three contributions gives the upper bound $4 (p_1 + p_2 + p_3) = 4$. 
Since bi-separable quantum states necessarily give the bipartite local-real distribution \eqref{PBLR}, the quantum state must be genuinely tripartite entangled to violate the Svetlichny's BLR bound \eqref{SLRbound}. 
As is the Mermin observable, the Svetlichny observable is symmetric under the permutation of the three qubits.

It has been shown that quantum mechanics can violate the BLR bound.
Interestingly, quantum mechanics has a non-trivial bound 
\be
| \bra {\cal B}_{\rm S} \ket_{\rm QM} | \,\leq \, 4 \sqrt{2}\,.
\ee
For example, the quantum limit $4 \sqrt{2}$ is reached by the GHZ state by taking e.g.\ $(A_1, A_2) = (\sigma_x, \sigma_y)$, $(B_1, B_2) = (\sigma_x \pm \sigma_y)/\sqrt{2}$ and $(C_1, C_2) = (\sigma_x, \sigma_y)$.
The quantum limit is smaller than the algebraic maximum of \eqref{Svetlichny}, which is 8, achieved when all eight terms of in Eq.\ \eqref{Svetlichny} take the same value $+1$ (or $-1$). 
This means the Svetlichny observable is effective in testing quantum mechanics.

An intriguing question is whether or not all non-FLR distributions violate the Mermin inequality \eqref{MLRbound}. 
It has been shown that the answer to this question is negative \cite{laskowski}. 
In fact, there exists a tighter inequality than Eq.\ \eqref{MLRbound}.
Such an inequality can be obtained by adding two more spin measurement axes to Alice and Bob, denoted by $A_3, A_4$ and $B_3, B_4$, respectively \cite{WU2003262}.
The so-called tight $4 \times 4 \times 2$ observable is defined by \cite{WU2003262, laskowski}
\be
{\cal B}_{442} = 
\left[
A_{1} (B_{1}+B_{2})+A_{2} (B_{1}-B_{2}) 
\right] (C_{1}+C_{2}) 
\,+\, 
\left[
A_{3}(B_{3}+B_{4})+A_{4}(B_{3}-B_{4})
\right]
(C_{1}-C_{2})\,. 
\label{B442}
\ee

For the fully local-real theory, one of the two terms vanishes since the outcomes of $C_1$ and $C_2$ are predetermined and either $+1$ or $-1$.
Within the non-vanishing term, one of the two terms in the squared bracket vanishes because of the same argument for $B_1$ and $B_2$.
In the end, only one term of the form $A_i (B_1 \pm B_2)(C_1 \pm C_2)$ survives, which leads to the upper bound \cite{laskowski}
\be
| \bra {\cal B}_{442} \ket_{\rm FLR} | \, \leq \, 4 \,.
\label{B4_FLR}
\ee
In fact, this bound is tight \cite{laskowski}, meaning that all non-FLR correlations violate the above inequality. 

On the other hand, quantum mechanics has a non-trivial bound (see the appendix)
\be
| \bra {\cal B}_{442} \ket_{\rm QM} | \, \leq \, 8\,.
\label{B4_QM}
\ee
Note that if all terms in Eq.\ \eqref{B442} were conveniently adjusted to take values of $+1$ or $-1$, the quantity $| \bra {\cal B}^\prime_{442} \ket |$ would saturate its algebraic bound of 16. 
It can be shown (see Appendix \ref{sec:NS}) that this bound is achievable within the No-signalling framework. 
Understanding this type of bound is important, as a violation in space-like configurations would require a careful examination/reconsideration of relativistic causality and quantum mechanics.

Unlike the Mermin and Svetlichny observables, ${\cal B}_{442}$ is not symmetric under the permutation of qubits because only Charlie has two measurement axes (it is symmetric under the exchange of Alice and Bob, though).
By giving the four measurement axes to Charlie and assigning the two measurement axes to Alice or Bob, one can define similar observables as
\bea
{\cal B}_{244} &=& 
\left[
C_{1} (B_{1}+B_{2})+C_{2} (B_{1}-B_{2}) 
\right] (A_{1}+A_{2}) 
\,+\, 
\left[
C_{3}(B_{3}+B_{4})+C_{4}(B_{3}-B_{4})
\right]
(A_{1}-A_{2})\,,
\nn \\
{\cal B}_{424} &=& 
\left[
A_{1} (C_{1}+C_{2})+A_{2} (C_{1}-C_{2}) 
\right] (B_{1}+B_{2}) 
\,+\, 
\left[
A_{3}(C_{3}+C_{4})+A_{4}(C_{3}-C_{4})
\right]
(B_{1}-B_{2})\,. 
\nn \\
\label{B244}
\eea
These observables also have the properties corresponding to Eqs.\ \eqref{B4_FLR} and \eqref{B4_QM}.
In our numerical analysis for the vector-type four-fermion interaction in section \ref{sec:analysis_vec}, we use the 
symmetrically extended value defined by maximum over the three tight observables:
\be
{\cal B}_{442}^{\rm sym} \,\equiv\, \max \left[
{\cal B}_{244},
{\cal B}_{424},
{\cal B}_{442}
\right]\,.
\label{B4sym}
\ee

\begin{figure}[t!]
\centering
\includegraphics[scale=.33]{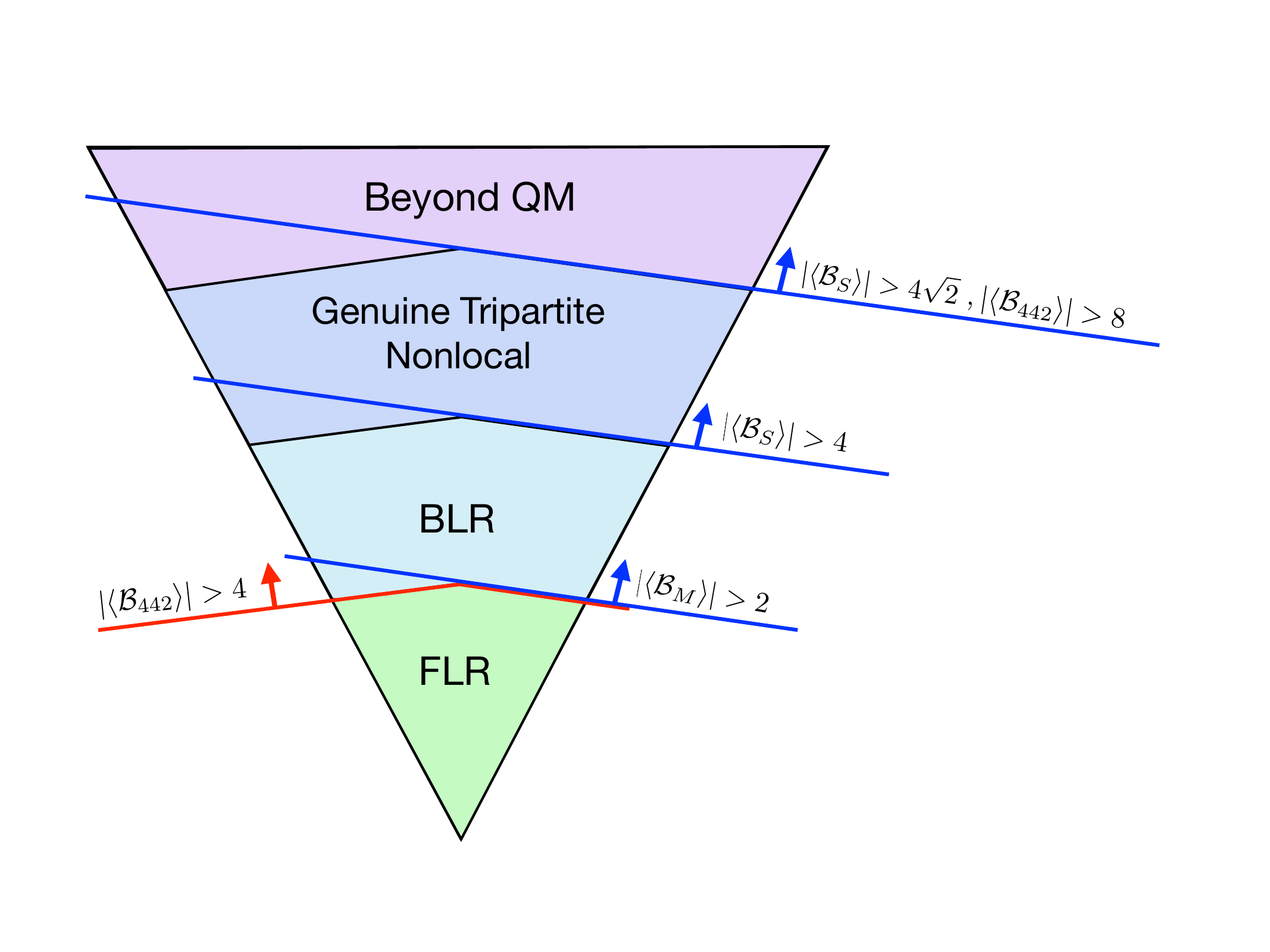}
\caption{\label{fig:diag}
\small Different types of non-localities and their detections in the three-qubit system. 
It is seen that the 442 inequality functional with the value 4 is tight, that is, it describes the boundary of the FLR polytope. 
}
\end{figure}

We summarise different types of nonlocal correlations and their detection with the Bell-type inequalities in Fig.\ \ref{fig:diag}. 

\subsection{Optimisation of the measurement axes}

The Bell-type operators generally depend on the measurement axes, and Bell-type inequalities hold for all choices of measurement axes. 
To detect non-locality efficiently, all measurement axes should be optimised so that the expectation value of the Bell-type observable is maximised for a given state.   

Let $\vec A_I$, $\vec B_I$ and $\vec C_J$ be the unit vectors pointing to the spin measurement axes of Alice, Bob and Charlie, respectively, where 
$I \in \{ 1,2 ~(3,4) \}$,\footnote{This notation means that we shall consider either set $\{ 1,2 \}$ or $\{ 1,2,3,4 \}$.} and $J \in \{ 1,2 \}$ label different choices of measurement axes. 
For given measurement axes ($\vec A_I$, $\vec B_I$ and $\vec C_J$), the Bell-type operators mentioned above are written as 
\be
{\cal B}_{\rm M} = \sum_{ijk} (\overline {\cal B}_{\rm M})_{ijk} [\sigma_i \otimes \sigma_j \otimes \sigma_k],
~~~~
{\cal B}_{\rm S} = \sum_{ijk} (\overline {\cal B}_{\rm S})_{ijk} [\sigma_i \otimes \sigma_j \otimes \sigma_k],
~~~~
{\cal B}_{442} = \sum_{ijk} (\overline {\cal B}_{442})_{ijk} [\sigma_i \otimes \sigma_j \otimes \sigma_k],
\ee
where $\sigma_i$ ($i=1,2,3$) are the Pauli matrices and
the rank-3 tensors are defined as
\be
[ {\overline {\cal B}}_{\rm M} ]_{ijk} 
~=~
[ \vec A_1 ]_i [ \vec B_1 ]_j [ \vec C_2 ]_k +
[ \vec A_1 ]_i [ \vec B_2 ]_j [ \vec C_1 ]_k +
[ \vec A_2 ]_i [ \vec B_1 ]_j [ \vec C_1 ]_k -
[ \vec A_2 ]_i [ \vec B_2 ]_j [ \vec C_2 ]_k 
\label{BbarM}
\ee
\bea
[ \overline {\mathcal B}_{\rm S} ]_{ijk} 
&=&
[ \vec A_1 ]_i [ \vec B_1 ]_j [ \vec C_1 ]_k +
[ \vec A_1 ]_i [ \vec B_1 ]_j [ \vec C_2 ]_k +
[ \vec A_1 ]_i [ \vec B_2 ]_j [ \vec C_1 ]_k +
[ \vec A_2 ]_i [ \vec B_1 ]_j [ \vec C_1 ]_k 
\nn \\
&-&
[\vec A_2]_i [\vec B_2]_j [\vec C_2]_k -
[\vec A_2]_i [\vec B_2]_j [\vec C_1]_k -
[\vec A_2]_i [\vec B_1]_j [\vec C_2]_k -
[\vec A_1]_i [\vec B_2]_j [\vec C_2]_k 
\label{BbarS}
\eea
\bea
[ \overline {\cal B}_{442} ]_{ijk} 
&=&
\left(
[ \vec A_{1} ]_i [ \vec B_{1}+ \vec B_{2} ]_j + [ \vec A_{2} ]_i [ \vec B_{1} - \vec B_{2} ]_j 
\right) [ \vec C_{1} + \vec C_{2} ]_k
\nn \\
&+& 
\left(
[ \vec A_{3} ]_i [ \vec B_{3} + \vec B_{4} ]_j + [ \vec A_{4}]_i [ \vec B_{3} - \vec B_{4}]_j
\right)
[ \vec C_{1} - \vec C_{2} ]_k\,,
\label{Bbar4}
\eea
where $[\vec A_I]_i$, $[\vec B_I]_j$ and $[\vec C_J]_k$ ($i,j,k \in \{ 1,2,3 \}$) are the components of the measurement axis vectors.

A general three-qubit state, $\rho$, 
can be expanded as
\be
\rho = \frac{1}{8} \sum_{\m, \n, \r} T_{ {\m} {\n} {\r}} \, \big[ \sigma_{\m} \otimes \sigma_{\n} \otimes \sigma_{\r}
\big],
\ee 
with $\m, \n,\r \in\{ 0,1,2,3 \}$,
$T_{ {\m} {\n} {\r}} \in {\mathbb R}$ and $\sigma^0 = {\mathbf 1}$.
The Bloch vectors of qubits A, B and C
correspond to 
$T_{i00}$,
$T_{0i0}$
and
$T_{00i}$,
respectively,
with $i,j,k \in \{ 1,2,3\}$.
The two-qubit correlation 
(AB), (AC) and (BC)
are given by
$T_{ij0}$,
$T_{i0j}$
and
$T_{0ij}$, respectively.
Finally, 
\be
T_{ijk} = {\rm Tr}\big( 
\rho \cdot [ \sigma_i \otimes \sigma_j \otimes \sigma_k ]
\big)
\label{Tensor}
\ee
is the three-qubit correlation. 

The expectation values of Bell-type operators are given by
\bea
&& \bra {\cal B}_{\rm M} \ket_\rho \,=\,
{\rm Tr} \big[ {\cal B}_{\rm M} \rho \big]
\,=\,
\sum_{i,j,k} [\overline {\cal B}_{\rm M}]_{ijk} T_{ijk},
\nn \\ 
&& \bra {\cal B}_{\rm S} \ket_\rho \,=\,
{\rm Tr} \big[ {\cal B}_{\rm S} \rho \big]
=
\sum_{i,j,k} [\overline {\cal B}_{\rm S}]_{ijk} T_{ijk},
\nn \\ 
&& \bra {\cal B}_{442} \ket_\rho \,=\,
{\rm Tr} \big[ {\cal B}_{442} \rho \big]
=
\sum_{i,j,k} [\overline {\cal B}_{442}]_{ijk} T_{ijk},
\label{BBB}
\eea
Only the three-qubit correlation $T_{ijk}$ appears in these quantities. 

For a given quantum state $\rho$ (or equivalently a $T_{ijk}$),
we would like to optimise the measurement axis vectors, 
$\vec A_I$, $\vec B_I$ and $\vec C_J$, such that the Bell-type observable is maximised. 
For the Mermin and Svetlichny observables, the expectation can be expressed as
\be
\bra {\cal B}_{{\rm M}/{\rm S}} \ket_\rho
= 
(\vec A_1 \cdot \vec D^{(1)}_{{\rm M}/{\rm S}})
+
(\vec A_2 \cdot \vec D^{(2)}_{{\rm M}/{\rm S}})\,,
\ee
with
\be
[\vec D^{(1)}_{\rm M}]_i \,\equiv\, \sum_{j,k}
T_{ijk} 
\left( [ \vec B_1 ]_j [\vec C_2]_k + [ \vec B_2 ]_j [\vec C_1]_k \right)\,,
\ee
\be
[\vec D^{(2)}_{\rm M}]_i \,\equiv\, \sum_{j,k}
T_{ijk} \left( [ \vec B_1 ]_j [\vec C_1]_k - [ \vec B_2 ]_j [\vec C_2]_k \right)\,,
\ee
\be
[\vec D^{(1)}_{\rm S}]_i \,\equiv\, \sum_{j,k}
T_{ijk} \left( 
[ \vec B_1 ]_i [ \vec C_1 ]_k +
[ \vec B_1 ]_i [ \vec C_2 ]_k +
[ \vec B_2 ]_i [ \vec C_1 ]_k -
[\vec B_2]_i [\vec C_2]_k 
\right)\,,
\ee
\be
[\vec D^{(2)}_{\rm S}]_i \,\equiv\, \sum_{j,k}
T_{ijk} \left( 
[ \vec B_1 ]_i [ \vec C_1 ]_k -
[\vec B_2]_i [\vec C_2]_k -
[\vec B_2]_i [\vec C_1]_k -
[\vec B_1]_i [\vec C_2]_k 
\right)\,.
\ee
The expectation values $\bra {\cal B}_{\rm M} \ket_\rho$
and $\bra {\cal B}_{\rm S} \ket_\rho$ are optimised by taking $\vec {A}_1$ and $\vec {A}_2$ aligned with 
$\vec {D}^{(1)}_{ {\rm M}/{\rm S} }$ and $\vec {D}^{(2)}_{ {\rm M}/{\rm S} }$, respectively; 
$\vec A_1 = \vec D^{(1)}_{ {\rm M}/{\rm S} } /| \vec D^{(1)}_{ {\rm M}/{\rm S} } |$ and 
$\vec A_2 = \vec D^{(2)}_{ {\rm M}/{\rm S} } /| \vec D^{(2)}_{ {\rm M}/{\rm S} } |$.
The resulting values are given by
\be
\bra {\cal B}_{ {\rm M}/{\rm S} } \ket_\rho = \left| \vec D^{(1)}_{ {\rm M}/{\rm S} } \right| + \left| \vec D^{(2)}_{ {\rm M}/{\rm S} } \right|\,.
\label{B2}
\ee
In the following analysis, we numerically optimise four unit vectors, $\vec B_{1/2}$ and $\vec C_{1/2}$, to maximize the Mermin and Svetlichny observables expressed in Eq.\ \eqref{B2}.

For the tight $4 \times 4 \times 2$ observable, the optimisation is more involved as it has ten independent measurement axes, $\vec A_I$, $\vec B_I$ ($I \in \{ 1,\cdots,4\}$) and $\vec C_J$ ($J \in \{ 1,2 \}$).
We developed a semi-analytical approach to optimise the tight $4 \times 4 \times 2$ observable, which is detailed in Appendix \ref{sec:app}.     
In this approach, the $\bra \B4 \ket$ is optimised analytically for all degrees of freedom except for one polar angle and two azimuthal angles. 
The numerical optimisation can be performed for the remaining three degrees of freedom, which is relatively straightforward.

\section{Three-Body Kinematics and Spin States}  
\label{sec:kin}

We consider a three-body decay, $X \to A B C$,
at the rest frame of particle $X$.
We assume all particles are spin-1/2 fermions\footnote{Strictly speaking, particle $B$ in the final state is antifermion.} and particle $X$ has the mass $m$,
while the other particles are massless\footnote{
The assumption that the final state fermions are massless greatly simplifies the spin structure of the three-fermion final state. However, measuring the polarisation of truly massless fermions poses significant experimental challenges. In practice, the final state fermions can be treated as effectively massless, meaning their masses are negligible compared to the characteristic mass scale of the decaying particle $X$. This approximation holds, for instance, in the top quark decay channels $t \to b \tau^+ \nu$ and $t \to b \bar s c$, where the fermion masses are small relative to $m_t$. Polarisation measurements of the $\tau$ lepton \cite{Hagiwara:1989fn, L3:1994hzc} and the $b$- and $c$-quarks \cite{Kats:2023zxb, ALEPH:1995aqx, OPAL:1998wmk, DELPHI:1999hkl} in this highly boosted regime have been both theoretically explored and experimentally performed.
On a different note, Ref.\cite{Abel:1992kz} highlights the difficulty of testing quantum non-locality at colliders via Bell-type inequalities, primarily due to the experimental inaccessibility of non-commuting observables in the final state. A detailed assessment of the experimental reconstructibility of the quantities shown in Fig.\ref{fig:diag} is left to future work.}.
We specify the momenta of the final state particles ($p_A$, $p_B$, $p_C$) as follows.
First, we take the $z$-axis in the direction of ${\bf p}_A$:
$p_A^\m = p_A(1,0,0,1)$.
The $x$ and $y$ axes are chosen such that the $y$ axis is perpendicular to the decay plane and the $p_B$ has a positive $x$ component. 
The opening angles between A-B and A-C are denoted by $\theta_B$ and $\theta_C$ ($0 \leq \theta_B, \theta_C \leq \pi$), respectively. 
With these angles, we can write
$p_B^\m = p_B(1, \sin \theta_B , 0, \cos \theta_B)$
and
$p_C^\m = p_C(1, -\sin \theta_C , 0, \cos \theta_C)$
with $0 \le \theta_B, \theta_C \le \pi$.

The energy-momentum conservation gives the constraints
\bea
p_A + p_B + p_C &=& m\,,
\nonumber \\
p_A + p_B \cos \theta_B + p_C \cos \theta_C &=& 0\,,
\nonumber \\
p_B \sin \theta_B - p_C \sin \theta_C &=& 0\,.
\eea
The solution is
\be
p_A = - \, m D \sin(\theta_B + \theta_C)\,,
~~~
p_B = m D \sin \theta_C\,,
~~~
p_C = m D \sin \theta_B \,,
\label{psol}
\ee
with
$D = [\sin \theta_B + \sin \theta_C -
\sin(\theta_B + \theta_C) ]^{-1}$.
In this way, the kinematics is completely determined by specifying two decay angles $\theta_B$ and $\theta_C$. 
The positivity of $p_A$ in Eq.\ \eqref{psol} restricts the sum of these angles in the range, $\pi \le \theta_B + \theta_C \le 2 \pi$.

\begin{figure}[t!]
\centering
\includegraphics[scale=.4]{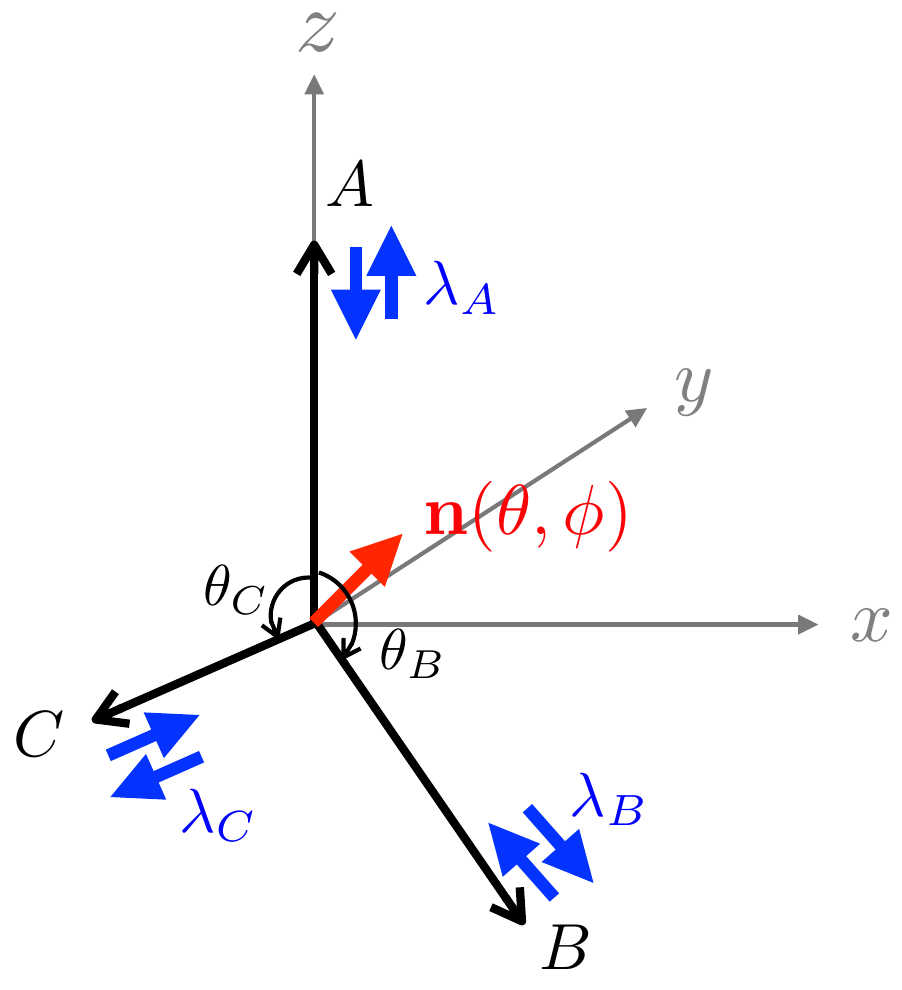}
\caption{\small The momentum and spin configuration.
The momentum of A is fixed to the $z$-direction.
At the rest frame of X, the decay plane is aligned with the $x$-$z$ plane and the two opening angles, A-B and A-C, are given by $\theta_B$ and $\theta_C$, respectively,
with $0 \le \theta_B, \theta_C \le \pi$ and $\pi \le \theta_B + \theta_C \le 2 \pi$.
The spin direction of X is given by
${\bf n}=(\sin \theta \cos\phi, \sin \theta \sin\phi, \cos\theta)$.
}
\label{angle}
\end{figure}

We denote the helicities of outgoing particles by $\lam_A,\lam_B$ and $\lam_C$ and the spin polarisation of the initial particle by ${\bf n} = (\sin \theta \cos\phi, \sin \theta \sin\phi, \cos\theta)$  (see Fig.\ \ref{angle}).
We write $| {\bf n} \ket$ for the asymptotic state of the initial state with the polarization
${\bf n}$ at time $t = - \infty$.
This state is evolved with the $S$-matrix into an ``out'' state at $t = + \infty$.  
The state is then expanded in terms of the complete set of the Fock space at $t = + \infty$
\be
\hat {\bf 1} = \sum_{f}
\left[ 
\left(
\prod_{i \in f} \int  
d \Pi_i
\right) | f \ketbra f | \right]\,,~~~~
d \Pi_i = \frac{d^3 {\bf p}_i}{(2 \pi)^3 2 E_i}  \,,
\label{compset}
\ee
where $f$ denotes single and multiparticle final states with fixed polarisations and $i$ is a particle in $f$. 
We select a particular decay mode, $X \to ABC$ 
\be
S | {\bf n} \ket \,= \,
\sum_{f} 
\left[ \left(
\prod_{i \in f} \int  
d \Pi_i \right)
| f \ketbra f | \right]
S | {\bf n} \ket
\, \ni \,
\sum_{\lambda} \int d \Pi_{\rm ABC} {\cal M}^{\bf n}_{\lambda, p} | {\bf \lam}, p \ket,
\ee
where $\lambda$ and $p$ represent the helicities and momenta of three outgoing particles collectively, i.e.\ $\lambda = \{ \lam_A, \lam_B, \lam_C\}$
and $p = \{ p_A^\mu,p_B^\mu,p_C^\mu \}$.
The Lorentz invariant three-particle phase space is represented as 
\be
d \Pi_{\rm ABC} = \frac{d^3 {\bf p}_A d^3 {\bf p}_B d^3 {\bf p}_C}{(2 \pi)^9 8 E_A E_B E_C}
(2 \pi)^4 \delta^4( p_X - p_A - p_B - p_C )\,,
\ee
and ${\cal M}^{\bf n}_{\lambda, p}$ is the matrix element, which is related to the transition amplitude 
$
\bra \lambda, p | i {\cal T} | {\bf n} \ket 
= i {\cal M}^{\bf n}_{\lambda, p} (2 \pi)^4 \delta^4 ( p_X - p_A - p_B - p_C ) 
$
with ${\cal T}$ being the $T$-matrix defined as $S = 1 + i {\cal T}$.

In the following analysis, we are interested in the spin state at a given phase space point
\be
| \Psi \ket \,\propto\,  \sum_{\lam_A,\lam_B,\lam_C} 
{\cal M}_{ \lam_A,\lam_B,\lam_C}^{\bf n} | \lam_A,\lam_B,\lam_C \ket\,.
\ee
In this expression, we suppressed the momentum label in the final state kets and dropped the common proportionally factor $d \Pi_{\rm ABC}$.
This is a pure state of three qubits, in which, after properly normalising, all mathematical formulae introduced in the previous sections can be used.

\section{Non-locality and Entanglement in Three-Body Decays}
\label{sec:analysis}

\subsection{Four-fermion Interactions}

There are $16$ non-redundant Lorentz structures formed from bilinear combinations of Dirac spinors $\bar{\psi} \Gamma \psi$ with
\begin{equation}
\Gamma = \left \{ {\bf1 },\gamma^5, \gamma^\mu, \gamma^\mu \gamma^5, \sigma^{\mu \nu} \right \}~,
\end{equation}
where $\g^\m$ is the Dirac $\gamma$ matrices, $\g^5 \equiv i \g^0 \g^1 \g^2 \g^3$
and $\sigma^{\m \n} \equiv \frac{i}{2} [\g^\m, \g^\n]$.
We combine two of these fermion bilinears to form Lorentz invariant four-fermion operators.
Such operators are classified into the following three types:
\begin{description}
\item[Scalar:]
\be
[ \bar \psi_A (c_S + i c_A \g_5) \psi_X ]
[ \bar \psi_C (d_S + i d_A \g_5) \psi_B ]\,,
\label{Lscalar}
\ee

\item[Vector:]
\be 
[ \bar \psi_A \gamma_\m (c_L P_L + c_R P_R) \psi_X ]
[ \bar \psi_C \gamma^\m (d_L P_L + d_R P_R) \psi_B ]\,,
\label{Lvector}
\ee

\item[Tensor:]
\be
[ \bar \psi_A (c_M + i c_E \g_5) \s^{\m \n} \psi_X ]
[ \bar \psi_C (d_M + i d_E \g_5) \s_{\m \n} \psi_B ]\,,\
\label{Ltensor}
\ee

\end{description}
where $c_S, c_A, d_S, d_A, c_L, c_R, d_L,d_R, c_M, c_E, d_M, d_E \in {\mathbb R}$ are coupling constants. 
For vector interactions, instead of using $\gamma^\mu$ and $\gamma^\mu \gamma_5$, we organise the bilinaers with the chiral projection operators $P_{R/L} \equiv \frac{1}{2}(1 \pm\gamma^5)$.  
We included a fermion bilinear for tensor interactions with $\gamma_5 \sigma^{\mu \nu}$.
Although this bilinear is not independent, it can represent a Lorentz construction of the type $\e^{\m \n \r \s}
[ \bar \psi_A \s_{\m \n} \psi_X ]
[ \bar \psi_C \s_{\r \s} \psi_B ]$
through the identity $i \g_5 \s^{\m \n} = - \frac{1}{2} \e^{\m \n \r \s} \s^{\r \s}$. 
In fact, one can show
\bea
&&[ \bar \psi_1 (c_M + i c_E \g_5) \s^{\m \n} \psi_0 ]
[ \bar \psi_3 (d_M + i d_E \g_5) \s_{\m \n} \psi_2 ]
\nonumber \\
&&~~~~~~~~=\,
\a [ \bar \psi_1 \s^{\m \n} \psi_0 ]
[ \bar \psi_3 \s_{\m \n} \psi_2 ]
- \frac{\b}{2}
\e^{\m \n \r \s}
[ \bar \psi_1 \s_{\m \n} \psi_0 ]
[ \bar \psi_3 \s_{\r \s} \psi_2 ]
\,,
\eea
with $\a = c_M d_M - c_E d_E$
and $\b = c_M d_E + c_E d_M$.
The number of parameters is reduced by half in the latter expression.
However, we prefer to work with the former expression since the coupling convention is more aligned with the other (scalar and vector) cases and the final expression becomes neat as we will see in the following section. 

Note that in the four-fermion operators above, the three final state fermions $A$, $B$ and $C$ do not appear symmetrically: only the spinor index of $A$ is contracted with that of the decaying particle $X$ to form a Lorentz singlet. 
As a result, the spin state of the final state is not symmetric under permutations of the three fermions, as will be shown below.
Consequently, the precise value of the tight $4 \times 4 \times 2$ observable depends on how the measurement axes are assigned to the three particles, i.e., which fermions are given two axes and which are given four. 
However, since this observable detects the boundary between the FLR and BLR regions exactly (see Fig.~\ref{fig:diag}), the specific assignment does not affect our classification of the state’s nonlocal character (i.e., whether it lies in the FLR region or not).
Throughout our analysis, when computing the $\B4$ observable, we assign four measurement axes to particles $A$ and $B$, and two axes to particle $C$.

\subsection{Scalar interaction}

For a given phase space point $(\theta_B, \theta_C)$ and the initial spin ${\bf n}(\theta, \phi)$,
the matrix element of $X \to ABC$ can be calculated from the scalar interaction \eqref{Lscalar} as \cite{Sakurai:2023nsc}
\bea
&& {\cal M}^{\bf n}_{\lam_A,\lam_B,\lam_C} 
\propto
2 \sqrt{2 m p_A p_B p_C} \cdot 
s \tfrac{\theta_B + \theta_C}{2} 
\cdot
\Big[\,
-\, c d \cdot \d^-_{\lam_A} \d^-_{\lam_B} \d^-_{\lam_C}   
\cdot e^{i \phi}  s \tfrac{\theta}{2} 
\,+\,
 c d^* \cdot \d^-_{\lam_A} \d^+_{\lam_B} \d^+_{\lam_C}   
\cdot e^{i \phi}  s \tfrac{\theta}{2} 
\nn \\
&&~~~~~~~~~~~ ~~~~~~~~~~~~~~~~ 
-\,
 c^* d \cdot \d^+_{\lam_A} \d^-_{\lam_B} \d^-_{\lam_C}   
\cdot c \tfrac{\theta}{2} 
\,+\,
 c^* d^* \cdot \d^+_{\lam_A} \d^+_{\lam_B} \d^+_{\lam_C}   
\cdot c \tfrac{\theta}{2} 
\Big] \,,
\eea
In this expression, we defined $c = c_S + i c_A$ and $d = d_S + i d_A$ and took $|c| = |d| = 1$ as we are not interested in the overall scale of the amplitude. 
We also introduced shorthand notations 
$c \a \equiv \cos \a$ and $s \a \equiv \sin \a$ are used.
As can be seen, the state depends only on the spin polarisation of the initial particle, ${\bf n}(\theta, \phi)$, and independent of the decay angles, $\theta_B$ and $\theta_C$.
The above expression implies there are only four helicity assignments producing non-zero amplitudes. 
The corresponding normalised spin state can be written as\footnote{ 
The absence of other polarisation states, such as $|++- \rangle$, follows from the assumption that the final state fermions $A$, $B$, and $C$ are massless. 
If these particles instead have small but nonzero masses, additional polarisation components would contribute to Eq.\ \eqref{psi_state_sc}. 
Nevertheless, as long as their masses remain negligible compared to the characteristic mass scale of the decaying particle $X$, the right-hand side of Eq.\ \eqref{psi_state_sc} provides a good approximation to the true final spin state $|\Psi\rangle$.
}
\be
| \Psi \ket \,=\, M_{LL} |--- \ket + M_{LR} |-++ \ket 
+
M_{RL} |+-- \ket + M_{RR} |+++ \ket,
\label{psi_state_sc}
\ee
with
$M_{LL} = -\frac{c d}{|cd|\sqrt{2}} \cdot e^{i \phi}  s \tfrac{\theta}{2}$,
$M_{LR} = \frac{c d^*}{|cd|\sqrt{2}} \cdot e^{i \phi}  s \tfrac{\theta}{2}$,
$M_{RL} =  - \frac{c^* d}{|cd|\sqrt{2}} \cdot c \tfrac{\theta}{2}$
and
$M_{RR} = \frac{c^* d^*
}{|cd|\sqrt{2}} \cdot c \tfrac{\theta}{2}$.
This state can be factorised in the following form 
\bea
| \Psi \ket 
\,=\,
\frac{1}{|cd|}
\big[  c e^{i \phi} s \tfrac{\theta}{2}
 | - \ket_A +
 c^* c \tfrac{\theta}{2}
| + \ket_A 
\big] \otimes
\frac{1}{\sqrt{2}}
 \big[ d | ++ \ket_{BC} - d^* | -- \ket_{BC}
\big]\,,
\label{bisep}
\eea
which implies that the state is bi-separable for particle A and subsystem BC.
This structure stems from the form of the four-fermion operator \eqref{Lscalar}, in which the spinors for A are factorised from those for B and C. 

Since the state is factorised as in Eq.\ \eqref{bisep}, A is unentangled with B and C individually and also collectively: 
\be
{\cal C}_{AB} = {\cal C}_{AC} = {\cal C}_{A(BC)} = 0,
\label{1_not_ent}
\ee
Particles B and C are, on the other hand, maximally entangled:
\be
{\cal C}_{BC} = 1 \,.
\ee
The monogamy relations \eqref{monogamy} implies 
B and C must also be maximally entangled with the rest of the system:
\be
{\cal C}_{B(AC)} = {\cal C}_{C(AB)} = 1,
\ee
and the three-tangle is vanishing, $\tau = 0$.
Because the state is bi-separable, 
the GTE measure vanishes
\be
F_3 = 0\,.
\ee

For the bi-separable state \eqref{bisep},
the three-particle correlation tensor \eqref{Tensor} is factorised as
\be
T_{ijk} = V_i \otimes U_{jk}\,,
\label{Tscalar}
\ee
with
\bea
 V_i &=& (\sin \theta \cos \phi, \sin \theta \sin \phi, \cos \theta) \,,
\nonumber \\
U_{jk} &=& 
\begin{pmatrix}
- \cos 2 \delta & - \sin 2 \delta & 0 \\
- \sin 2 \delta  & \cos 2 \delta & 0 \\
0 & 0 & 1
\end{pmatrix}\,,
\label{VU}
\eea
where $\delta$ is a CP phase, 
$d \equiv e^{i \delta}$.
Notice that $\vec V$ coincides with the initial spin polarisation ${\bf n}(\theta, \phi)$. 
The matrix $U$ has two eigenvalues, $\pm1$.
The eigenvector corresponding to the $-1$ eigenvalue is
$\vec u_- = (\cos \frac{\delta}{2}, \sin \frac{\delta}{2}, 0)$.
The two orthonormal eigenvectors corresponding to the $+1$ eigenvalue can be expressed as
$\vec u_+^{(1)} = (-\sin \frac{\delta}{2}, \cos \frac{\delta}{2}, 0)$
and 
$\vec u_+^{(2)} = (0,0, 1)$.
The one-parameter family of normalised eigenvectors with the $+1$ eigenvalue can be written as $\vec u_+ (\alpha) = \cos \alpha \cdot \vec u_+^{(1)} + \sin \alpha \cdot \vec u^{(2)}$.

For the state with the correlation matrix, Eqs. \eqref{Tscalar}, \eqref{VU}, the Mermin, Svetlichny and tight $4 \times 4 \times 2$ observables can be optimised analytically. 
Plugging Eqs.\ \eqref{Tscalar} and \eqref{VU} into 
Eqs.\ \eqref{BbarM}, \eqref{BbarS}, \eqref{Bbar4} and \eqref{BBB}, we see that all terms are proportional to $\vec A_I \cdot \vec V$. 
This means the observables are maximised by taking $\vec A_I = \vec V = {\bf n}(\theta, \phi)$ for all $I \in \{ 1,2~(3,4)\}$.
After fixing $\vec A_I$ in this way, the observables are reduced to 
\bea
\bra {\cal B}_{\rm M} \ket_\rho &=&
\vec B_1 \cdot U \cdot \left( \vec C_1 + \vec C_2 \right)
+
\vec B_2 \cdot U \cdot \left( \vec C_1 - \vec C_2 \right)
\nn \\
\bra {\cal B}_{\rm S} \ket_\rho &=&
2 \left[ \vec B_1 \cdot U \cdot \vec C_1 
-
\vec B_2 \cdot U \cdot \vec C_2
\right]
\nn \\
\bra {\cal B}_{442} \ket_\rho &=&
2 \left[ 
\vec B_1 \cdot U \cdot \left( \vec C_1 + \vec C_2 \right)
+
\vec B_3 \cdot U \cdot \left( \vec C_1 - \vec C_2 \right)
\right]
\label{BBB_2}
\eea
For the Mermin and tight 442 observables, we introduce a pair of orthonormal vectors $\left( \vec C_+, \vec C_- \right)$ as
\be
\vec C_1 + \vec C_2 = 2 \bar c \vec C_+,~~~~ 
\vec C_1 - \vec C_2 = 2 \bar s \vec C_-,
\ee
with $|\vec C_\pm| = 1$, $\vec C_+ \cdot \vec C_- = 0$ and  
$\bar c^2 + \bar s^2 = 1$.
For these new vectors, the observables are written as 
\bea
\bra {\cal B}_{\rm M} \ket_\rho &=&
2 \left[ \bar c \left(\vec B_1 \cdot U \cdot \vec C_+ \right)
+
\bar s \left( \vec B_2 \cdot U \cdot \vec C_- \right)
\right]
\nn \\
\bra {\cal B}_{442} \ket_\rho &=&
4 \left[ \bar c \left(\vec B_1 \cdot U \cdot \vec C_+ \right)
+
\bar s \left( \vec B_3 \cdot U \cdot \vec C_- \right)
\right]
\eea

There are several ways to arrange the two curly brackets in an observable simultaneously to its maximum magnitude $\pm 1$.
For example, $\vec C_\pm$, $\vec B_1$, $\vec B_2$ and $\vec B_3$ can be taken as eigenvectors of the $U$ matrix, making sure that $\vec C_+$ and $\vec C_-$ are orthogonal. 
Once the two curly brackets are optimised to the values either $+1$ or $-1$, then $\bar c$ and $\bar s$ are set either to $\frac{1}{\sqrt{2}}$ or $-\frac{1}{\sqrt{2}}$.
This completes the optimisation, and the resulting values are
\be
\bra \BM \ket_\rho \,=\, 2 \sqrt{2},~~~
\bra {\cal B}_{442} \ket_\rho \,=\, 4 \sqrt{2}\,.
\ee
We see that the expectation of these observables exceeds the fully local-real bounds, $\bra \BM \ket_{\rm FLR} \leq 2$ and $\bra {\cal B}_{442} \ket_{\rm FLR} \leq 4$, while they do not saturate the quantum mechanical bounds, $\eBM_{\rm QM}^{\rm max} = 4$ and $\eBB4_{\rm QM}^{\rm max} = 8$. 

The similar consideration can be applied to the case where the two measurement axes are assigned to Alice.
From an explicit calculation, we also find $\bra {\cal B}_{244} \ket_\rho = 4 \sqrt{2}$, leading to $\bra {\cal B}_{442}^{\rm sym} \ket_\rho = 4 \sqrt{2}$

The optimisation of the Svetlichny observable is more straightforward.  
In the expression \eqref{BBB_2} one can take $\vec B_1 = \vec{C}_1 = \vec u_+(\alpha) $,
and 
$\vec B_2 = - \vec{C}_2 = \vec u_+(\alpha) $
or 
$\vec B_2 = \vec{C}_2 = \vec u_-$.
Then, we find the value of the optimised Svetlichny observable as
\be
\bra \BS \ket_\rho \,=\, 4\,.
\ee
This saturates but does not violate the bi-partite local-real bound, as the state is bi-separable.

\subsection{Vector interaction}
\label{sec:analysis_vec}

The matrix element of $X \to ABC$ for the vector interaction \eqref{Lvector} is found as 
\cite{Sakurai:2023nsc}
\bea
\label{M_vec}
&& {\cal M}^{\bf n}_{\lam_A,\lam_B,\lam_C} 
\propto
4 \sqrt{2 m p_A p_B p_C} 
\nonumber \\ 
&&~~\,\,
\cdot \Big[\, 
\d_{\lam_A}^- \d_{\lam_B}^+ \d_{\lam_C}^- \cdot 
c_L d_L  
s \tfrac{\theta_C}{2} \left[
c \tfrac{\theta}{2} c \tfrac{\theta_B}{2} 
+ e^{i \phi} s \tfrac{\theta}{2} s \tfrac{\theta_B}{2}
\right] 
\,-\,
\d_{\lam_A}^- \d_{\lam_B}^- \d_{\lam_C}^+
\cdot c_L d_R
s \tfrac{\theta_B}{2} \left[
 c \tfrac{\theta}{2} c \tfrac{\theta_C}{2} 
- e^{i \phi} s \tfrac{\theta}{2} s \tfrac{\theta_C}{2}
\right]
\nonumber \\
&&~~+
\d_{\lam_A}^+ \d_{\lam_B}^+ \d_{\lam_C}^-
\cdot c_R d_L
s \tfrac{\theta_B}{2} \left[
 c \tfrac{\theta}{2} s \tfrac{\theta_C}{2} 
+ e^{i \phi} s \tfrac{\theta}{2} c \tfrac{\theta_C}{2}
\right]
\,+\,
\d_{\lam_A}^+ \d_{\lam_B}^- \d_{\lam_C}^+
\cdot c_R d_R
s \tfrac{\theta_C}{2} \left[ 
 c \tfrac{\theta}{2} s \tfrac{\theta_B}{2} 
- e^{i \phi}  s \tfrac{\theta}{2} c \tfrac{\theta_B}{2}
\right] \Big] \,.
\nonumber \\
\eea
These matrix elements imply the final spin state
\be
| \Psi \ket \,=\,
M_{LL} |-+- \ket + M_{LR} |--+ \ket
+ M_{RL} |++- \ket + M_{RR} |+-+ \ket\,,
\label{psi_state}
\ee
with
$(M_{LL}, M_{LR}, M_{RL}, M_{RR}) = (
{\cal M}^{\bf n}_{-+-},
{\cal M}^{\bf n}_{--+},
{\cal M}^{\bf n}_{++-}
{\cal M}^{\bf n}_{+-+})/{\cal N}$
and $|{\cal N}|^2 = \sum_{\lam_A, \lam_D, \lam_C} $ \\ $| {\cal M}^{\bf n}_{\lam_A,\lam_B,\lam_C} |^2$.
In the limit of chiral interactions, the above state reduces to a bi-separable or a fully separable state.
For example, if $c_R = 0$ in the interaction term 
\eqref{Lvector}, the final state is bi-separable:
\be
| \Psi \ket \propto
| - \ket \otimes \left\{
d_L  
s \tfrac{\theta_C}{2} \left[
c \tfrac{\theta}{2} c \tfrac{\theta_B}{2} 
+ e^{i \phi} s \tfrac{\theta}{2} s \tfrac{\theta_B}{2}
\right] | + - \ket
\,-\,
d_R
s \tfrac{\theta_B}{2} \left[
 c \tfrac{\theta}{2} c \tfrac{\theta_C}{2} 
- e^{i \phi} s \tfrac{\theta}{2} s \tfrac{\theta_C}{2}
\right]
| - + \ket
\right\}\,.
\ee
On the other hand, for $d_R = 0$, it becomes a fully separable state
\be
| \Psi \ket \propto
\left\{
c_L 
s \tfrac{\theta_C}{2} \left[
c \tfrac{\theta}{2} c \tfrac{\theta_B}{2} 
+ e^{i \phi} s \tfrac{\theta}{2} s \tfrac{\theta_B}{2}
\right] | - \ket
+
c_R 
s \tfrac{\theta_B}{2} \left[
 c \tfrac{\theta}{2} s \tfrac{\theta_C}{2} 
+ e^{i \phi} s \tfrac{\theta}{2} c \tfrac{\theta_C}{2}
\right] | + \ket
\right\}
\otimes | + - \ket\,.
\ee

From explicit calculations, we find 
\bea
{\cal C}_{AB} = {\cal C}_{AC} = 0\,,~~
{\cal C}_{BC} =
2 | M_{LL} M_{LR}^* + M_{RL} M_{RR}^* |\,.
\label{Cij_vec}
\eea
In the assumed form of interaction \eqref{Lvector}, particle A is never entangled individually with B and C.
For one-to-other entanglement, we obtain
\bea
{\cal C}_{A(BC)} &=&
2 \big| M_{RR} M_{LL} - M_{LR} M_{RL} \big|\,,
\nonumber \\
{\cal C}_{B(AC)} &=& {\cal C}_{C(AB)} 
\nonumber \\
&=&
2 \sqrt{\left( | M_{LL} |^2 + | M_{RL} |^2 \right)
\left( | M_{LR} |^2 + | M_{RR} |^2 \right)}
\,.
\label{Cijk_vec}
\eea
Since all three one-to-other concurrence measures are non-vanishing in general,  
the GTE measure $F_3$ is also non-vanishing in that case.

The three-tangle can be calculated by
\be
\tau \equiv {\cal C}^2_{i(jk)} - [ {\cal C}^2_{ij
} + {\cal C}^2_{ik} ],
\ee
for any choice of $\{ i,j,k\} \in \{ A,B,C\}$ with 
$i \neq j \neq k \neq i$. 
Since ${\cal C}_{AB} = {\cal C}_{AC} = 0$, it follows 
\be
\tau = {\cal C}^2_{A(BC)}\,.
\ee

Non-zero components of the three-particle correlation tensor are found as
\bea
&& T_{333} \,=\, |M_{LL}|^2 + |M_{LR}|^2
- |M_{RL}|^2 - |M_{RR}|^2\,,
\nonumber \\
&& T_{133} \,=\, - 2 {\rm Re}[ M_{RL}^* M_{LL} + M_{RR}^* M_{LR}  ]\,,
\nonumber \\
&& T_{233} \,=\, - 2 {\rm Im}[ M_{RL}^* M_{LL} + M_{RR}^* M_{LR}  ]\,,
\nonumber \\
&& T_{311} \,=\, T_{322} \,=\, 2 {\rm Re}[ M_{RR}^* M_{RL} - M_{LR}^* M_{LL}  ]\,,
\nonumber \\
&& T_{312} \,=\, - T_{321} \,=\, 2 {\rm Im}[ M_{RR}^* M_{RL} - M_{LR}^* M_{LL}  ]\,,
\nonumber \\
&& T_{111} \,=\, T_{122} \,=\, 2 {\rm Re}[ M_{RR}^* M_{LL} + M_{RL}^* M_{LR}  ]\,,
\nonumber \\
&& T_{222} \,=\, T_{211} \,=\,2 {\rm Im}[ M_{RR}^* M_{LL} + M_{RL}^* M_{LR}  ]\,,
\nonumber \\
&& T_{221} \,=\, - T_{212} \,=\,2 {\rm Re}[ M_{RR}^* M_{LL} - M_{RL}^* M_{LR}  ]\,,
\nonumber \\
&& T_{112} \,=\, - T_{121} \,=\,2 {\rm Im}[ M_{RR}^* M_{LL} - M_{RL}^* M_{LR}  ]\,,
\eea
which allows us to calculate the Bell-type observables.
As the final spin state \eqref{psi_state} is not symmetric under qubit permutation, in this subsection, we focus on the symmetrically extended tight $4 \times 4 \times 2$ observable, $\B4^{\rm sym}$, defined in Eq.\ \eqref{B4sym}.

\begin{figure}[t!]
\centering
\includegraphics[scale=.45]{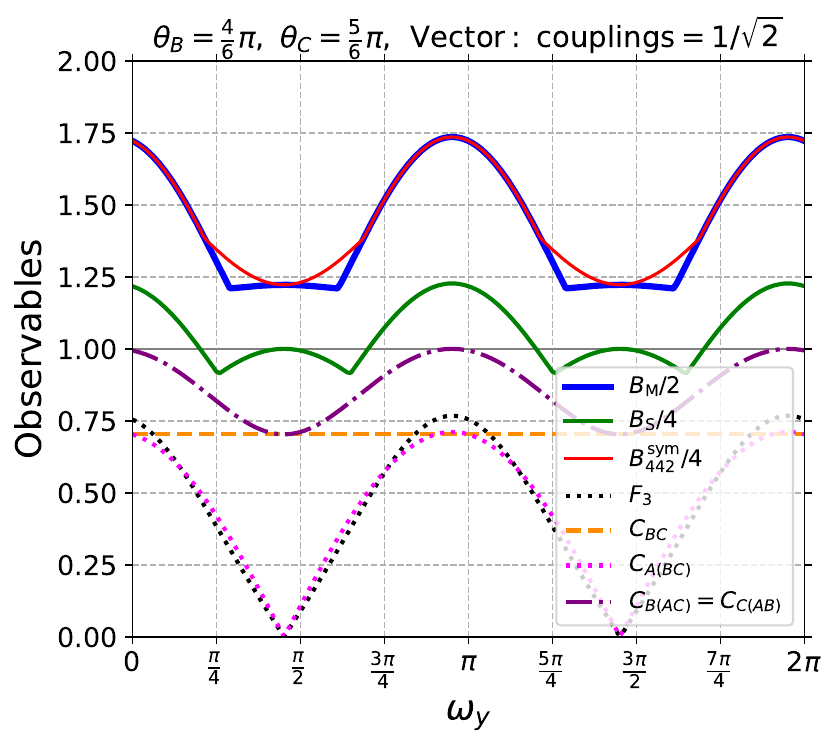}
\includegraphics[scale=.45]{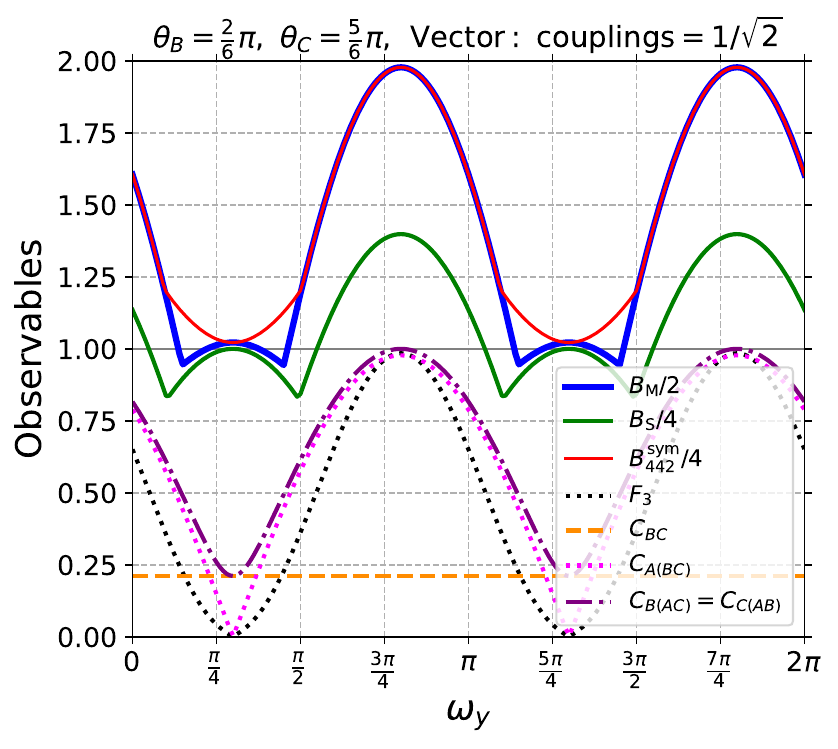}
\includegraphics[scale=.45]{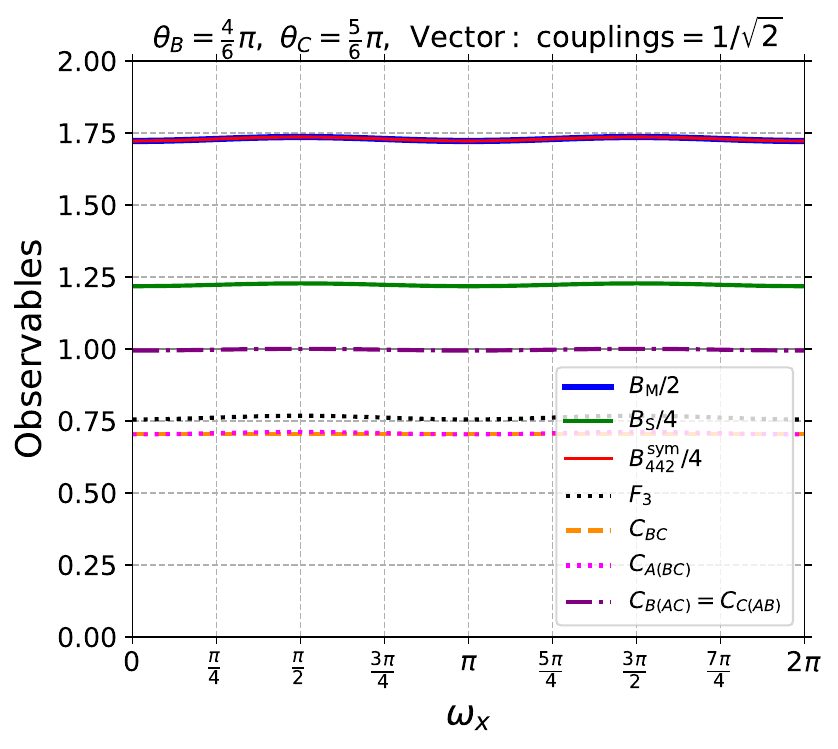}
\includegraphics[scale=.45]{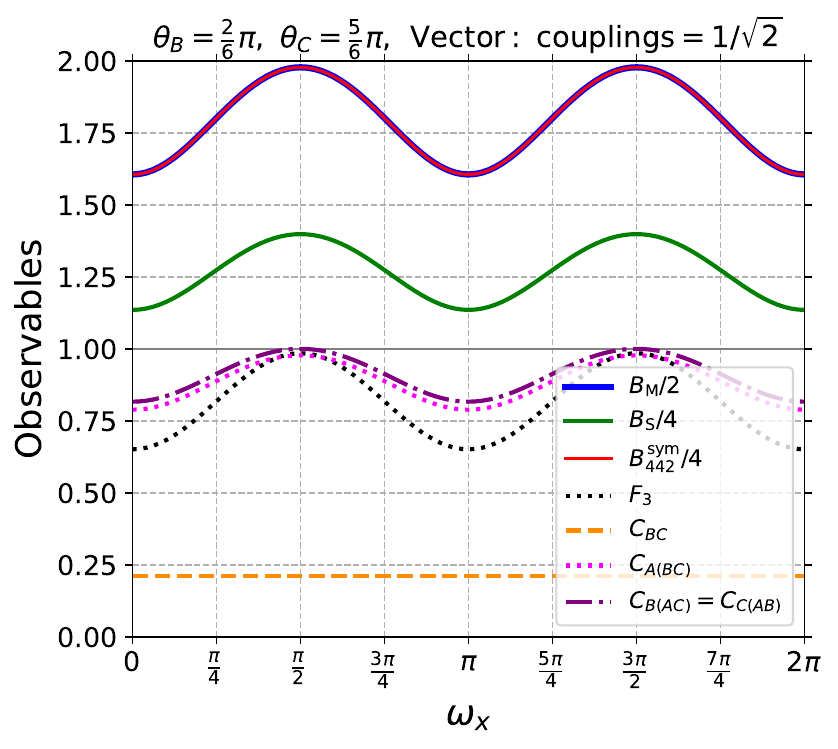}
\caption{\label{fig:1d_vec}
\small
The dependence of various entanglement measures,
$F_3$ (black dotted),
${\cal C}_{A(BC)}$ (magenta dotted) and
${\cal C}_{B(AC)} = {\cal C}_{C(AB)}$ (purple dashed-dotted),
and the Bell-type observables,
${\eBM}/2$ (blue solid),
${\eBS}/4$ (green solid)
and 
$\eB4/4$ (red solid),
on the initial spin direction ${\bf n}$, 
assuming the vector interaction in Eq.\ \eqref{Lvector}.
The maximum values of the corresponding local-real theories normalise the values of Bell-type observables.  
In the left and right columns, the decay angles are fixed at $(\theta_B, \theta_C) = (\frac{4 \pi}{6}, \frac{5 \pi}{6})$
and $(\frac{2 \pi}{6}, \frac{5 \pi}{6})$, respectively. 
In the upper (lower) panels, the initial spin is rotated about the $y$ ($x$) axis clockwise, and the horizontal axis indicates the rotation angle $\omega_y$ ($\omega_x$). 
}
\end{figure}

Fig.\ \ref{fig:1d_vec} illustrates how various entanglement measures and Bell-type observables respond to changes in the initial spin direction, ${\bf n}$.
The horizontal axes of the plots represent the angle between the $z$-axis and ${\bf n}$.
In the upper (lower) panels, ${\bf n}$ rotates about the $y$ ($x$) -axes clackwise with the rotation angle $\omega_y$ ($\omega_x$).  
In each plot, the decay angles, $\theta_B$ and $\theta_C$, are fixed to some reference values.  
In the left and right panels, we take $(\theta_B, \theta_C) = (\frac{4}{6} \pi, \frac{5}{6} \pi)$
and $(\frac{2}{6} \pi, \frac{5}{6} \pi)$, respectively.
In all plots, the coupling constants are taken so that the four-fermion interaction \eqref{Lvector} is vector-like, that is, $c_L = c_R = d_L = d_R =\frac{1}{\sqrt{2}}$.

The entanglement measures shown in the plots are the GTE measure $F_3$ (black dotted), 
a non-vanishing one-to-one concurrence ${\cal C}_{BC}$ (orange dashed),
two one-to-other concurrences, ${\cal C}_{A(BC)}$ (magenta dotted)
and ${\cal C}_{B(AC)} = {\cal C}_{C(AB)}$ (purple dashed-dotted).  
The presented values of Bell-type observables are divided by the corresponding local-real theory bounds, 2, 4 and 4, for ${\eBM}$ (blue solid), $\eBS$ (green solid) and $\eB4$ (red solid), respectively.
If those curves are above 1, it indicates the quantum state is nonlocal, violating the corresponding local-real bounds. 

In all plots in Fig.\ \ref{fig:1d_vec}, we observe that the one-to-one concurrence ${\cal C}_{BC}$ is independent of the initial spin axis ${\bf n}$, whereas all the other observables depend on ${\bf n}$ in nontrivial ways.
We see that $\eB4$ coincides with $2 \eBM$ over large regions of $\omega_y$ in the upper plots and the entire region of $\omega_x$ in the lower plots.
In this region (with $\eB4 = 2 \eBM$), we observe $\bra {\cal B}_{244} \ket = \bra {\cal B}_{424} \ket = \bra {\cal B}_{424} \ket = \eB4$.  
In the other region (with $\eB4 \neq 2 \eBM$), the three $4 \times 4 \times 2$ observables do not agree in general, though their differences are typically a few per cent or less.

In the upper panels, the one-to-other concurrence measures, ${\cal C}_{A(BC)}$ and ${\cal C}_{B(AC)}={\cal C}_{C(AB)}$, exhibit two peaks and two troughs over the course of the entire evolution of ${\bf n}$ around the $y$-axis.
Their peak and trough positions in $\omega_y$ also agree between ${\cal C}_{A(BC)}$ and ${\cal C}_{B(AC)}={\cal C}_{C(AB)}$.
The GTE measure $F_3$ also behaves the same way, and it vanishes at the trough positions together with ${\cal C}_{B(AC)}={\cal C}_{C(AB)}$. 
The behaviours of Bell-type observables in the same plots are more complicated. 
The ${\eBM}$ and $\eBS$ exhibit two global maxima, two local maxima and four global minima. 
Additionally, their values display non-smooth transitions at certain values of $\omega_y$.
We see that $\eBM$ and $\eB4$ violate fully local-real bounds for all $\omega_y$.  
On the other hand, we observe that Svetlichny's BLR bound is not violated in two regions of $\omega_y$.

In the lower panel, where the initial spin is rotated about the $x$-axis, the change of various entanglement measures and Bell-type observables is milder. 
All observables, except for a constant ${\cal C}_{BC}$, exhibit two peaks and two troughs at the same locations of $\omega_x$ among all observables. 
In all regions of $\omega_x$, the FLR bounds are violated; $\bra \BM \ket_\rho > 2$ and $\eB4 \ket > 4$.
The Svetlichny's BLR bound is also violated in the left plot's $\omega_x$ regions.
In the right plot, on the other hand, the bound is saturated at two points, $\omega_x = 0$ and $\pi$.

\begin{figure}[t!]
\centering
\includegraphics[scale=.33]{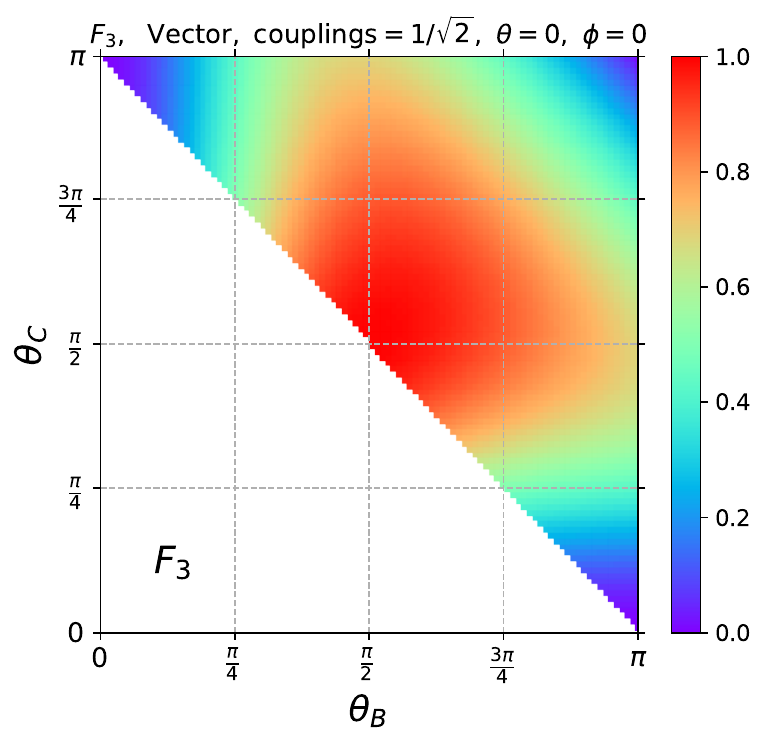}
\includegraphics[scale=.33]{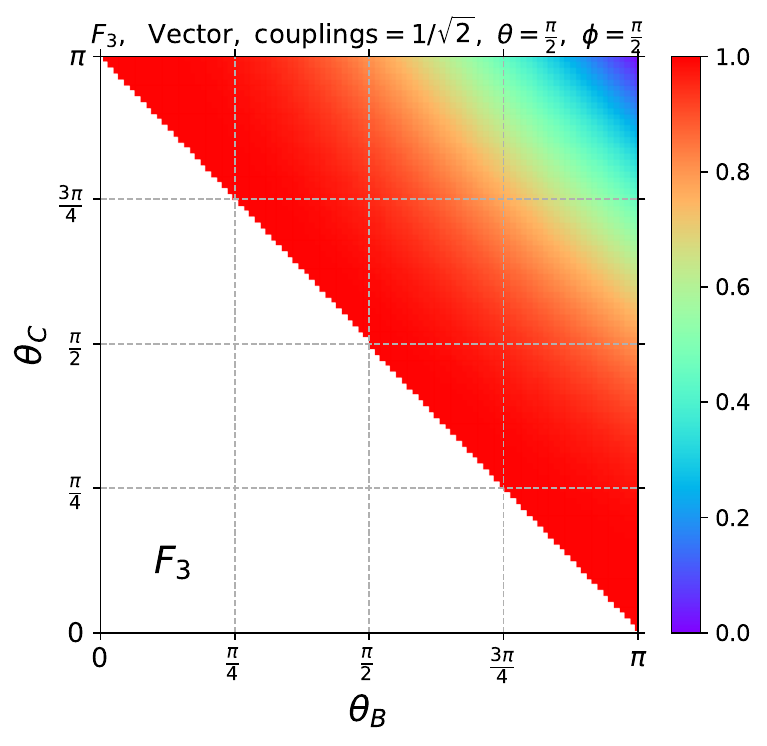}
\includegraphics[scale=.33]{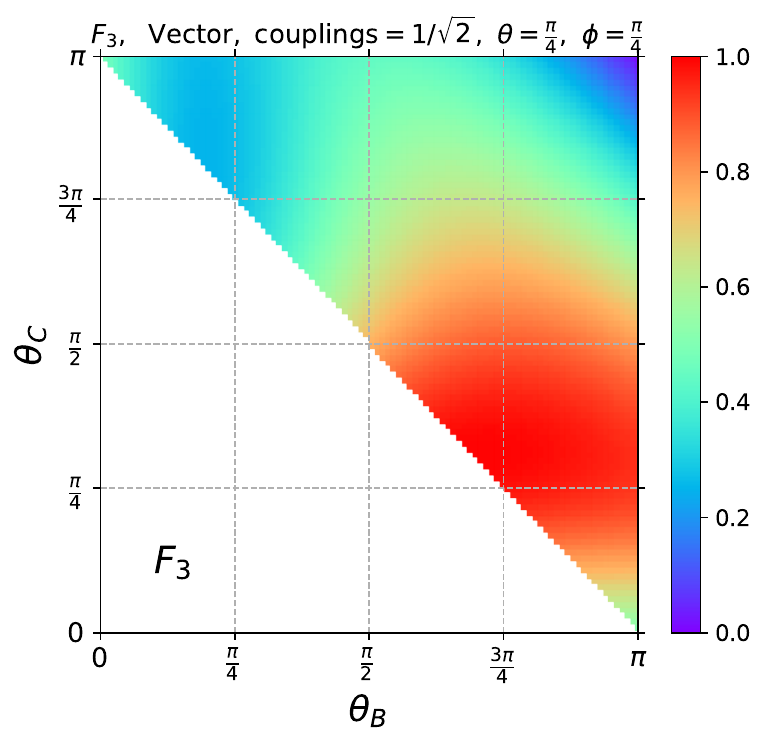}
\includegraphics[scale=.33]{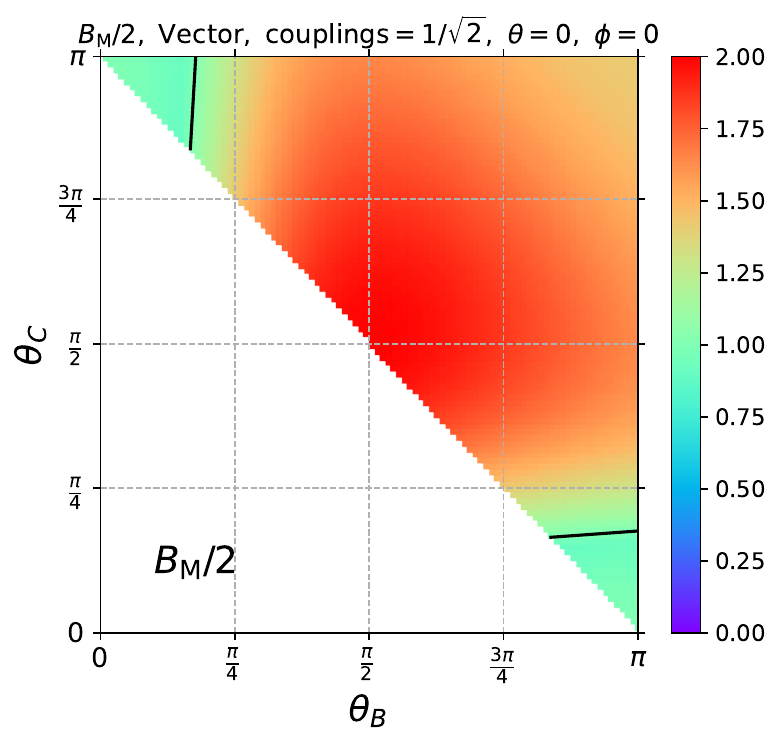}
\includegraphics[scale=.33]{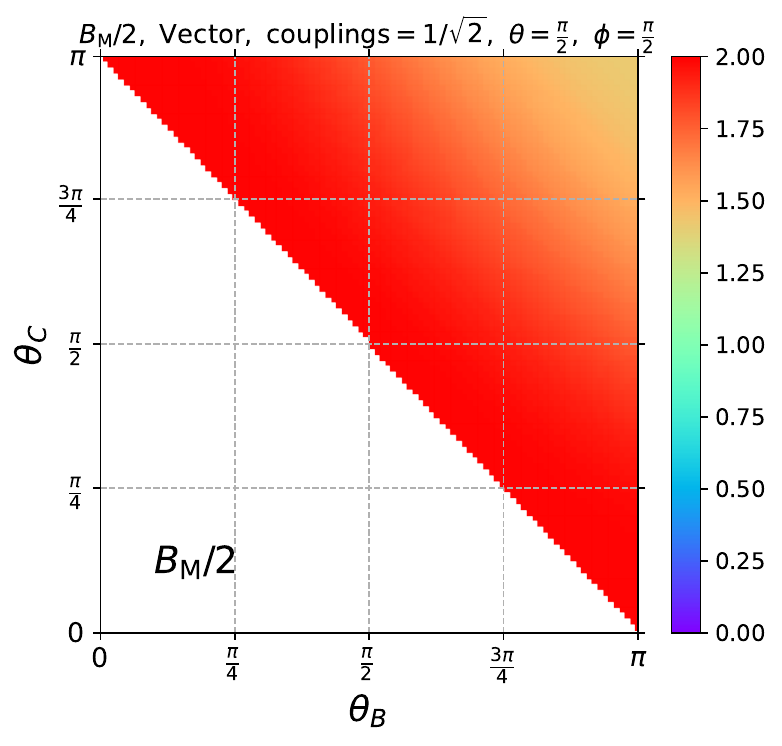}
\includegraphics[scale=.33]{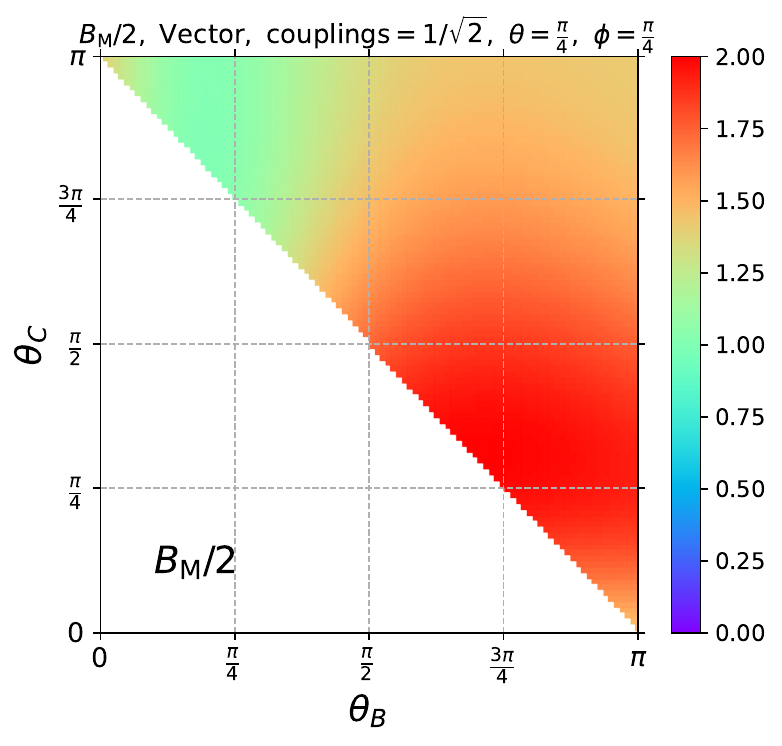}
\includegraphics[scale=.33]{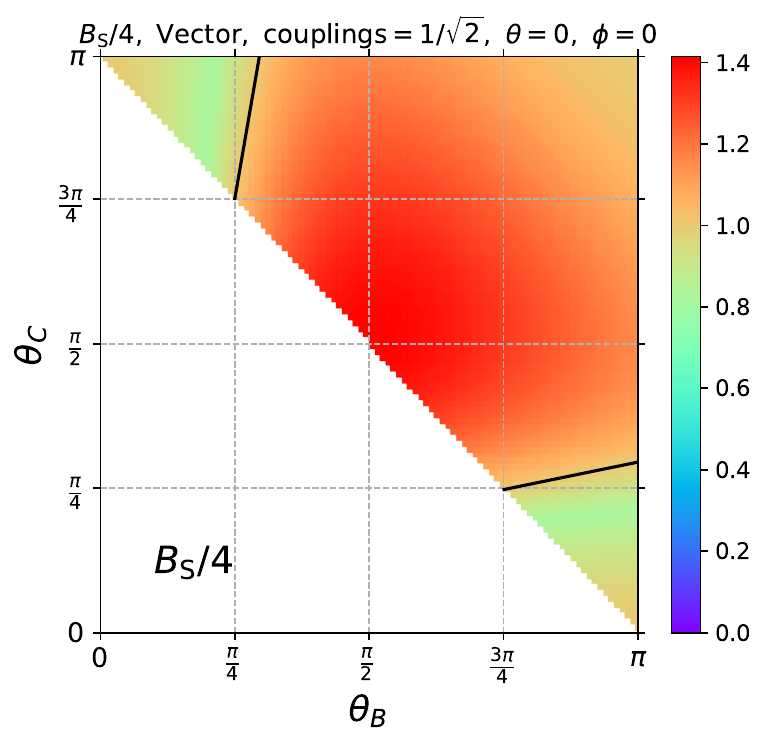}
\includegraphics[scale=.33]{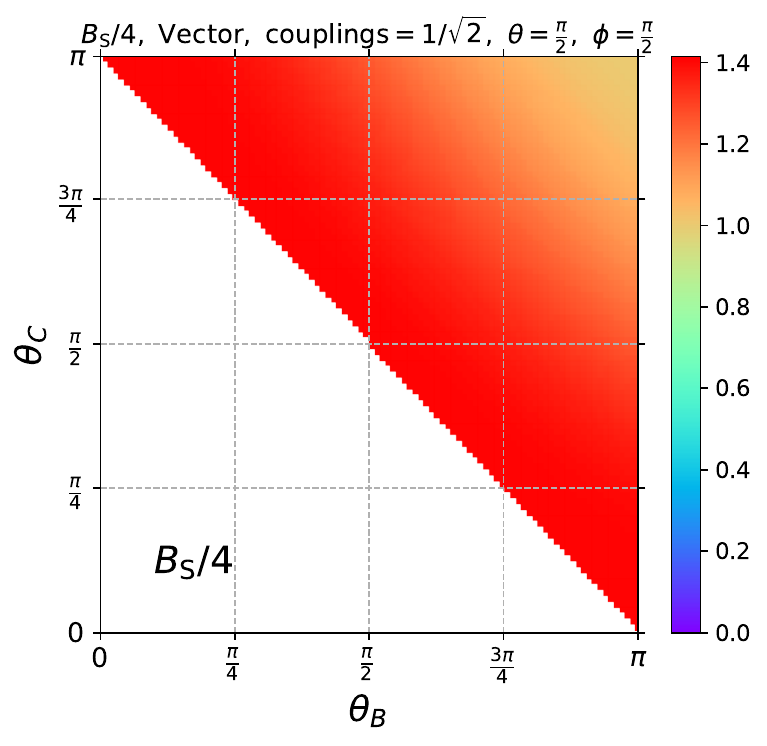}
\includegraphics[scale=.33]{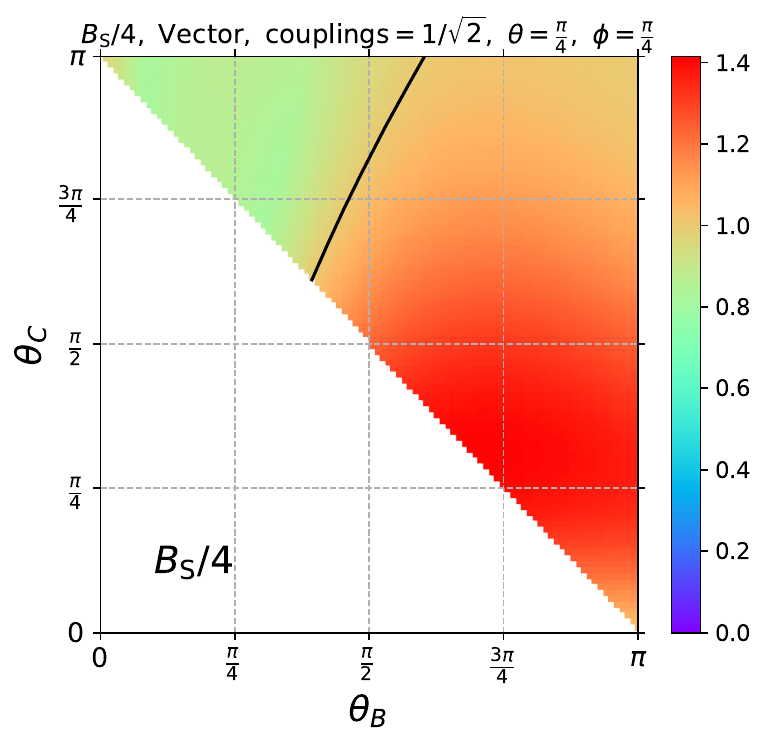}
\includegraphics[scale=.33]{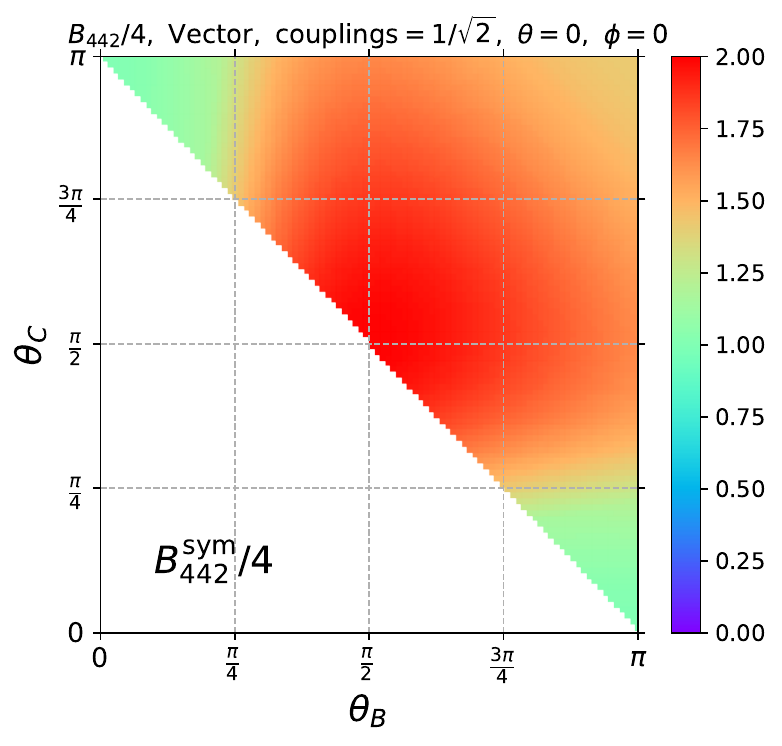}
\includegraphics[scale=.33]{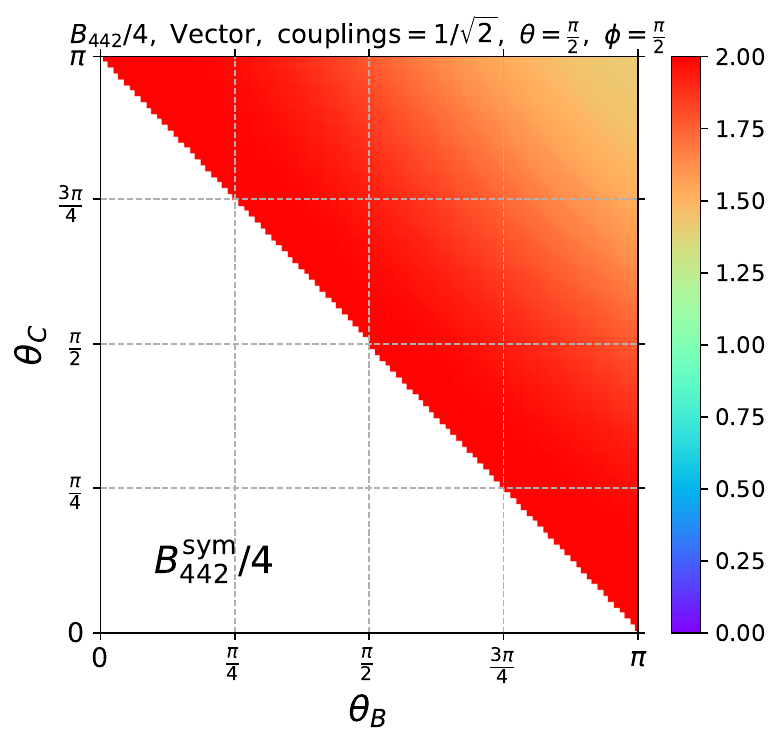}
\includegraphics[scale=.33]{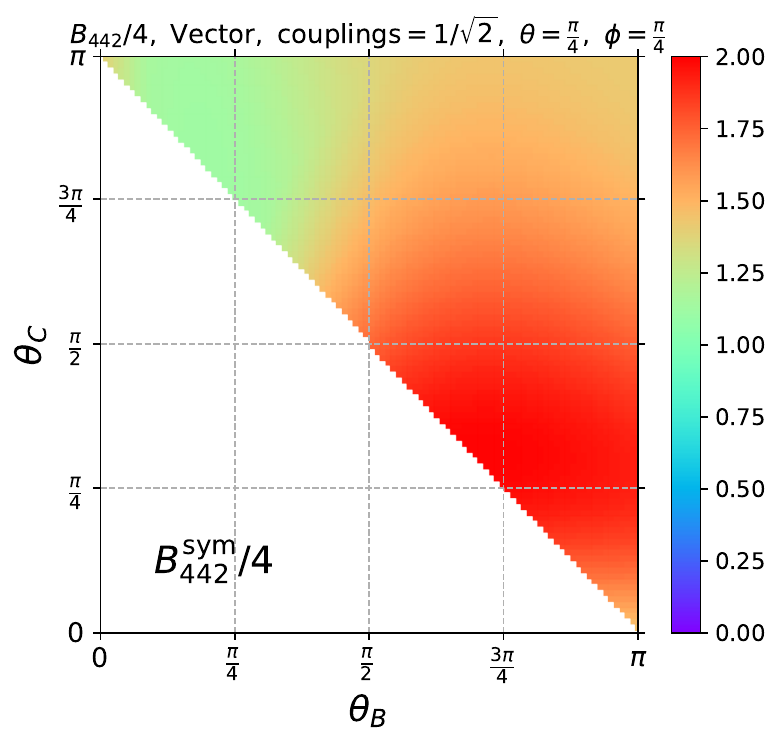}
\caption{\label{fig:2d_vec}
\small 
The dependence of the GTE measure, $F_3$, (the first line), $\eBM$ (the second line), $\eBS$ (the third line) and $\eB4$ (the fourth line) on the decay angles $\theta_B$ and $\theta_C$, evaluated assuming the vector interaction Eq.\ \eqref{Lvector}.
In the left, middle and right columns, the initial spin is fixed as ${\bf n} \propto {\bf e}_z$, ${\bf e}_y$ and $({\bf e}_x + {\bf e}_y + {\bf e}_z)$, respectively.
For the Bell-type observables, the colour indicates the expectation values of the observables normalised by their local-real theory maxima. 
Namely, the colour indicates $\eBM/2$, $\eBS/4$ and $\eB4/4$ for the plots in the second, third and fourth lines, respectively.  
The maximum of the colour bar is set at the corresponding quantum mechanical maximum values, $F_3^{\rm max} = 1$, 
$\eBM_{\rm QM}^{\rm max} = 4$,
$\eBS_{\rm QM}^{\rm max} = 4\sqrt{2}$ and
$\eB4_{\rm QM}^{\rm max} = 8$.
The black solid curves appearing in some of the $\eBM$ and $\eBS$ plots indicate the saturation of the corresponding local-real bounds, $\eBM = 2$
and $\eBS = 4$.
}
\end{figure}

In Fig.\ \ref{fig:2d_vec}, we show the dependence on the GTE measure, $F_3$ (the top panels), and various Bell-type observables, $\bra \BM \ket/2$ (the second panels),
$\eBS/4$ (the third panels) and $\eB4/4$ (the fourth panels), on the decay angles, $\theta_B$ and $\theta_C$.
In the left and middle columns, the initial spin is fixed in the directions of the $z$ ($\theta=\phi=0$) and $y$ ($\theta=\phi=\frac{\pi}{2}$) -axes, respectively.  
In the right column, $\bf n$ is taken in the direction of ${\bf e}_x + {\bf e}_y + {\bf e}_z$, corresponding to $\theta = \phi = \frac{\pi}{4}$. 
The coupling constants are fixed as $c_L = c_R = d_L = d_R =\frac{1}{\sqrt{2}}$.
The black contours, which appear in some plots, represent the boundary of the corresponding local-real bounds, $\bra \BM \ket = 2$ and $\bra \BS \ket = 4$.
The lower-left half of the plot is empty, as the kinematics in this region are unphysical. 

In the left panels with ${\bf n} = {\bf e}_z$, all observables are invariant under the exchange, $\theta_B \leftrightarrow \theta_C$, as this operation is related to the $\pi$ rotation about the $z$ axis.  
$F_3$ vanishes at three points $(\theta_B, \theta_C) = (\pi, 0)$, $(0, \pi)$ and $(\pi, \pi)$ and reaches the maximum value one at $(\theta_B, \theta_C) = (\frac{\pi}{2}, \frac{\pi}{2})$.
A similar behaviour is observed for all Bell-type observables. 
Their expectation values are maximised around $(\theta_B, \theta_C) = (\frac{\pi}{2}, \frac{\pi}{2})$ and minimised around $(\theta_B, \theta_C) = (0, \frac{\pi}{2})$ and $(0, \frac{\pi}{2})$.

In the bottom left plot, we observe that the tight $4 \times 4 \times 2$ bound is always violated in the entire regions of the decay angles. 
This indicates that the three-particle correlations are non-FLR type in all decay angles.   
On the other hand, the non-violation of the Mermin inequality is observed at the top-left and bottom-right corners. 
This means the Mermin inequality fails to detect non-FLR correlations in these regions, as $\B4$ gives a tighter bound than $\BM$. 
We also see that Svetlichny's bound is violated in some regions at the top-left and bottom-right corners. 
In these regions, the three-particle correlations may be describable by a bipartite local-real theory. 

In the middle column with ${\bf n} = {\bf e}_y$, all observables are again symmetric under $\theta_B \leftrightarrow \theta_C$, as these configurations are related by the parity symmetry. 
We also see that all observables depend only on the combination $\theta_B + \theta_C$.
In the top plot, we see that $F_3$ reaches the maximum value one at $\theta_B + \theta_C = \pi$ 
and vanishese at $\theta_B = \theta_C = \pi$.
The same dependency is observed for the Bell-type observables. 
We observe that the fully local-real bounds, $\BM \leq 2$ and $\B4 \leq 4$, are violated in all decay angles. 
The bipartite local bound, $\BS \leq 4$, is also violated in all angles except at $\theta_B = \theta_C = \pi$.  
At this top right corner, the Svetlichny bound is saturated, $\BS = 4$.  
In fact, the bound cannot be violated as $F_3$ vanishes at this point. 

In the right column with ${\bf n} \propto ({\bf e}_x+{\bf e}_y+{\bf e}_z)$, the observables are not symmetric under $\theta_B \leftrightarrow \theta_C$ as expected, because the initial spin is not on the ($z$-$y$) plane (the set of fixed points under $x \to -x$).  
All observables have the maximum around $(\theta_B, \theta_C) \sim (\frac{3\pi}{4}, \frac{\pi}{4} )$.
The GTE measure, $F_3$, exhibts the global minimum, $F_3 = 0$, at $(\theta_B, \theta_C) = (\pi,\pi)$.
$F_3$ also has a local minimum around $\theta_B \sim \frac{\pi}{4}$.
The Bell-type observables take the minimum values in the region corresponding to the local minimum of $F_3$.
While the fully local-real bounds are violated in all decay angles, non-violation of the bipartite local-real bound is observed at the vicinity of $F_3$'s local minimum (the top-left corner). 
The bound is saturated, $\eBS = 4$, at the top-right corner, $\theta_B = \theta_C= \pi$, as the state is not genuinely tripartite  entangled, $F_3 = 0$, at this point.

\subsection{Tensor interaction}

The matrix elements of $X \to ABC$ for the tensor-type interaction \eqref{Ltensor} are given by 
\cite{Sakurai:2023nsc}
\bea
&& {\cal M}^{\bf n}_{\lam_A,\lam_B,\lam_C} 
\propto
- 8 \sqrt{2 m p_1 p_2 p_3} 
\cdot 
\Big[ \, c^* d^* \cdot \d^+_{\lam_A} \d^+_{\lam_B} \d^+_{\lam_C}   
\cdot 
[ 2 e^{i \phi} s \tfrac{\theta}{2} s \tfrac{\theta_B}{2} s \tfrac{\theta_C}{2}
-
c \tfrac{\theta}{2} s \tfrac{\theta_B - \theta_C}{2} ] 
\nonumber \\
&&
~~~~~~~~~~~~~~~~~~~~~~~~~~~~~~~~~~~~~~~~+\,
 c d \cdot \d^-_{\lam_A} \d^-_{\lam_B} \d^-_{\lam_C}   
\cdot 
[ e^{i \phi} s \tfrac{\theta}{2} s \tfrac{\theta_B - \theta_C}{2} 
+
2 c \tfrac{\theta}{2} 
s \tfrac{\theta_B}{2} s \tfrac{\theta_C}{2}
]
\Big] \,,
\label{M_tens}
\eea
where we defined $c = c_M + i c_E$, $d = d_M + i d_E$ and took $|c| = |d| = 1$.
The corresponding state is found as
\be
| \Psi \ket \,=\,
M_{R} |+++ \ket + M_{L} |--- \ket
\,,
\label{psi_state3}
\ee
with
$(M_{R}, M_{L}) = ({\cal M}^{\bf n}_{+++}, {\cal M}^{\bf n}_{---})/{\cal N}$
and
$|{\cal N}|^2 = |{\cal M}^{\bf n}_{+++}|^2 + |{\cal M}^{\bf n}_{---}|^2$.
This state interpolates between fully separable states,
$|+++\ket$ and $|---\ket$,
and the maximally entangled GHZ state,
$|{\rm GHZ} \ket = (|+++\ket + |---\ket)/\sqrt{2}$.

As in the GHZ state, all one-to-one entanglement measures vanish for this state
\be
{\cal C}_{AB} =
{\cal C}_{AC} =
{\cal C}_{BC} = 0 \,.
\label{Cij_tens}
\ee
On the other hand, one-to-other entanglements are non-vanishing and universal
\be
{\cal C}_{A(BC)} = {\cal C}_{B(AC)} = {\cal C}_{C(AB)} =
2 |M_R M_L|\,.
\ee
In this symmetric case, the GTE measure $F_3$ is given by the square of the one-to-other concurrence (as mentioned below Eq.\ \eqref{F3})
\be
F_3 = {\cal C}^2_{i(jk)} = 4 |M_R M_L|^2\,.
\ee

The final spin state $| \Psi \ket$ in Eq.\ \eqref{psi_state3} is symmetric under qubit permutation, and therefore we do not need to distinguish ${\cal B}_{442}$ from ${\cal B}_{442}^{\rm sym}$, as they coincide.

Many components of the three-particle correlation tensor, $T_{ijk}$, 
vanish apart from the following ones:
\bea
&& T_{111} = 
- T_{122} = - T_{212} = - T_{221} =
2 {\rm Re}[ M_R^* M_L ]\,,
\nonumber \\ 
&& T_{222} = 
- T_{112} = - T_{121} = - T_{211}
= 2 {\rm Im}[ M_R^* M_L ]\,,
\nonumber \\
&& T_{333} = |M_R^2| - |M_L^2| \,.
\eea

\begin{figure}[t!]
\centering
\includegraphics[scale=.45]{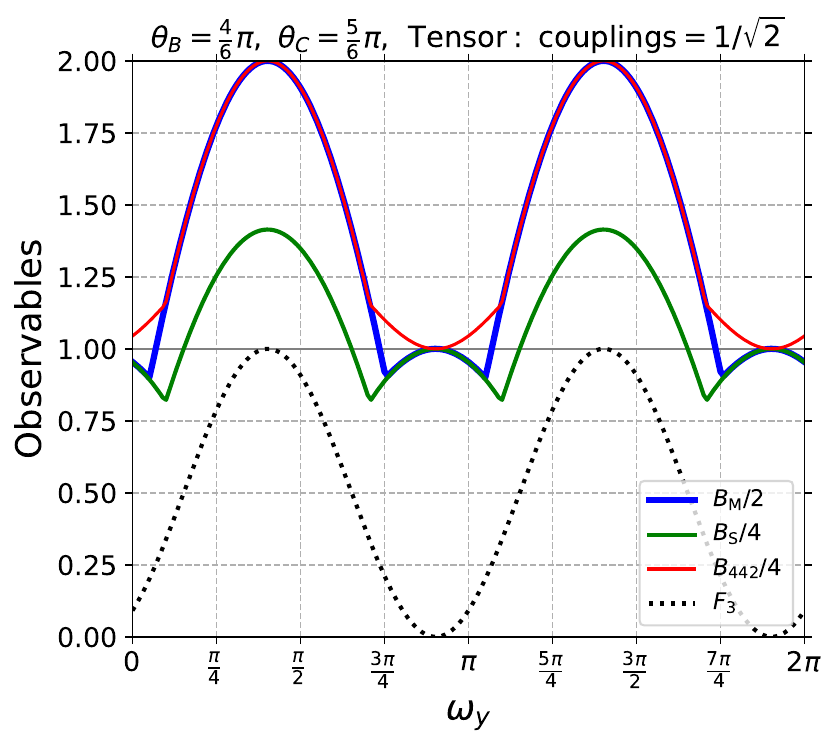}
\includegraphics[scale=.45]{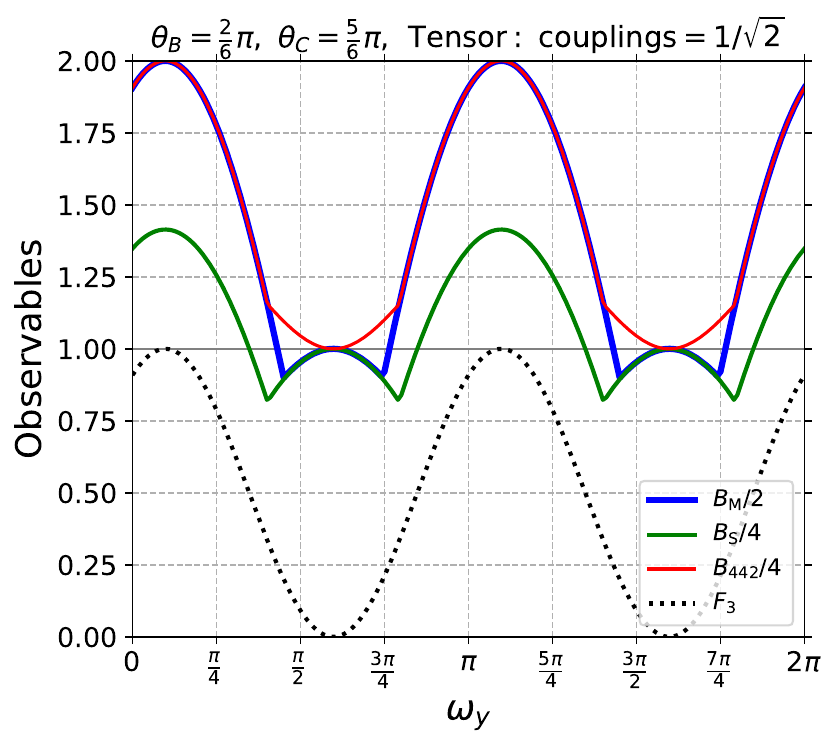}
\includegraphics[scale=.45]{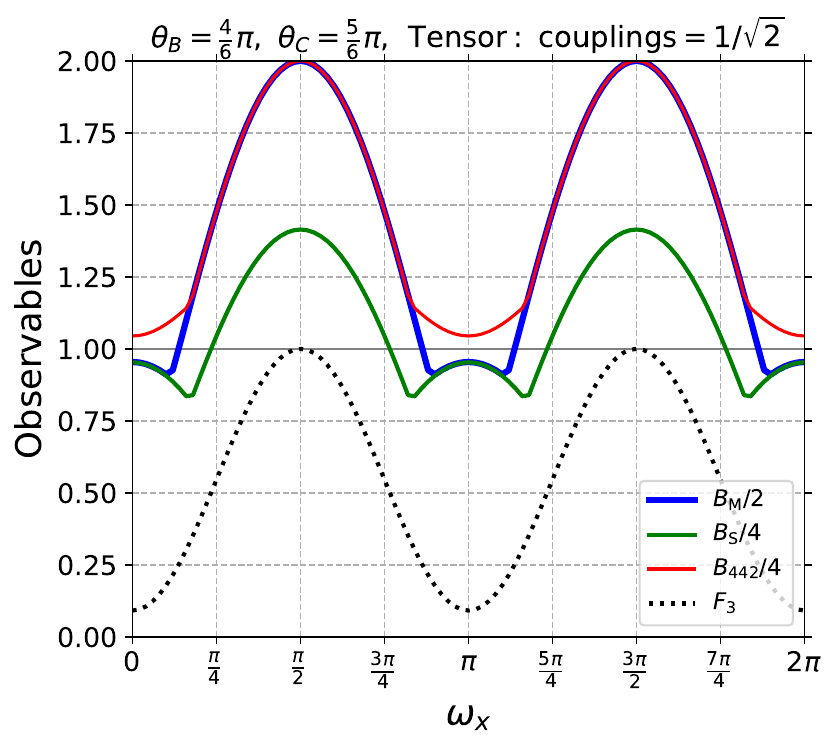}
\includegraphics[scale=.45]{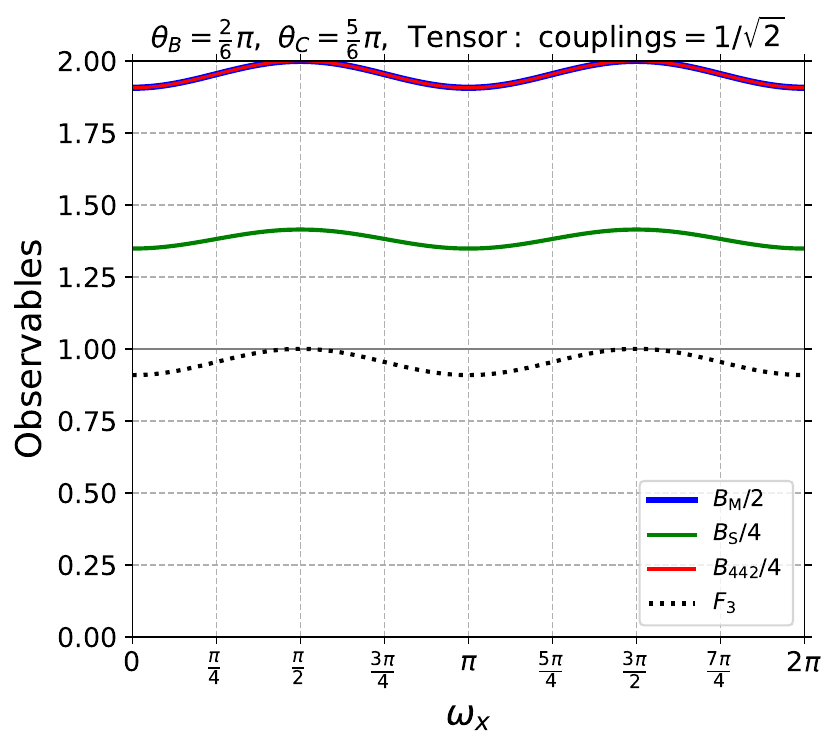}
\caption{\label{fig:1d_tens}
\small
The dependence of the GTE measure,
$F_3$ (black dotted) and Bell-type observables,
${\eBM}/2$ (blue solid),
${\eBS}/4$ (green solid)
and 
$\eBB4/4$ (red solid),
on the initial spin direction ${\bf n}$, 
assuming the tensor-type interaction in Eq.\ \eqref{Ltensor}.
The convention of the figure is the same as of Fig.\ \ref{fig:1d_vec}. 
}
\end{figure}

Fig.\ \ref{fig:1d_tens} illustrates the response of the GTE measure, $F_3$ (black dotted), and  Bell-type observables, $\eBM/2$ (blue solid), 
$\eBS/4$ (green solid) and $\eBB4/4$ (red solid), to changes in the initial spin direction, ${\bf n}$, assuming the tensor-type interaction in Eq.\ \eqref{Ltensor}.
The couplings are fixed as $c_M = c_E = d_M = d_E =\frac{1}{\sqrt{2}}$.
The convention of the figure is the same as of Fig.\ \ref{fig:1d_vec}.

In the upper panels, where the ${\bf n}$ rotates about the $y$-axis clockwise with the rotation angle $\omega_y$, $F_3$ reaches the maximum value one at two positions and exhibits zero GTE at two places. 
Around the two points with $F_3 = 1$, three Bell-type observables also peak. 
In particular, the Mermin observable saturates its quantum mechanical maximum $\eBM = 4$ at these points. At the vicinity of the points with $F_3 = 0$, $\eBM$ and $\eBS$ go below their local-real bounds.
However, we see that in the region where the Mermin inequality is not violated, the tight $4 \times 4 \times 2$ bound is violated except for the two points with $F_3 = 0$.
At these two points, the bound is saturated, $\eBB4 = 4$.
This means in all $\omega_y$, there exists a spin correlation experiment whose result cannot be described by the fully local-real theory, except for the two points in $\omega_y$.    
On the other hand, the experimental outcome may be explained by some bipartite local-real theory in the region where $\eBS \leq 4$.
Another interesting observation is that $\eBS = 2 \eBM$ holds in large regions of $\omega_x$.  

In the lower panels, the initial spin is rotated around the $x$-axis by an angle $\omega_x$. 
In the lower-right plot, where $(\theta_B, \theta_C) = \left(\frac{4\pi}{6}, \frac{5\pi}{6}\right)$, the behaviour of the observables resembles that in the upper plots. 
However, a key difference is that $F_3$ never reaches zero, indicating that the states remain genuinely tripartite  entangled (GTE) throughout. 
Additionally, $\eBB4$ never touches the line with $\eBB4/4 = 1$. 
In this family of configurations, there is not a fully local-real theory that can account for all outcomes of three-particle correlation measurements.

In the lower-right plot, with $(\theta_B, \theta_C) = \left(\frac{2\pi}{6}, \frac{5\pi}{6}\right)$, the behavior of the observables is notably different. The relation $\eBS = 2 \eBM$ holds across all values of $\omega_x$. 
Furthermore, both the FLR and BLR bounds are violated throughout the entire range of $\omega_x$.

\begin{figure}[t!]
\centering
\includegraphics[scale=.33]{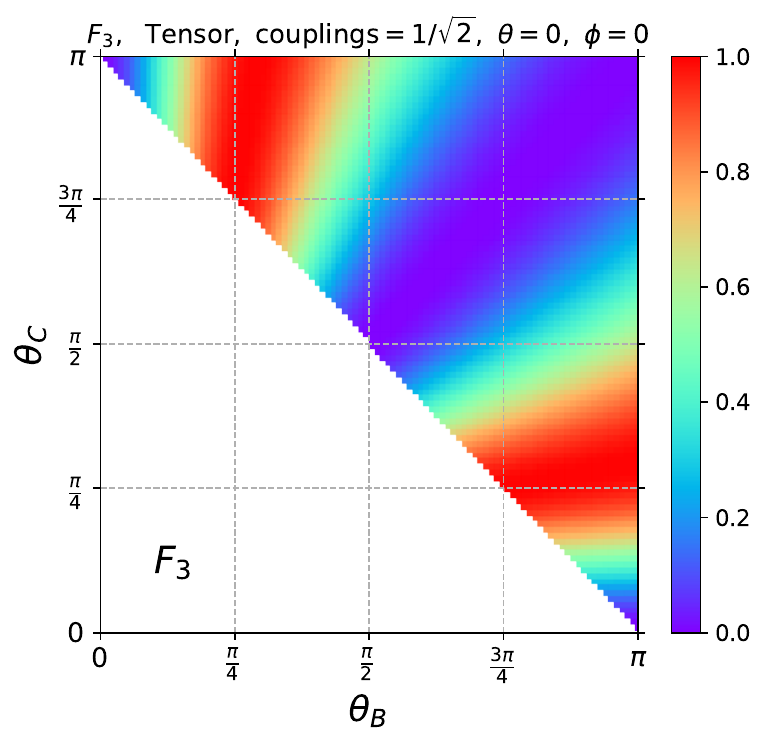}
\includegraphics[scale=.33]{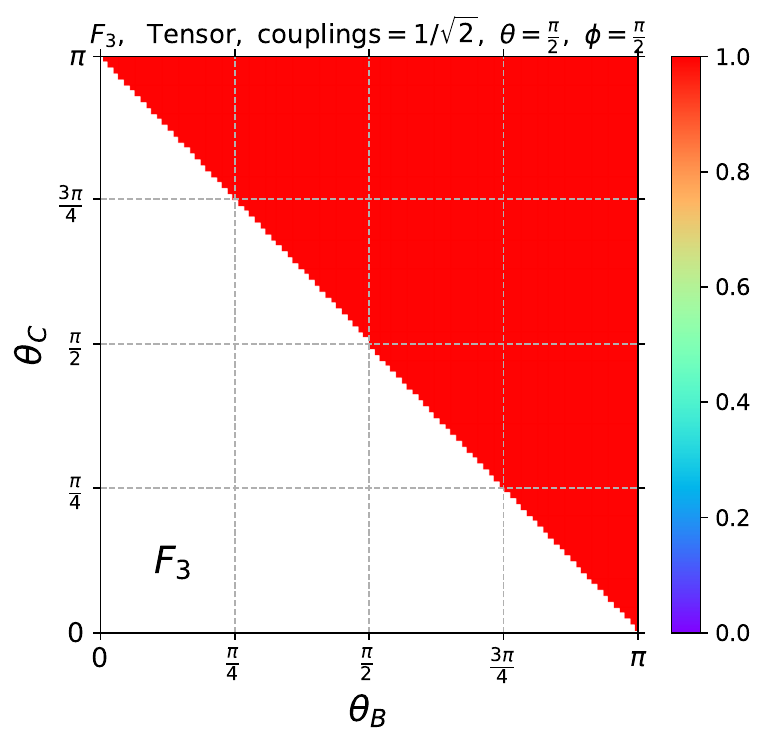}
\includegraphics[scale=.33]{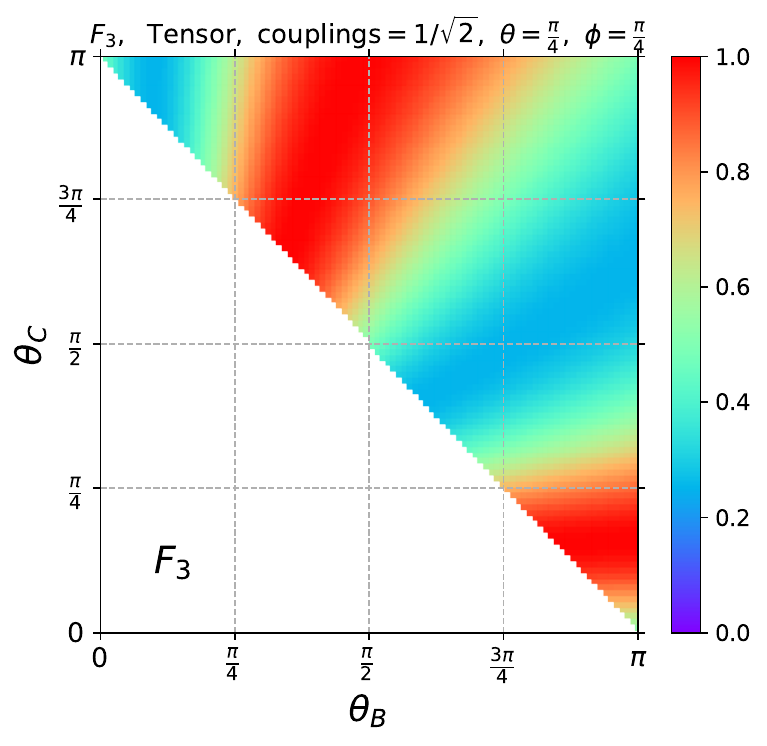}
\includegraphics[scale=.33]{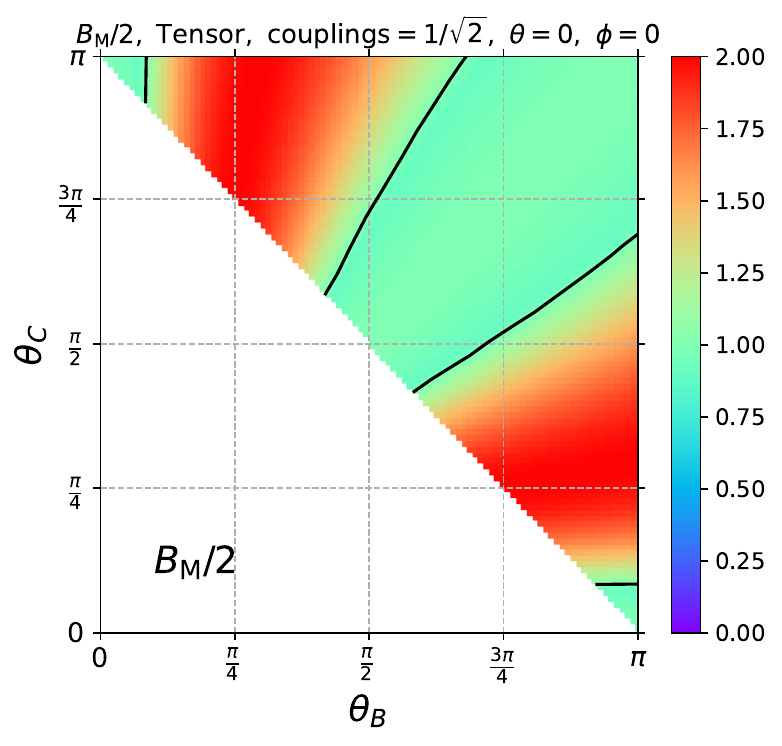}
\includegraphics[scale=.33]{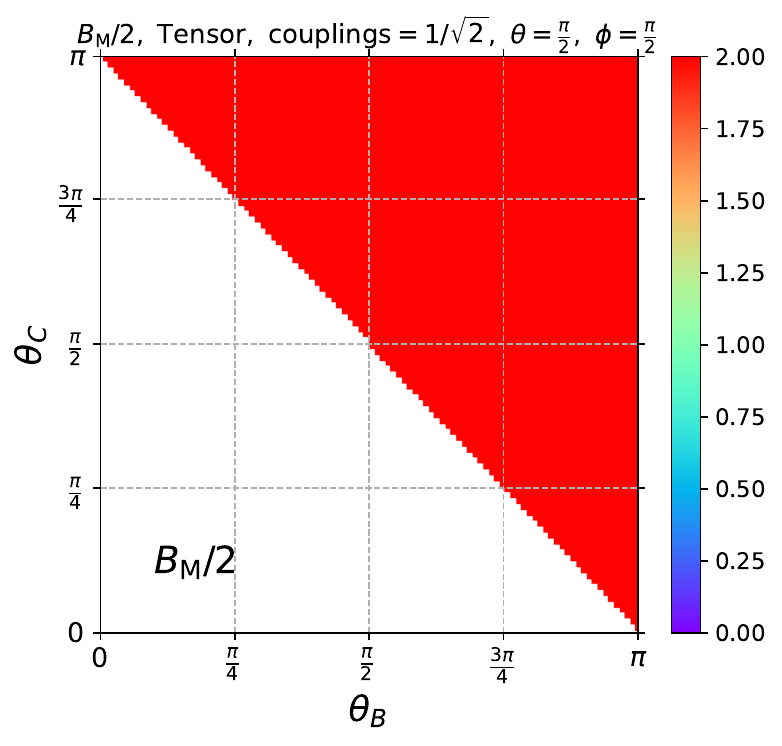}
\includegraphics[scale=.33]{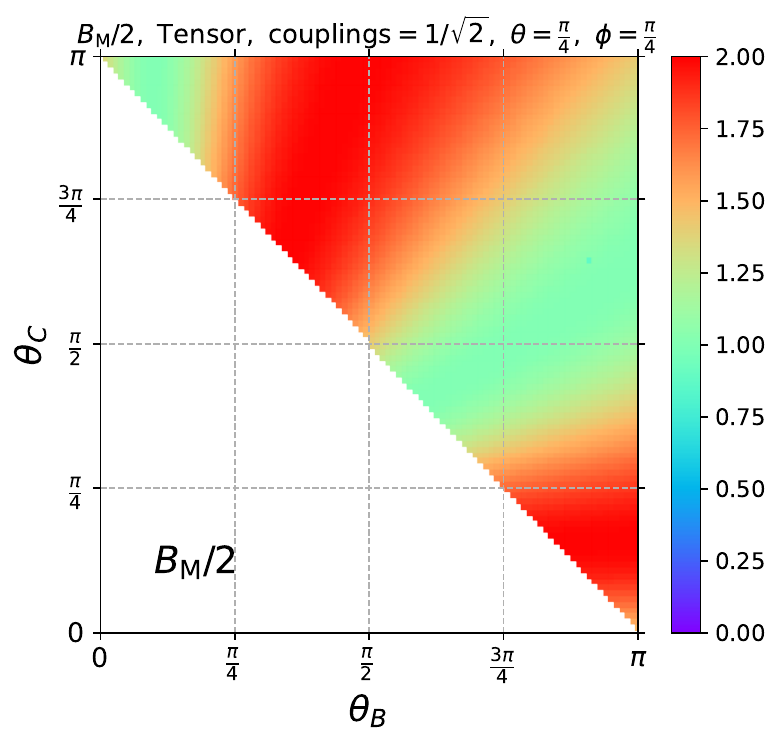}
\includegraphics[scale=.33]{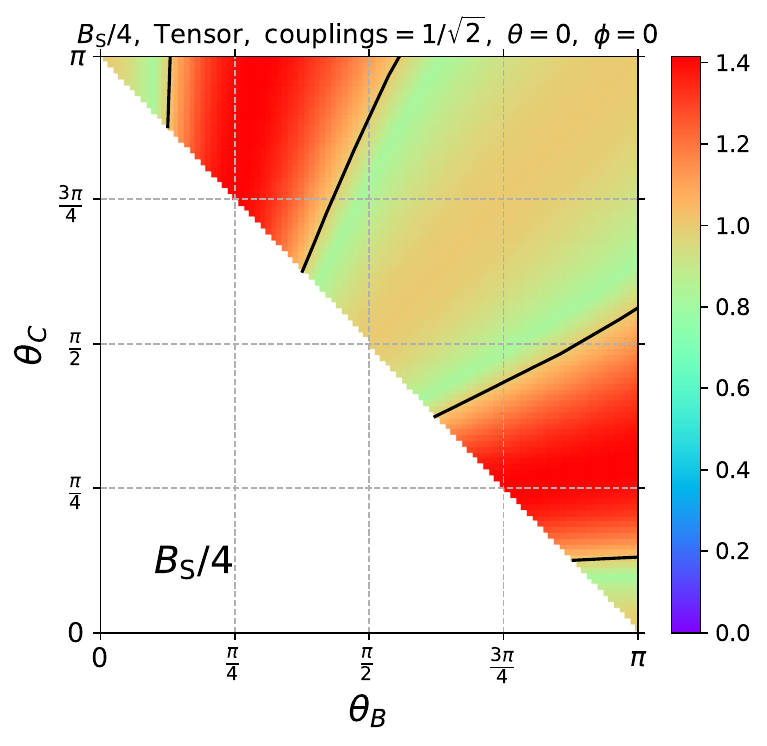}
\includegraphics[scale=.33]{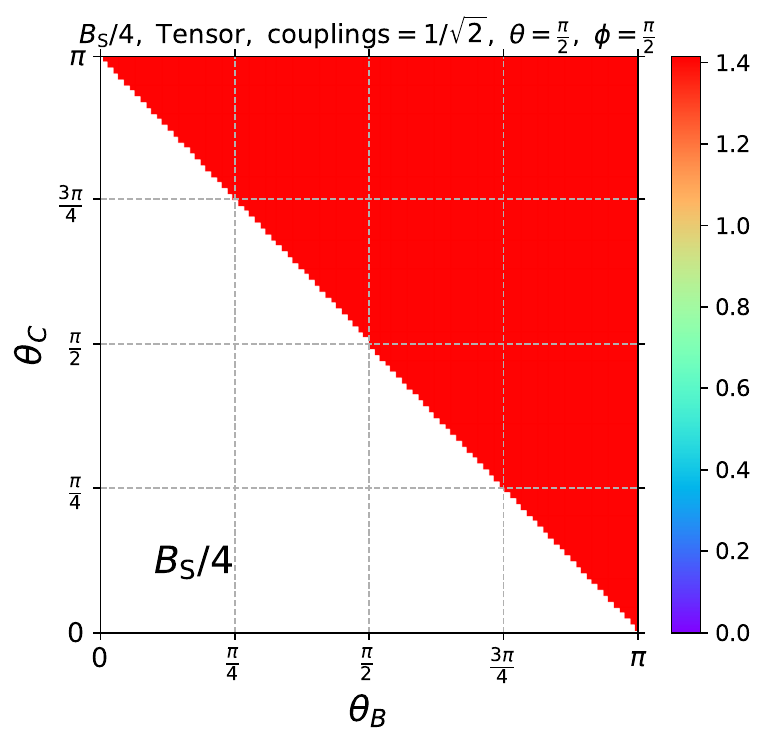}
\includegraphics[scale=.33]{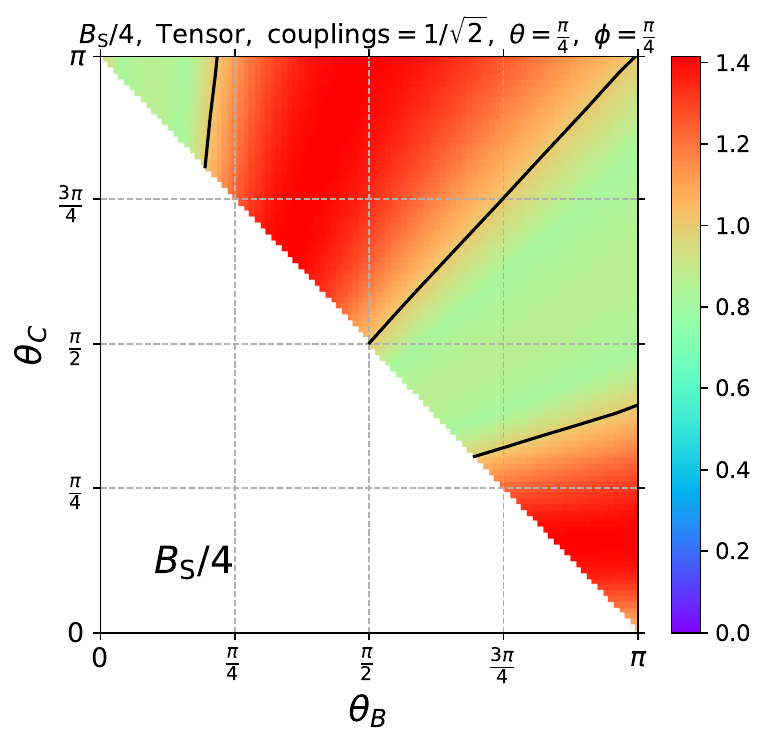}
\includegraphics[scale=.33]{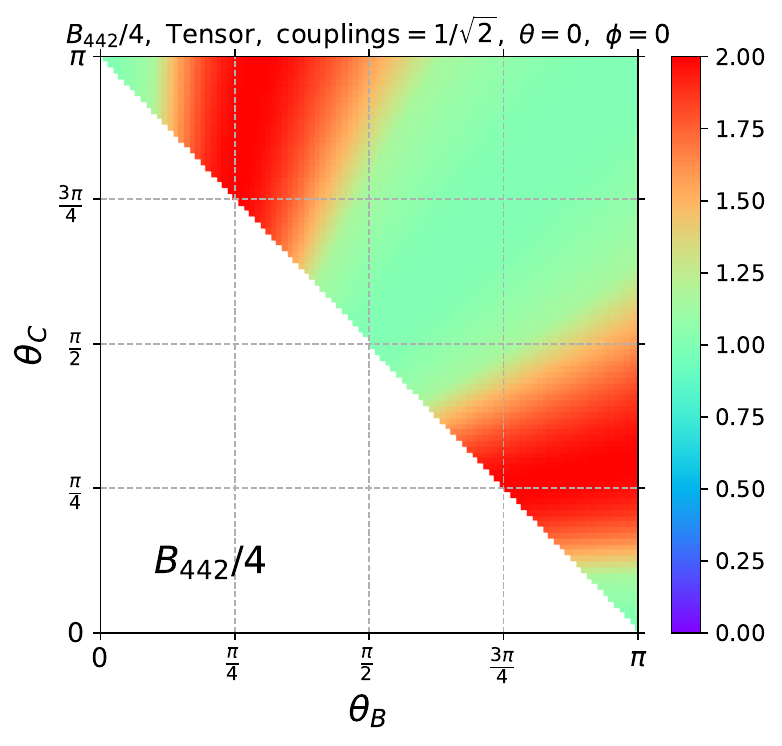}
\includegraphics[scale=.33]{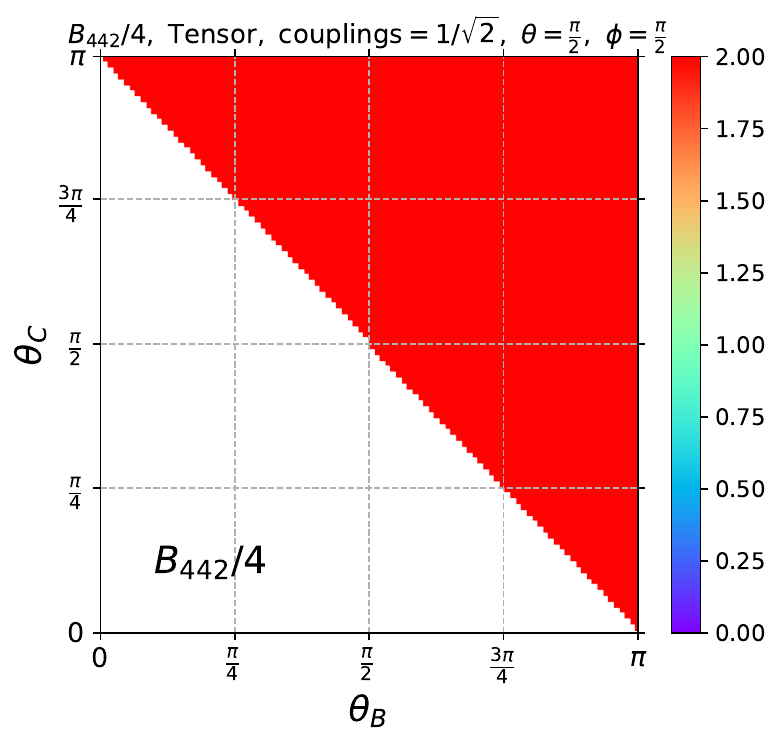}
\includegraphics[scale=.33]{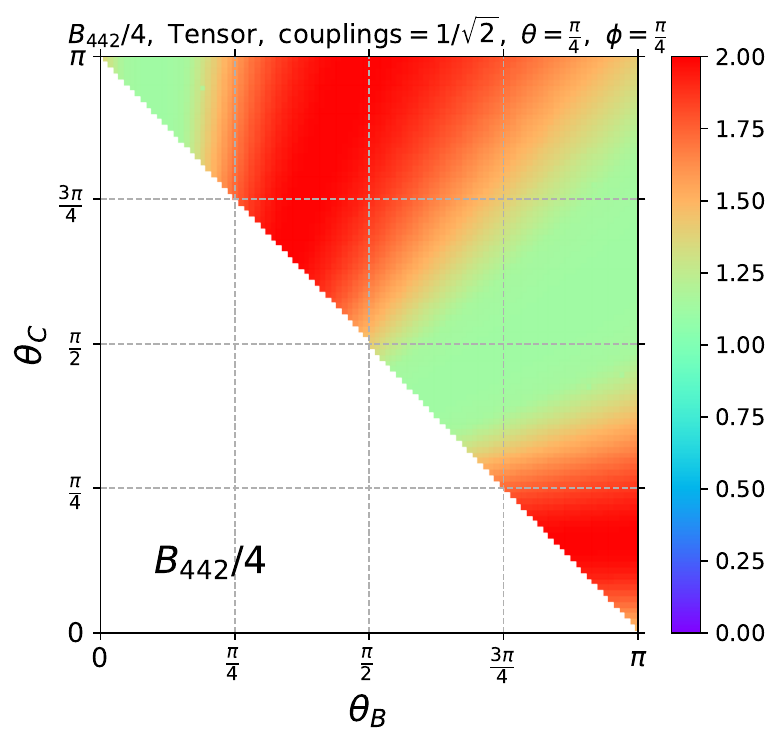}
\caption{\label{fig:2d_tens}
\small 
The dependence of the GTE measure, $F_3$, (the first line), $\eBM$ (the second line), $\eBS$ (the third line) and $\eBB4$ (the fourth line) on the decay angles $\theta_B$ and $\theta_C$, evaluated assuming the tensor-type interaction in Eq.\ \eqref{Ltensor}.
The convention of the figure is the same as 
of Fig.\ \ref{fig:2d_vec}.
}
\end{figure}

Fig.\ \ref{fig:2d_tens} presentes the dependence of $F_3$ (the first line), $\eBM$ (the second line), $\eBS$ (the third line) and $\eBB4$ (the fourth line) on the decay angles $\theta_B$ and $\theta_C$, evaluated assuming the tensor-type interaction in Eq.\ \eqref{Ltensor}.
The convention of the figure is the same as of Fig.\ \ref{fig:2d_vec}

In the left column, where the initial spin is aligned along the $z$-axis ($\theta = \phi = 0$), the plots exhibit symmetry under $\theta_B \leftrightarrow \theta_C$, as this operation corresponds to a rotation about the $z$-axis. 
The GME measure $F_3$ vanishes at the top-left [($\theta_B, \theta_C) = (0, \pi)]$ and bottom-right [$(\theta_B, \theta_C) = (\pi, 0)$)] corners, as well as along the line defined by $\theta_B = \theta_C$. 
At these points and along this line, the tight $4 \times 4 \times 2$ bound is saturated, $\eBB4 = 4$. 
Outside these regions, no non-violation of the FLR bound is observed for $\eBB4$.
In contrast, the Mermin inequality is violated within finite regions where $F_3$ takes small values. 
In these regions with $\eBB4 > 4$, the Mermin observable fails to detect the nonlocality associated with the fully local-real theory.
In regions where $F_3$ takes small values, the Svetlichny bound, $\eBS \leq 4$, is not violated. 
Conversely, in the two regions around $\theta_B \sim \frac{\pi}{4}$ and $\theta_C \sim \frac{\pi}{4}$, the GTE measure reaches its maximum value, $F_3 = 1$. 
Within these regions, the Bell-type observables also attain their maximum values, saturating their quantum mechanical bounds: $\eBM_{\rm QM}^{\rm max} = 4$,
$\eBS_{\rm QM}^{\rm max} = 4\sqrt{2}$
and
$\eBB4_{\rm QM}^{\rm max} = 8$.

In the middle column, the initial spin is fixed as ${\bf n} = {\bf e}_y$ ($\theta=\phi=\frac{\pi}{2}$).
In this configuration, the quantum state \eqref{M_tens} reduces to the maximally entangled GHZ state, 
$|\Psi \ket = [|+++ \ket - i |---\ket]
\sqrt{2}$ independently of the decay angles. 
As a result, $F_3$ and all Bell-type observables take the quantum mechanical maximum values across the $(\theta_B, \theta_C$) plane. 

We take the initial spin to be aligned along $({\bf e}_x + {\bf e}_y + {\bf e}_z)$, correspondng to $\theta = \phi = \frac{\pi}{2}$, in the right column.
The behaviour of the observables is similar to the case with ${\bf n} = {\bf e}_z$ (shown in the left column), as two strips appear where the observables reach their quantum mechanical maximum values.
However, there are notable differences. 
First, the response of the observables is asymmetric under the exchange $\theta_B \leftrightarrow \theta_C$. 
Second, $F_3$ does not vanish anywhere in the ($\theta_B, \theta_C$) plane. 
Additionally, the Mermin inequality is consistently violated across the entire plane, indicating that we cannot have a fully local-realistic theory that can account for all experimental results about the three-particle correlations for any decay configuration.
On the other hand, there are two regions in the decay plane where Svetlichny’s bound is not violated. 
In these regions, a bipartite local-real theory may exist that explains all experimental results involving three-particle correlations.

\section{Conclusion}
\label{sec:concl}

In this work, we investigated a three-body entanglement and non-locality arising from the decay of a massive fermion into three massless spin-1/2 particles.  
Employing a framework based on general four-fermion interactions, we have analysed the resulting correlations through the lens of several Bell-type inequalities, including the Mermin inequality and the tight $4 \times 4 \times 2$ inequality.  
Our analysis reveals the complex interplay between entanglement and non-locality in this three-particle system.

We demonstrated that the scalar interaction leads to bi-separable states exhibiting maximal bipartite entanglement but lacking genuine tripartite entanglement or non-locality. 
In contrast, the vector interaction generates genuinely tripartite entangled states, which violate both fully local-real and bipartite local-real bounds under specific conditions. 
The tensor interaction produces states that can saturate the quantum bounds of both Mermin and Svetlichny inequalities, highlighting the rich structure of three-partite entanglement.

Our numerical optimization of measurement axes, particularly for the tight $4 \times 4 \times 2$ inequality, has yielded insights into the detection capabilities of different Bell-type inequalities for various initial spin and decay configurations. 
We observe regions where the Mermin inequality fails to detect non-locality while the $4 \times 4 \times 2$ inequality successfully does.   
This highlights the need for properly choosing observables for experimental tests of quantum mechanics in multipartite systems.

The results presented in this paper advance our understanding of non-locality in three-particle systems. 
The presented analysis, combining theoretical and numerical optimization of the measurement axes, provides a useful framework for future experimental investigations of three-body non-locality in particle physics, ultimately contributing to a deeper understanding of the fundamental principles of quantum mechanics.  
Further research should focus on extending these techniques to more complex decay channels with higher spins and exploring the implications for specific particle decay processes.

\section{Acknowledgements}
We thank the organizers of two workshops: the first \textit{Quantum Observables for Collider Physics}, November 2023, funded by the  Galileo Galilei Institute for Theoretical Physics of the \textit{Istituto Nazionale di Fisica Nucleare}, and the second \textit{Quantum Tests in Collider Physics}, October 2024, supported by the Institute for Particle Physics Phenomenology, Durham, and Merton College, Oxford and for hosting several of the authors during the preparation of this manuscript.
SSA and PH acknowledge support from the National Science Centre in Poland under the research grant Maestro (2021/42/A/ST2/00356).

\appendix

\section{Saturating an algebraic bound by No-Signalling correlations}
\label{sec:NS}

\subsection{General remarks on No-Signalling correlations}
A natural question that has been considered is what physical statistics are allowed in a possible post-quantum theory. 
In particular --- what physical constraints should they obey? The most natural constraints are so-called {\it No-Signalling conditions} \cite{PopescuRohrlich94}, where a change of a setting in one laboratory can not change and output statistics in other laboratories. 
The only additional assumption is that the laboratories are space-like separated, and there is no further dependence on space-time \cite{PopescuRohrlich94}. 
Remarkably, there is also an extended formalism called {\it relativistically causal} \cite{Grunhaus,Horodecki-Ramanathan-2019}, which explicitly depends on the space-time variables and, in principle, allows for so-called jamming of correlations. 
However, any conceivable extension of the Standard Model seems to fit the no-signalling post-quantum formalism, so this is one that we consider here. 
In the case of bipartite statistics, the no-signalling conditions are 
formulated as follows (\cite{PopescuRohrlich94}):
\begin{eqnarray}
&&\sum_{a}P(a,b|A_i,B_{j}) = \sum_{a}P(a,b|A_{i'},B_j) := P(b|B_j) \ \ \textnormal{ for all } \  i, i' \nonumber , \\
&&\sum_{b}P(a,b|A_i,B_{j}) = \sum_{b}P(a,b|A_i,B_{j'}) := P(a|A_i) \ \ \textnormal{for all} \ j,  j'.
\end{eqnarray} 

Similarly, for the three observers, we have analogous conditions (cf. \cite{Bell-nonlocality}):

\begin{eqnarray}
&&\sum_{a}P(a,b,c|A_i,B_{j},C_{k}) = \sum_{a}P(a,b,c|A_{i'},B_j,C_{k}) := P(b,c|B_j,C_{k}) \  \textnormal{ for all} \  i, i' \nonumber ,\\
&&\sum_{b}P(a,b,c|A_i,B_{j},C_{k}) =\sum_{b}P(a,b,c|A_i,B_{j'},C_{k}) := P(a,c|A_i, C_{k}) \ \ \textnormal{for all} \ j, j' \nonumber , \\
&&\sum_{c}P(a,b,c|A_i,B_{j},C_{k}) =\sum_{c}P(a,b,c|A_i,B_j,C_{k'}) := P(a,b|A_i, B_{j}) \ \ \textnormal{for all} \ k, k' .
\label{3-nosignalling}
\end{eqnarray} 
The above scheme can be extended to many parties as well \cite{Bell-nonlocality}. 
The interpretation is clear: local changes in settings can not alter remote statistics. 

\subsection{PR-boxes: no-signalling correlations saturating the 
CHSH inequality}

Below, we shall recall the behaviour of the CHSH inequality from the perspective of the no-signalling framework  (see \cite{PopescuRohrlich94,Bell-nonlocality}).
In comparison to the main text, we shall change the indices of 
the values of the CHSH operator \eqref{CHSH}.
Instead of $A_i$, $B_j$ with $i,j=1,2$ we shall use the binary values 
$A_x$, $B_y$ with $x,y=0,1$ ($x(i)=i-1$, $y(i)=i-1$)
which gives:
\begin{equation}
{\cal B}_{\rm CHSH} = A_0 (B_0 + B_1) + A_0 (B_0 - B_1)\,. 
\label{CHSH-b}
\end{equation}
Additionally, with a slight abuse of notation, we shall label the outcomes $A_x, B_y \in \{ +1, -1\}$ using the indices $a,b \in \{0, 1\}$, such that the outcomes are expressed as $A_x = (-1)^a$ and $B_y = (-1)^b$.
Given the family of statistics  ${\boldsymbol{P}}=\{ P(a,b|A_x,B_{y}) \}$ with binary inputs and the above binary enumeration of the 
results, one can write the mean value of  \eqref{CHSH-b} as
\begin{equation}
\begin{split}
\langle {\cal B}_{\rm CHSH} \rangle_{\boldsymbol{P}} &= \left\langle{A_0B_0}\right\rangle_{\boldsymbol{P}} +\left\langle{A_0B_1}\right\rangle_{\boldsymbol{P}}
+\left\langle{A_1B_0}\right\rangle_{\boldsymbol{P}} - \left\langle{A_1B_1}\right\rangle_{\boldsymbol{P}},\\
&=\sum_{x,y = 0,1} (-1)^{xy} \left\langle{A_x B_y}\right\rangle_{\boldsymbol{P}}.
\label{CHSH-2}
\end{split}
\end{equation}
However, we know that because of the binary character of the observables $A_x$ and $B_y$, one has $\left\langle{A_xB_y}\right\rangle_{\boldsymbol{P}} = \sum_{a b} (-1)^{a\oplus b} P(a,b|A_x, B_y)$. 
Therefore, we have
\begin{align}
    \left\langle{\cal{B}_{\rm CHSH}}\right\rangle_{\boldsymbol{P}} &= \sum_{a,b,x,y=0,1} (-1)^{a \oplus b} (-1)^{xy}
P(a,b|A_x,B_{y}) 
\,. 
\end{align}
Here, $\oplus$ means addition modulo 2 (for example, $0 \oplus 1=1$ but $1 \oplus 1=0$). 

The following no-signalling statistics ${\boldsymbol{P}^{s}}=\{ P^{s}(a,b|A_x,B_{y}) \}$ ($s=0,1$), called {\it Popescu-Rohrlich (PR) boxes} \cite{PopescuRohrlich94}
\begin{equation}\label{PR-boxes}
P^{s}(a,b|A_x,B_{y}) =
\begin{cases}
\frac{1}{2} & \text{$ a \oplus b \oplus s = xy$} \\
0 & \text{otherwise}
\end{cases},
\end{equation}
are known to reach the algebraic maximum value of CHSH, namely,
\begin{equation}
\langle {\cal B}_{\rm CHSH} \rangle_{\boldsymbol{P}^{s}} = (-1)^{s} 4 \,.
\end{equation}
$\boldsymbol{{P}}^{0}$ and $\boldsymbol{{P}}^{1}$ are often called 
PR-box and anti-PR-box, respectively. 
It is easy to see that those statistics are no-signalling since their marginals are fully random:
\begin{eqnarray}
&& \sum_{a=0,1}P^{s}(a,b|A_x,B_{y})=P(b)=\frac{1}{2}\nonumber, \\
&& \sum_{b=0,1}P^{s}(a,b|A_x,B_{y})=P(a)=\frac{1}{2}. 
\label{2-nosignalling}
\end{eqnarray}
so they {\it do not depend not only on
the distant settings (which is required by no-signalling conditions) but also on the local ones}. 
If one defines the completely random (therefore manifestly no-signalling)
bipartite box 
\begin{equation}\label{noise}
P^{*}(a,b|A_x,B_{y}) = \frac{1}{4}\,, \ \ \textnormal{for all} \ a,b,x,y \in \{ 0,1 \} \,,
\end{equation}
then the quantum extremal values can be achieved by a special convex combination  
${\boldsymbol{P}_{Q}^{s}}= (1-\frac{1}{\sqrt{2}}) \boldsymbol{P}^{*} + \frac{1}{\sqrt{2}} \boldsymbol{P}^{s}$
which reproduces the statistics of measurements in a maximally entangled quantum state $|\Psi\rangle=\frac{1}{\sqrt{2}} (|0 1 \rangle + |10\rangle )$ with the measurement settings
$(\theta_{0}^{A},\theta_{1}^{A}) =(\frac{\pi}{4}, -\frac{\pi}{4})$ and $(\theta_{0}^{B}, \theta_{1}^{B}) = (0, \frac{\pi}{2})$ with $\theta^{A/B}_{x/y}$ being the angle from the $z$-axis on the $x$-$z$ plane.
In short:
 \begin{equation}
\langle {\cal B}_{\rm CHSH} \rangle_{\boldsymbol{P}_{Q}^{s}} = (-1)^{s}2\sqrt{2}.
\end{equation}
The remarkable property, which we shall need later, is that 
unbiased mixture of the PR and anti-PR boxes produces a maximally random one
\begin{equation}
\boldsymbol{P}^{*}= \frac{1}{2} \boldsymbol{P}^{0 \oplus r} + \frac{1}{2} \boldsymbol{P}^{1\oplus r}.
\label{property}
\end{equation}
Here, we added a trivially arbitrary bit $r=0,1$ (which either permutes or not the boxes in the mixture) for the sake of future analysis.

\textcolor{blue}{}
\subsection{No-signalling correlations saturating 
 algebraic bound of ${\cal B}_{442}$}

For the tight $4 \times 4 \times 2$ observable, there are four settings on Alice and Bob's sites
$A_{i}$ ($i=1,2,3,4$) and  $B_{j}$ ($j=1,2,3,4$) and two at Charlie's site $C_{k}$ $(k=1,2)$. 
We shall use again a binary mapping for the latter; $z=k-1$ getting the mapping $C_k = (C_1,C_2) \to (C_0,C_1) = C_z$, 
and binary expansions for the former two;
$A_i = (A_1,A_2,A_3,A_4) \to (A_{00},A_{01},A_{10},A_{11}) = A_{u u'}$ and
$B_j = (B_1,B_2,B_3,B_4) \to (B_{00},B_{01},B_{10},B_{11}) = B_{v v'}$.
Generally, the inequality will now involve the triples $A_{uu'}B_{vv'}C_{z}$. 
This gives us eventually the observable ${\cal B}_{442}$ in the form 
\begin{equation}
{\cal B}_{442} = [A_{00} (B_{00} + B_{01}) + A_{01} (B_{00} - B_{01})](C_0 + C_1) 
+ [A_{10} (B_{10} + B_{11}) + A_{11} (B_{10} - B_{11})](C_0 - C_1).
\label{CHSH-1}
\end{equation}

Note that given any tripartite box $\boldsymbol{P}= \{ P(a,b,c|A_{uu'},B_{vv'},C_{z}) \} $ the mean value of the above inequality would be 
 \begin{equation}
\langle {\cal B}_{442}\rangle_{\boldsymbol{P}}=
\sum_{a,b,c,u,u',v,v',z=0,1} \delta_{uv}(-1)^{a \oplus b \oplus c } (-1)^{u'v'}(-1)^{uvz} P(a,b,c|A_{uu'},B_{vv'},C_{z}).
\label{B442P}
 \end{equation}
Let us consider the following tripartite boxes $\boldsymbol{\tilde{P}}$
\begin{equation}\label{3-party-PR}
\tilde{P}(a,b,c|A_{uu'},B_{vv'},C_{z}) =
\begin{cases}
\frac{1}{4} & \text{$ a \oplus b \oplus c \oplus uvz= u'v'$} \\
0 & \text{otherwise}
\end{cases}
\end{equation}
One can check that $\boldsymbol{\tilde{P}}$ makes the expectation value of ${\cal B}_{442}$ reach its algebraic maximum, i.e.\footnote{There are other distributions that lead to the algebraic maximum of $\langle {\cal B}_{442} \rangle$.  
For example, one can consider similar distributions as \eqref{3-party-PR} but changing the condition for $\tfrac{1}{4}$ to $a \oplus b \oplus c \oplus uz = u' v'$
or $a \oplus b \oplus c \oplus vz = u' v'$.
This can be understood because $\delta_{uv} (-1)^{uvz} = \delta_{uv} (-1)^{uz} = \delta_{uv} (-1)^{vz}$ in Eq.\ \eqref{B442P}. }
\begin{equation}
\langle {\cal B}_{442}\rangle_{\boldsymbol{\tilde{P}}} = 16\,.
\end{equation} 
This can be easily proven by using the fact that $\boldsymbol{\tilde{P}}$ can be expressed in terms of bipartite Popescu-Rohrlich boxes $\boldsymbol{P}^{s}$ in Eq.\ \eqref{PR-boxes}, namely, 
\begin{equation}
\tilde{P}(a,b,c|A_{uu'},B_{vv'},C_{z})= \frac{1}{2}P^{c \oplus uvz}(a,b,|A_{u'}, B_{v'}).
\label{relation-to-PR}
\end{equation}
The last thing is to prove that $\boldsymbol{\tilde{P}}$ is no-signalling. This, however, is immediate. 
First, exploiting (\ref{property}), $r=uvz$, and with (\ref{relation-to-PR}) we get $\sum_{c=0,1} \tilde{P}(a,b,c|A_{uu'},B_{vv'},C_{z})=P^*(a,b,| A_{u'}, B_{v'}) = \frac{1}{4}$, i.e.\ it is independent on all the input values $u,u'$, $v,v'$, $z$.
Second, exploiting (\ref{2-nosignalling}) with $s=c\oplus uvz$ together with (\ref{relation-to-PR}), we again get  
$\sum_{a=0,1} \tilde{P}(a,b,c|A_{uu'},B_{vv'},C_{z})=\sum_{b=0,1} \tilde{P}(a,b,c|A_{uu'},B_{vv'},C_{z})$ \\ $=\frac{1}{4}$, which are completely independent on all the inputs.


\section{A semi-analytical optimisation for ${\cal B}_{442}$}
\label{sec:app}

The tight $4 \times 4 \times 2$ observable, ${\cal B}_{442}$, in \eqref{B442} assumes in total ten measurement axes for three observers. 
In order to maximise its sensitivity in detecting the corresponding non-locality, one must optimise these measurement axes.  
A numerical optimisation over ten axes is computationally expansive, and we, therefore, adopt a semi-analytical optimisation described below.  
In this procedure, we analytically optimise the measurement axes up to a pair of orthonormal vectors. 
The $\bra {\cal B}_{442} \ket_\rho$ can be maximised numerically over the remaining degrees of freedom. 

Our task is to maximise the expectation value of the tight $4 \times 4 \times 2$ observable
\bea
&& \bra {\cal B}_{442} \ket_\rho 
\,=\, \sum_{i,j,k} \, T_{ijk} \Big\{
\left[
[ \vec A_{1} ]_i [ \vec B_{1}+ \vec B_{2} ]_j + [ \vec A_{2} ]_i [ \vec B_{1} - \vec B_{2} ]_j 
\right] [ \vec C_{1} + \vec C_{2} ]_k
\nn \\
&& ~~~~+ 
\left[
[ \vec A_{3} ]_i [ \vec B_{3} + \vec B_{4} ]_j + [ \vec A_{4}]_i [ \vec B_{3} - \vec B_{4}]_j
\right]
[ \vec C_{1} - \vec C_{2} ]_k 
\Big\}
\,,
\label{B442_a}
\eea
over all measurement axes represented by ten unit vectors, $\vec A_I$, $\vec B_I$ ($I \in \{ 1,2,3,4 \}$) and $\vec C_J$ ($J = 1,2$).

We define pairs of orthonormal vectors, $\vec C_\pm$, ($\vec C_+ \cdot \vec C_- = 0$) 
as
\be
\vec C_1 + \vec C_2 \equiv 2 c_\gamma \vec C_+, 
~~~~
\vec C_1 - \vec C_2 \equiv 2 s_\gamma \vec C_-,
\ee
where $| \vec C_\pm |^2 = 1$ implies $c_\gamma^2 + s_\gamma^2 = 1$.
Similarly, we define 
\bea
&&\vec B_1 + \vec B_2 \equiv 2 c_\delta \vec D_+, 
~~~~
\vec B_1 - \vec B_2 \equiv 2 s_\delta \vec D_-,
\nn \\
&&\vec B_3 + \vec B_4 \equiv 2 c_\epsilon \vec E_+, 
~~~~
\vec B_3 - \vec B_4 \equiv 2 s_\epsilon \vec E_-,
\eea
with $| \vec D_\pm |^2 = | \vec E_\pm |^2 = 1$,
$ \vec D_+ \cdot \vec D_- =  \vec E_+ \cdot \vec E_-  = 0$
and $c_\delta^2 + s_\delta^2 =c_\epsilon^2 + s_\epsilon^2 =1$.
We further define rank-2 tensors ($3 \times 3$ matrices) $T_\pm$ as
\be
[T_\pm]_{ij} = \sum_k T_{ijk} [ \vec C_\pm ]_k,
\ee
In terms of these new vectors and tensors, Eq.\ \eqref{B442_a} can be written as
\be
\bra {\cal B}_{442} \ket_\rho = 
4 \left[ c_\gamma X + s_\gamma Y \right]
\ee
with
\bea
X &=& c_\delta \left( \vec A_1 \cdot T_+ \cdot \vec D_+ \right)
+ s_\delta \left( \vec A_2 \cdot T_+ \cdot \vec D_- \right) \,,
\nn \\
Y &=&
c_\epsilon \left( \vec A_3 \cdot T_- \cdot \vec E_+ \right)
+ s_\epsilon \left( \vec A_4 \cdot T_- \cdot \vec E_- \right) \,.
\eea
For given $T_{ijk}$, $\vec A_I$, $\vec B_I$, and $\vec C_{\pm}$, $X$ and $Y$ are fixed, and one can optimise $c_\gamma$ and $s_\gamma$ as
\be
c_\gamma = \frac{X}{\sqrt{X^2 + Y^2}},
~~~~
s_\gamma = \frac{Y}{\sqrt{X^2 + Y^2}},
\ee
which leads to $\bra {\cal B}_{442} \ket_\rho = 
4 \sqrt{ X^2 + Y^2 }$.
This quantity is further optimised over $c_\delta,s_\delta,c_\epsilon$ and $s_\epsilon$ with  
\bea
&& c_\delta = \frac{\vec A_1 \cdot T_+ \cdot \vec D_+}{
\sqrt{\left[ \vec A_1 \cdot T_+ \cdot \vec D_+ \right]^2 
+
\left[ \vec A_2 \cdot T_+ \cdot \vec D_- \right]^2
}},
~~~
s_\delta = \frac{ \vec A_2 \cdot T_+ \cdot \vec D_- }{
\sqrt{\left[ \vec A_1 \cdot T_+ \cdot \vec D_+ \right]^2 
+
\left[ \vec A_2 \cdot T_+ \cdot \vec D_- \right]^2
}},
\nn \\
&& c_\epsilon = 
\frac{\vec A_3 \cdot T_- \cdot \vec E_+}{
\sqrt{\left[ \vec A_3 \cdot T_- \cdot \vec E_+ \right]^2 
+
\left[ \vec A_4 \cdot T_- \cdot \vec E_- \right]^2
}},
~~~
s_\epsilon = 
\frac{\vec A_4 \cdot T_- \cdot \vec E_-}{
\sqrt{\left[ \vec A_3 \cdot T_- \cdot \vec E_+ \right]^2 
+
\left[ \vec A_4 \cdot T_- \cdot \vec E_- \right]^2
}},
\nn \\
\eea
which results in
\be
\bra {\cal B}_{442} \ket_\rho 
\,=\,
4 \sqrt{
\left[ \vec A_1 \cdot T_+ \cdot \vec D_+ \right]^2 
+
\left[ \vec A_2 \cdot T_+ \cdot \vec D_- \right]^2
+
\left[ \vec A_3 \cdot T_- \cdot \vec E_+ \right]^2 
+
\left[ \vec A_4 \cdot T_- \cdot \vec E_- \right]^2
}\,.
\ee
For given $T_{ijk}$, $\vec D_\pm$, $\vec E_\pm$ and $\vec C_\pm$, we maximise the first term by taking $\vec A_1$ in the direction of $T_+ \cdot \vec D_+$.
The other terms can be maximised in the same way, that is, we take
\be
\vec A_1 = \frac{ T_+ \cdot \vec D_+ }{ \left| T_+ \cdot \vec D_+ \right|  }, 
~~~
\vec A_2 = \frac{ T_+ \cdot \vec D_- }{ \left| T_+ \cdot \vec D_- \right|  }, 
~~~
\vec A_3 = \frac{ T_- \cdot \vec E_+ }{ \left| T_- \cdot \vec E_+ \right|  }, 
~~~
\vec A_4 = \frac{ T_- \cdot \vec E_- }{ \left| T_- \cdot \vec E_- \right|  }.
\ee
After these optimisation processes, $\bra {\cal B}_{442} \ket_\rho $ can be written as
\be
\bra {\cal B}_{442} \ket_\rho 
= 4 \sqrt{ 
\left[ \vec D_+ \cdot U_+ \cdot \vec D_+ \right]
+
\left[ \vec D_- \cdot U_+ \cdot \vec D_- \right]
+
\left[ \vec E_+ \cdot U_- \cdot \vec E_+ \right]
+
\left[ \vec E_- \cdot U_- \cdot \vec E_- \right]
},
\label{B442_b}
\ee
with $U_\pm \equiv (T_\pm)^T \cdot T_\pm$.

Finally for given $T_{ijk}$ and $\vec C_\pm$, 
we optimise the above expression over two pairs of orthonormal vectors $(\vec D_+, \vec D_-)$ and 
$(\vec E_+, \vec E_-)$.
We first note that $U_\pm$ are real symmetric matrices.
Therefore, their eigenvalues are non-negative, and the eigenvectors corresponding to different eigenvalues are orthogonal.  
This means Eq.\ \eqref{B442_b} can be maximised by taking $(\vec D_+, \vec D_-)$ to be the eigenvectors corresponding to the two largest eigenvalues of $U_+$, $\lambda_1^+$ and $\lambda_2^+$.
Similarly, $(\vec E_+, \vec E_-)$ should be taken as the eigenvectors corresponding to the two largest eigenvalues of $U_-$, $\lambda_1^-$ and $\lambda_2^-$.  
As a result, we have  
\be
\bra {\cal B}_{442} \ket_\rho
= 4 \sqrt{ \lambda_1^+ + \lambda_2^+ + \lambda_1^- + \lambda_2^- }  \,.
\label{B442_c}
\ee
Note that this expression has some similarity to the analytical 
formula for the optimised ${\cal B}_{\rm CHSH}$ derived in \cite{HORODECKI1995340}.
For a given state, $\rho$ (or equivalently $T_{ijk}$), 
the matrices $U_\pm$ are functions of a pair of orthonormal vectors $(\vec C_+, \vec C_-)$.
One, therefore, still has to optimise the above expression over $\vec C_\pm$.
Since a pair of orthonormal vectors can be parametrised by one polar angle, $0 \leq \bar \theta \leq \pi$ 
and two azimuthal angles, $0 \leq \phi_1, \phi_2 \leq 2 \pi$ by  
\be
\vec C_+
=
\begin{pmatrix}
\sin \bar \theta \cos \phi_1 \\
\sin \bar \theta \sin \phi_1 \\
\cos \bar \theta
\end{pmatrix},
~~~~
\vec C_-
=
\begin{pmatrix}
\cos \bar \theta \cos \phi_1  \cos \phi_2- \sin \phi_1 \sin \phi_2  
 \\
\cos \bar \theta \sin \phi_1 \cos \phi_2 + \cos \phi_1 \sin \phi_2 
\\
- \sin \bar \theta \cos \phi_2 
\end{pmatrix},
\ee
the expression \eqref{B442_c} can be maximised over these three angles. 
Compared to an optimisation of the initial expression \eqref{B442_a} over ten measurement axes, the numerical optimisation over three angles is computationally much less expensive. 

Since the elements of the spin correlation matrix cannot exceed one and $|\vec{C}_{\pm}| = 1$, the maximum eigenvalue of the $U_\pm$ matrix is bounded by 1.
This and the expression \eqref{B442_c} leads to the quantum upper bound 
\be
\bra {\cal B}_{442} \ket_{\rm QM} \, \leq \, 8\,.
\ee

\bibliographystyle{JHEP}
\bibliography{refs} 

\providecommand{\href}[2]{#2}\begingroup\raggedright\begin{thebibliography}{10}

\bibitem{Schrödinger_1935}
E.~Schrödinger, \emph{Discussion of probability relations between separated
  systems}, \href{https://doi.org/10.1017/S0305004100013554}{\emph{Mathematical
  Proceedings of the Cambridge Philosophical Society} {\bfseries 31} (1935)
  555–563}.

\bibitem{Einstein:1935rr}
A.~Einstein, B.~Podolsky and N.~Rosen, \emph{{Can quantum mechanical
  description of physical reality be considered complete?}},
  \href{https://doi.org/10.1103/PhysRev.47.777}{\emph{Phys. Rev.} {\bfseries
  47} (1935) 777}.

\bibitem{PhysicsPhysiqueFizika}
J.S.~Bell, \emph{On the einstein podolsky rosen paradox},
  \href{https://doi.org/10.1103/PhysicsPhysiqueFizika.1.195}{\emph{Physics
  Physique Fizika} {\bfseries 1} (1964) 195}.

\bibitem{Clauser:1969ny}
J.F.~Clauser, M.A.~Horne, A.~Shimony and R.A.~Holt, \emph{{Proposed experiment
  to test local hidden variable theories}},
  \href{https://doi.org/10.1103/PhysRevLett.23.880}{\emph{Phys. Rev. Lett.}
  {\bfseries 23} (1969) 880}.

\bibitem{Freedman:1972zza}
S.J.~Freedman and J.F.~Clauser, \emph{{Experimental Test of Local
  Hidden-Variable Theories}},
  \href{https://doi.org/10.1103/PhysRevLett.28.938}{\emph{Phys. Rev. Lett.}
  {\bfseries 28} (1972) 938}.

\bibitem{Aspect:1981nv}
A.~Aspect, P.~Grangier and G.~Roger, \emph{{Experimental realization of
  Einstein-Podolsky-Rosen-Bohm Gedankenexperiment: A New violation of Bell's
  inequalities}}, \href{https://doi.org/10.1103/PhysRevLett.49.91}{\emph{Phys.
  Rev. Lett.} {\bfseries 49} (1982) 91}.

\bibitem{Weihs:1998gy}
G.~Weihs, T.~Jennewein, C.~Simon, H.~Weinfurter and A.~Zeilinger,
  \emph{{Violation of Bell's inequality under strict Einstein locality
  conditions}}, \href{https://doi.org/10.1103/PhysRevLett.81.5039}{\emph{Phys.
  Rev. Lett.} {\bfseries 81} (1998) 5039}
  [\href{https://arxiv.org/abs/quant-ph/9810080}{{\ttfamily
  quant-ph/9810080}}].

\bibitem{Hensen:2015ccp}
B.~Hensen et~al., \emph{{Loophole-free Bell inequality violation using electron
  spins separated by 1.3 kilometres}},
  \href{https://doi.org/10.1038/nature15759}{\emph{Nature} {\bfseries 526}
  (2015) 682} [\href{https://arxiv.org/abs/1508.05949}{{\ttfamily
  1508.05949}}].

\bibitem{Giustina:2015yza}
M.~Giustina et~al., \emph{{Significant-Loophole-Free Test of
  Bell\textquoteright{}s Theorem with Entangled Photons}},
  \href{https://doi.org/10.1103/PhysRevLett.115.250401}{\emph{Phys. Rev. Lett.}
  {\bfseries 115} (2015) 250401}
  [\href{https://arxiv.org/abs/1511.03190}{{\ttfamily 1511.03190}}].

\bibitem{Stevens:2015awv}
J.~Stevens et~al., \emph{{Strong Loophole-Free Test of Local Realism}},
  \href{https://doi.org/10.1103/PhysRevLett.115.250402}{\emph{Phys. Rev. Lett.}
  {\bfseries 115} (2015) 250402}.

\bibitem{Handsteiner:2016ulx}
J.~Handsteiner et~al., \emph{{Cosmic Bell Test: Measurement Settings from Milky
  Way Stars}},
  \href{https://doi.org/10.1103/PhysRevLett.118.060401}{\emph{Phys. Rev. Lett.}
  {\bfseries 118} (2017) 060401}
  [\href{https://arxiv.org/abs/1611.06985}{{\ttfamily 1611.06985}}].

\bibitem{Storz:2023jjx}
S.~Storz et~al., \emph{{Loophole-free Bell inequality violation with
  superconducting circuits}},
  \href{https://doi.org/10.1038/s41586-023-05885-0}{\emph{Nature} {\bfseries
  617} (2023) 265}.

\bibitem{Tsirelson}
B.S.~Cirel'son, \emph{Quantum generalizations of bell's inequality},
  \href{https://doi.org/10.1007/BF00417500}{\emph{Letters in Mathematical
  Physics} {\bfseries 4} (1980) 93}.

\bibitem{PhysRevA.63.062112}
R.A.~Bertlmann and B.C.~Hiesmayr, \emph{Bell inequalities for entangled kaons
  and their unitary time evolution},
  \href{https://doi.org/10.1103/PhysRevA.63.062112}{\emph{Phys. Rev. A}
  {\bfseries 63} (2001) 062112}.

\bibitem{Hiesmayr2012}
B.C.~Hiesmayr, A.~Di~Domenico, C.~Curceanu, A.~Gabriel, M.~Huber, J.-A.~Larsson
  et~al., \emph{{Revealing Bell's Nonlocality for Unstable Systems in High
  Energy Physics}},
  \href{https://doi.org/10.1140/epjc/s10052-012-1856-x}{\emph{Eur. Phys. J. C}
  {\bfseries 72} (2012) 1856}
  [\href{https://arxiv.org/abs/1111.4797}{{\ttfamily 1111.4797}}].

\bibitem{Afik:2020onf}
Y.~Afik and J.R.M.n.~de~Nova, \emph{{Entanglement and quantum tomography with
  top quarks at the LHC}},
  \href{https://doi.org/10.1140/epjp/s13360-021-01902-1}{\emph{Eur. Phys. J.
  Plus} {\bfseries 136} (2021) 907}
  [\href{https://arxiv.org/abs/2003.02280}{{\ttfamily 2003.02280}}].

\bibitem{Ashby-Pickering:2022umy}
R.~Ashby-Pickering, A.J.~Barr and A.~Wierzchucka, \emph{{Quantum state
  tomography, entanglement detection and Bell violation prospects in weak
  decays of massive particles}},
  \href{https://doi.org/10.1007/JHEP05(2023)020}{\emph{JHEP} {\bfseries 05}
  (2023) 020} [\href{https://arxiv.org/abs/2209.13990}{{\ttfamily
  2209.13990}}].

\bibitem{Barr:2021zcp}
A.J.~Barr, \emph{{Testing Bell inequalities in Higgs boson decays}},
  \href{https://doi.org/10.1016/j.physletb.2021.136866}{\emph{Phys. Lett. B}
  {\bfseries 825} (2022) 136866}
  [\href{https://arxiv.org/abs/2106.01377}{{\ttfamily 2106.01377}}].

\bibitem{Barr:2022wyq}
A.J.~Barr, P.~Caban and J.~Rembieli\'nski, \emph{{Bell-type inequalities for
  systems of relativistic vector bosons}},
  \href{https://doi.org/10.22331/q-2023-07-27-1070}{\emph{Quantum} {\bfseries
  7} (2023) 1070} [\href{https://arxiv.org/abs/2204.11063}{{\ttfamily
  2204.11063}}].

\bibitem{Afik:2022kwm}
Y.~Afik and J.R.M.n.~de~Nova, \emph{{Quantum information with top quarks in
  QCD}}, \href{https://doi.org/10.22331/q-2022-09-29-820}{\emph{Quantum}
  {\bfseries 6} (2022) 820} [\href{https://arxiv.org/abs/2203.05582}{{\ttfamily
  2203.05582}}].

\bibitem{Aguilar-Saavedra:2022uye}
J.A.~Aguilar-Saavedra and J.A.~Casas, \emph{{Improved tests of entanglement and
  Bell inequalities with LHC tops}},
  \href{https://doi.org/10.1140/epjc/s10052-022-10630-4}{\emph{Eur. Phys. J. C}
  {\bfseries 82} (2022) 666}
  [\href{https://arxiv.org/abs/2205.00542}{{\ttfamily 2205.00542}}].

\bibitem{Aoude:2022imd}
R.~Aoude, E.~Madge, F.~Maltoni and L.~Mantani, \emph{{Quantum SMEFT tomography:
  Top quark pair production at the LHC}},
  \href{https://doi.org/10.1103/PhysRevD.106.055007}{\emph{Phys. Rev. D}
  {\bfseries 106} (2022) 055007}
  [\href{https://arxiv.org/abs/2203.05619}{{\ttfamily 2203.05619}}].

\bibitem{Severi:2022qjy}
C.~Severi and E.~Vryonidou, \emph{{Quantum entanglement and top spin
  correlations in SMEFT at higher orders}},
  \href{https://doi.org/10.1007/JHEP01(2023)148}{\emph{JHEP} {\bfseries 01}
  (2023) 148} [\href{https://arxiv.org/abs/2210.09330}{{\ttfamily
  2210.09330}}].

\bibitem{Fabbrichesi:2022ovb}
M.~Fabbrichesi, R.~Floreanini and E.~Gabrielli, \emph{{Constraining new physics
  in entangled two-qubit systems: top-quark, tau-lepton and photon pairs}},
  \href{https://doi.org/10.1140/epjc/s10052-023-11307-2}{\emph{Eur. Phys. J. C}
  {\bfseries 83} (2023) 162}
  [\href{https://arxiv.org/abs/2208.11723}{{\ttfamily 2208.11723}}].

\bibitem{Altakach:2022ywa}
M.M.~Altakach, P.~Lamba, F.~Maltoni, K.~Mawatari and K.~Sakurai, \emph{{Quantum
  information and CP measurement in
  H\textrightarrow{}\ensuremath{\tau}+\ensuremath{\tau}- at future lepton
  colliders}}, \href{https://doi.org/10.1103/PhysRevD.107.093002}{\emph{Phys.
  Rev. D} {\bfseries 107} (2023) 093002}
  [\href{https://arxiv.org/abs/2211.10513}{{\ttfamily 2211.10513}}].

\bibitem{Acin:2000cs}
A.~Acin, J.I.~Latorre and P.~Pascual, \emph{{Three party entanglement from
  positronium}}, \href{https://doi.org/10.1103/PhysRevA.63.042107}{\emph{Phys.
  Rev. A} {\bfseries 63} (2001) 042107}
  [\href{https://arxiv.org/abs/quant-ph/0007080}{{\ttfamily
  quant-ph/0007080}}].

\bibitem{Sakurai:2023nsc}
K.~Sakurai and M.~Spannowsky, \emph{{Three-Body Entanglement in Particle
  Decays}}, \href{https://doi.org/10.1103/PhysRevLett.132.151602}{\emph{Phys.
  Rev. Lett.} {\bfseries 132} (2024) 151602}
  [\href{https://arxiv.org/abs/2310.01477}{{\ttfamily 2310.01477}}].

\bibitem{Morales:2024jhj}
R.A.~Morales, \emph{{Tripartite entanglement and Bell non-locality in
  loop-induced Higgs boson decays}},
  \href{https://doi.org/10.1140/epjc/s10052-024-12921-4}{\emph{Eur. Phys. J. C}
  {\bfseries 84} (2024) 581}
  [\href{https://arxiv.org/abs/2403.18023}{{\ttfamily 2403.18023}}].

\bibitem{Subba:2024mnl}
A.~Subba and R.~Rahaman, \emph{{On bipartite and tripartite entanglement at
  present and future particle colliders}},
  \href{https://arxiv.org/abs/2404.03292}{{\ttfamily 2404.03292}}.

\bibitem{Aoude:2023hxv}
R.~Aoude, E.~Madge, F.~Maltoni and L.~Mantani, \emph{{Probing new physics
  through entanglement in diboson production}},
  \href{https://doi.org/10.1007/JHEP12(2023)017}{\emph{JHEP} {\bfseries 12}
  (2023) 017} [\href{https://arxiv.org/abs/2307.09675}{{\ttfamily
  2307.09675}}].

\bibitem{Bernal:2023ruk}
A.~Bernal, P.~Caban and J.~Rembieli\'nski, \emph{{Entanglement and Bell
  inequalities violation in $H\rightarrow ZZ$ with anomalous coupling}},
  \href{https://doi.org/10.1140/epjc/s10052-023-12216-0}{\emph{Eur. Phys. J. C}
  {\bfseries 83} (2023) 1050}
  [\href{https://arxiv.org/abs/2307.13496}{{\ttfamily 2307.13496}}].

\bibitem{Han:2023fci}
T.~Han, M.~Low and T.A.~Wu, \emph{{Quantum entanglement and Bell inequality
  violation in semi-leptonic top decays}},
  \href{https://doi.org/10.1007/JHEP07(2024)192}{\emph{JHEP} {\bfseries 07}
  (2024) 192} [\href{https://arxiv.org/abs/2310.17696}{{\ttfamily
  2310.17696}}].

\bibitem{Dong:2023xiw}
Z.~Dong, D.~Gon$\c{c}$alves, K.~Kong and A.~Navarro, \emph{{Entanglement and
  Bell inequalities with boosted $tt\textasciimacron{}$}},
  \href{https://doi.org/10.1103/PhysRevD.109.115023}{\emph{Phys. Rev. D}
  {\bfseries 109} (2024) 115023}
  [\href{https://arxiv.org/abs/2305.07075}{{\ttfamily 2305.07075}}].

\bibitem{Fabbrichesi:2023jep}
M.~Fabbrichesi, R.~Floreanini, E.~Gabrielli and L.~Marzola, \emph{{Stringent
  bounds on HWW and HZZ anomalous couplings with quantum tomography at the
  LHC}}, \href{https://doi.org/10.1007/JHEP09(2023)195}{\emph{JHEP} {\bfseries
  09} (2023) 195} [\href{https://arxiv.org/abs/2304.02403}{{\ttfamily
  2304.02403}}].

\bibitem{Maltoni:2024tul}
F.~Maltoni, C.~Severi, S.~Tentori and E.~Vryonidou, \emph{{Quantum detection of
  new physics in top-quark pair production at the LHC}},
  \href{https://doi.org/10.1007/JHEP03(2024)099}{\emph{JHEP} {\bfseries 03}
  (2024) 099} [\href{https://arxiv.org/abs/2401.08751}{{\ttfamily
  2401.08751}}].

\bibitem{Afik:2024uif}
Y.~Afik, Y.~Kats, J.R.M.n.~de~Nova, A.~Soffer and D.~Uzan, \emph{{Entanglement
  and Bell nonlocality with bottom-quark pairs at hadron colliders}},
  \href{https://arxiv.org/abs/2406.04402}{{\ttfamily 2406.04402}}.

\bibitem{Maltoni:2024csn}
F.~Maltoni, C.~Severi, S.~Tentori and E.~Vryonidou, \emph{{Quantum tops at
  circular lepton colliders}},
  \href{https://doi.org/10.1007/JHEP09(2024)001}{\emph{JHEP} {\bfseries 09}
  (2024) 001} [\href{https://arxiv.org/abs/2404.08049}{{\ttfamily
  2404.08049}}].

\bibitem{Aguilar-Saavedra:2024fig}
J.A.~Aguilar-Saavedra and J.A.~Casas, \emph{{Entanglement Autodistillation from
  Particle Decays}},
  \href{https://doi.org/10.1103/PhysRevLett.133.111801}{\emph{Phys. Rev. Lett.}
  {\bfseries 133} (2024) 111801}
  [\href{https://arxiv.org/abs/2401.06854}{{\ttfamily 2401.06854}}].

\bibitem{Aguilar-Saavedra:2024hwd}
J.A.~Aguilar-Saavedra, \emph{{A closer look at post-decay $t \bar t$
  entanglement}},
  \href{https://doi.org/10.1103/PhysRevD.109.096027}{\emph{Phys. Rev. D}
  {\bfseries 109} (2024) 096027}
  [\href{https://arxiv.org/abs/2401.10988}{{\ttfamily 2401.10988}}].

\bibitem{Aguilar-Saavedra:2024vpd}
J.A.~Aguilar-Saavedra, \emph{{Full quantum tomography of top quark decays}},
  \href{https://doi.org/10.1016/j.physletb.2024.138849}{\emph{Phys. Lett. B}
  {\bfseries 855} (2024) 138849}
  [\href{https://arxiv.org/abs/2402.14725}{{\ttfamily 2402.14725}}].

\bibitem{Aguilar-Saavedra:2024whi}
J.A.~Aguilar-Saavedra, \emph{{Tripartite entanglement in
  H\textrightarrow{}ZZ,WW decays}},
  \href{https://doi.org/10.1103/PhysRevD.109.113004}{\emph{Phys. Rev. D}
  {\bfseries 109} (2024) 113004}
  [\href{https://arxiv.org/abs/2403.13942}{{\ttfamily 2403.13942}}].

\bibitem{Grabarczyk:2024wnk}
R.~Grabarczyk, \emph{{An improved Bell-CHSH observable for gauge boson pairs}},
   \href{https://arxiv.org/abs/2410.18022}{{\ttfamily 2410.18022}}.

\bibitem{Carena:2023vjc}
M.~Carena, I.~Low, C.E.M.~Wagner and M.-L.~Xiao, \emph{{Entanglement
  suppression, enhanced symmetry, and a standard-model-like Higgs boson}},
  \href{https://doi.org/10.1103/PhysRevD.109.L051901}{\emph{Phys. Rev. D}
  {\bfseries 109} (2024) L051901}
  [\href{https://arxiv.org/abs/2307.08112}{{\ttfamily 2307.08112}}].

\bibitem{Kowalska:2024kbs}
K.~Kowalska and E.M.~Sessolo, \emph{{Entanglement in flavored scalar
  scattering}}, \href{https://doi.org/10.1007/JHEP07(2024)156}{\emph{JHEP}
  {\bfseries 07} (2024) 156}
  [\href{https://arxiv.org/abs/2404.13743}{{\ttfamily 2404.13743}}].

\bibitem{Chang:2024wrx}
S.~Chang and G.~Jacobo, \emph{{Consequences of minimal entanglement in bosonic
  field theories}},
  \href{https://doi.org/10.1103/PhysRevD.110.096020}{\emph{Phys. Rev. D}
  {\bfseries 110} (2024) 096020}
  [\href{https://arxiv.org/abs/2409.13030}{{\ttfamily 2409.13030}}].

\bibitem{Thaler:2024anb}
J.~Thaler and S.~Trifinopoulos, \emph{{Flavor Patterns of Fundamental Particles
  from Quantum Entanglement?}},
  \href{https://arxiv.org/abs/2410.23343}{{\ttfamily 2410.23343}}.

\bibitem{Low:2024mrk}
I.~Low and Z.~Yin, \emph{{An Area Law for Entanglement Entropy in Particle
  Scattering}},  \href{https://arxiv.org/abs/2405.08056}{{\ttfamily
  2405.08056}}.

\bibitem{Low:2024hvn}
I.~Low and Z.~Yin, \emph{{Entanglement Entropy is Elastic Cross Section}},
  \href{https://arxiv.org/abs/2410.22414}{{\ttfamily 2410.22414}}.

\bibitem{Altomonte:2024upf}
C.~Altomonte, A.J.~Barr, M.~Eckstein, P.~Horodecki and K.~Sakurai,
  \emph{{Prospects for quantum process tomography at high energies}},
  \href{https://arxiv.org/abs/2412.01892}{{\ttfamily 2412.01892}}.

\bibitem{Han:2024ugl}
T.~Han, M.~Low, N.~McGinnis and S.~Su, \emph{{Measuring Quantum Discord at the
  LHC}},  \href{https://arxiv.org/abs/2412.21158}{{\ttfamily 2412.21158}}.

\bibitem{Han:2025ewp}
T.~Han, M.~Low and Y.~Su, \emph{{Entanglement and Bell Nonlocality in $\tau^+
  \tau^-$ at the BEPC}},  \href{https://arxiv.org/abs/2501.04801}{{\ttfamily
  2501.04801}}.

\bibitem{Barr:2024djo}
A.J.~Barr, M.~Fabbrichesi, R.~Floreanini, E.~Gabrielli and L.~Marzola,
  \emph{{Quantum entanglement and Bell inequality violation at colliders}},
  \href{https://doi.org/10.1016/j.ppnp.2024.104134}{\emph{Prog. Part. Nucl.
  Phys.} {\bfseries 139} (2024) 104134}
  [\href{https://arxiv.org/abs/2402.07972}{{\ttfamily 2402.07972}}].

\bibitem{Abel:1992kz}
S.A.~Abel, M.~Dittmar and H.K.~Dreiner, \emph{{Testing locality at colliders
  via Bell's inequality?}},
  \href{https://doi.org/10.1016/0370-2693(92)90071-B}{\emph{Phys. Lett. B}
  {\bfseries 280} (1992) 304}.

\bibitem{ATLAS:2023fsd}
{\scshape ATLAS} collaboration, \emph{{Observation of quantum entanglement with
  top quarks at the ATLAS detector}},
  \href{https://doi.org/10.1038/s41586-024-07824-z}{\emph{Nature} {\bfseries
  633} (2024) 542} [\href{https://arxiv.org/abs/2311.07288}{{\ttfamily
  2311.07288}}].

\bibitem{CMS:2024pts}
{\scshape CMS} collaboration, \emph{{Observation of quantum entanglement in top
  quark pair production in proton-proton collisions at $\sqrt{s}$ = 13 TeV}},
  \href{https://arxiv.org/abs/2406.03976}{{\ttfamily 2406.03976}}.

\bibitem{CMS:2024zkc}
{\scshape CMS} collaboration, \emph{{Measurements of polarization and spin
  correlation and observation of entanglement in top quark pairs using
  lepton+jets events from proton-proton collisions at $\sqrt{s}$ = 13 TeV}},
  \href{https://arxiv.org/abs/2409.11067}{{\ttfamily 2409.11067}}.

\bibitem{Aguilar-Saavedra:2024mnm}
J.A.~Aguilar-Saavedra, \emph{{Toponium hunter\textquoteright{}s guide}},
  \href{https://doi.org/10.1103/PhysRevD.110.054032}{\emph{Phys. Rev. D}
  {\bfseries 110} (2024) 054032}
  [\href{https://arxiv.org/abs/2407.20330}{{\ttfamily 2407.20330}}].

\bibitem{Seki:2014cgq}
S.~Seki, I.Y.~Park and S.-J.~Sin, \emph{{Variation of Entanglement Entropy in
  Scattering Process}},
  \href{https://doi.org/10.1016/j.physletb.2015.02.028}{\emph{Phys. Lett. B}
  {\bfseries 743} (2015) 147}
  [\href{https://arxiv.org/abs/1412.7894}{{\ttfamily 1412.7894}}].

\bibitem{Peschanski:2016hgk}
R.~Peschanski and S.~Seki, \emph{{Entanglement Entropy of Scattering
  Particles}},
  \href{https://doi.org/10.1016/j.physletb.2016.04.063}{\emph{Phys. Lett. B}
  {\bfseries 758} (2016) 89}
  [\href{https://arxiv.org/abs/1602.00720}{{\ttfamily 1602.00720}}].

\bibitem{Peschanski:2019yah}
R.~Peschanski and S.~Seki, \emph{{Evaluation of Entanglement Entropy in High
  Energy Elastic Scattering}},
  \href{https://doi.org/10.1103/PhysRevD.100.076012}{\emph{Phys. Rev. D}
  {\bfseries 100} (2019) 076012}
  [\href{https://arxiv.org/abs/1906.09696}{{\ttfamily 1906.09696}}].

\bibitem{Horodecki:2009zz}
R.~Horodecki, P.~Horodecki, M.~Horodecki and K.~Horodecki, \emph{{Quantum
  entanglement}}, \href{https://doi.org/10.1103/RevModPhys.81.865}{\emph{Rev.
  Mod. Phys.} {\bfseries 81} (2009) 865}
  [\href{https://arxiv.org/abs/quant-ph/0702225}{{\ttfamily
  quant-ph/0702225}}].

\bibitem{Guhne:2008qic}
O.~G\"uhne and G.~T\'oth, \emph{{Entanglement detection}},
  \href{https://doi.org/10.1016/j.physrep.2009.02.004}{\emph{Phys. Rept.}
  {\bfseries 474} (2009) 1} [\href{https://arxiv.org/abs/0811.2803}{{\ttfamily
  0811.2803}}].

\bibitem{Friis_2018}
N.~Friis, G.~Vitagliano, M.~Malik and M.~Huber, \emph{Entanglement
  certification from theory to experiment},
  \href{https://doi.org/10.1038/s42254-018-0003-5}{\emph{Nature Reviews
  Physics} {\bfseries 1} (2018) 72–87}.

\bibitem{PhysRevA.40.4277}
R.F.~Werner, \emph{Quantum states with einstein-podolsky-rosen correlations
  admitting a hidden-variable model},
  \href{https://doi.org/10.1103/PhysRevA.40.4277}{\emph{Phys. Rev. A}
  {\bfseries 40} (1989) 4277}.

\bibitem{Vidal_2000}
G.~Vidal, \emph{Entanglement monotones},
  \href{https://doi.org/10.1080/09500340008244048}{\emph{Journal of Modern
  Optics} {\bfseries 47} (2000) 355–376}.

\bibitem{PhysRevA.54.3824}
C.H.~Bennett, D.P.~DiVincenzo, J.A.~Smolin and W.K.~Wootters, \emph{Mixed-state
  entanglement and quantum error correction},
  \href{https://doi.org/10.1103/PhysRevA.54.3824}{\emph{Phys. Rev. A}
  {\bfseries 54} (1996) 3824}.

\bibitem{Uhlmann1998}
A.~Uhlmann, \emph{Entropy and optimal decompositions of states relative to a
  maximal commutative subalgebra},
  \href{https://doi.org/10.1023/A:1009664331611}{\emph{Open Systems {\&}
  Information Dynamics} {\bfseries 5} (1998) 209}.

\bibitem{PhysRevLett.80.2245}
W.K.~Wootters, \emph{Entanglement of formation of an arbitrary state of two
  qubits}, \href{https://doi.org/10.1103/PhysRevLett.80.2245}{\emph{Phys. Rev.
  Lett.} {\bfseries 80} (1998) 2245}.

\bibitem{PhysRevD.35.3066}
G.~Svetlichny, \emph{Distinguishing three-body from two-body nonseparability by
  a bell-type inequality},
  \href{https://doi.org/10.1103/PhysRevD.35.3066}{\emph{Phys. Rev. D}
  {\bfseries 35} (1987) 3066}.

\bibitem{GHZ}
D.M.~Greenberger, M.A.~Horne and A.~Zeilinger, \emph{Going beyond bell's
  theorem}, .

\bibitem{Dur:2000zz}
W.~Dur, G.~Vidal and J.I.~Cirac, \emph{{Three qubits can be entangled in two
  inequivalent ways}},
  \href{https://doi.org/10.1103/PhysRevA.62.062314}{\emph{Phys. Rev. A}
  {\bfseries 62} (2000) 062314}
  [\href{https://arxiv.org/abs/quant-ph/0005115}{{\ttfamily
  quant-ph/0005115}}].

\bibitem{PhysRevLett.85.1560}
A.~Ac\'{\i}n, A.~Andrianov, L.~Costa, E.~Jan\'e, J.I.~Latorre and R.~Tarrach,
  \emph{Generalized schmidt decomposition and classification of
  three-quantum-bit states},
  \href{https://doi.org/10.1103/PhysRevLett.85.1560}{\emph{Phys. Rev. Lett.}
  {\bfseries 85} (2000) 1560}.

\bibitem{Coffman:1999jd}
V.~Coffman, J.~Kundu and W.K.~Wootters, \emph{{Distributed entanglement}},
  \href{https://doi.org/10.1103/PhysRevA.61.052306}{\emph{Phys. Rev. A}
  {\bfseries 61} (2000) 052306}
  [\href{https://arxiv.org/abs/quant-ph/9907047}{{\ttfamily
  quant-ph/9907047}}].

\bibitem{Osborne2006}
T.J.~Osborne and F.~Verstraete, \emph{General monogamy inequality for bipartite
  qubit entanglement},
  \href{https://doi.org/10.1103/physrevlett.96.220503}{\emph{Physical Review
  Letters} {\bfseries 96} (2006) }.

\bibitem{Jin2023}
Z.-X.~Jin, Y.-H.~Tao, Y.-T.~Gui, S.-M.~Fei, X.~Li-Jost and C.-F.~Qiao,
  \emph{Concurrence triangle induced genuine multipartite entanglement
  measure}, \href{https://doi.org/10.1016/j.rinp.2022.106155}{\emph{Results in
  Physics} {\bfseries 44} (2023) 106155}.

\bibitem{Guo_2022}
Y.~Guo, Y.~Jia, X.~Li and L.~Huang, \emph{Genuine multipartite entanglement
  measure}, \href{https://doi.org/10.1088/1751-8121/ac5649}{\emph{Journal of
  Physics A: Mathematical and Theoretical} {\bfseries 55} (2022) 145303}.

\bibitem{PhysRevLett.97.260502}
R.~Lohmayer, A.~Osterloh, J.~Siewert and A.~Uhlmann, \emph{Entangled
  three-qubit states without concurrence and three-tangle},
  \href{https://doi.org/10.1103/PhysRevLett.97.260502}{\emph{Phys. Rev. Lett.}
  {\bfseries 97} (2006) 260502}.

\bibitem{Osterloh:2025wwr}
A.~Osterloh, \emph{{The exact convex roof for GHZ-W mixtures for three qubits
  and beyond}},  \href{https://arxiv.org/abs/2501.07084}{{\ttfamily
  2501.07084}}.

\bibitem{PopescuRohrlich94}
S.~Popescu and D.~Rohrlich, \emph{Quantum nonlocality as an axiom},
  \href{https://doi.org/10.1007/BF02058098}{\emph{Foundations of Physics}
  {\bfseries 24} (1994) 379}.

\bibitem{Capasso:1973wt}
V.~Capasso, D.~Fortunato and F.~Selleri, \emph{{Sensitive observables of
  quantum mechanics}}, \href{https://doi.org/10.1007/BF00669912}{\emph{Int. J.
  Theor. Phys.} {\bfseries 7} (1973) 319}.

\bibitem{GISIN1991201}
N.~Gisin, \emph{Bell's inequality holds for all non-product states},
  \href{https://doi.org/https://doi.org/10.1016/0375-9601(91)90805-I}{\emph{Physics
  Letters A} {\bfseries 154} (1991) 201}.

\bibitem{GISIN199215}
N.~Gisin and A.~Peres, \emph{Maximal violation of bell's inequality for
  arbitrarily large spin},
  \href{https://doi.org/https://doi.org/10.1016/0375-9601(92)90949-M}{\emph{Physics
  Letters A} {\bfseries 162} (1992) 15}.

\bibitem{Cereceda_2002}
J.L.~Cereceda, \emph{Three-particle entanglement versus three-particle
  nonlocality},
  \href{https://doi.org/10.1103/physreva.66.024102}{\emph{Physical Review A}
  {\bfseries 66} (2002) }.

\bibitem{PhysRevLett.65.1838}
N.D.~Mermin, \emph{Extreme quantum entanglement in a superposition of
  macroscopically distinct states},
  \href{https://doi.org/10.1103/PhysRevLett.65.1838}{\emph{Phys. Rev. Lett.}
  {\bfseries 65} (1990) 1838}.

\bibitem{PhysRevLett.88.170405}
D.~Collins, N.~Gisin, S.~Popescu, D.~Roberts and V.~Scarani, \emph{Bell-type
  inequalities to detect true $\mathit{n}$-body nonseparability},
  \href{https://doi.org/10.1103/PhysRevLett.88.170405}{\emph{Phys. Rev. Lett.}
  {\bfseries 88} (2002) 170405}.

\bibitem{PhysRevA.70.060101}
P.~Mitchell, S.~Popescu and D.~Roberts, \emph{Conditions for the confirmation
  of three-particle nonlocality},
  \href{https://doi.org/10.1103/PhysRevA.70.060101}{\emph{Phys. Rev. A}
  {\bfseries 70} (2004) 060101}.

\bibitem{laskowski}
W.~Laskowski, T.~Paterek, M.~\ifmmode~\dot{Z}\else \.{Z}\fi{}ukowski and
  i.c.v.~Brukner, \emph{Tight multipartite bell's inequalities involving many
  measurement settings},
  \href{https://doi.org/10.1103/PhysRevLett.93.200401}{\emph{Phys. Rev. Lett.}
  {\bfseries 93} (2004) 200401}.

\bibitem{WU2003262}
X.-H.~Wu and H.-S.~Zong, \emph{A new bell inequality for three spin-half
  particle system},
  \href{https://doi.org/https://doi.org/10.1016/S0375-9601(02)01672-9}{\emph{Physics
  Letters A} {\bfseries 307} (2003) 262}.

\bibitem{Hagiwara:1989fn}
K.~Hagiwara, A.D.~Martin and D.~Zeppenfeld, \emph{{Tau Polarization
  Measurements at LEP and SLC}},
  \href{https://doi.org/10.1016/0370-2693(90)90120-U}{\emph{Phys. Lett. B}
  {\bfseries 235} (1990) 198}.

\bibitem{L3:1994hzc}
{\scshape L3} collaboration, \emph{{A Measurement of tau polarization at LEP}},
  \href{https://doi.org/10.1016/0370-2693(94)90316-6}{\emph{Phys. Lett. B}
  {\bfseries 341} (1994) 245}.

\bibitem{Kats:2023zxb}
Y.~Kats and D.~Uzan, \emph{{Prospects for measuring quark polarization and spin
  correlations in $b\overline{b }$ and $c\overline{c }$ samples at the LHC}},
  \href{https://doi.org/10.1007/JHEP03(2024)063}{\emph{JHEP} {\bfseries 03}
  (2024) 063} [\href{https://arxiv.org/abs/2311.08226}{{\ttfamily
  2311.08226}}].

\bibitem{ALEPH:1995aqx}
{\scshape ALEPH} collaboration, \emph{{Measurement of Lambda(b) polarization in
  Z decays}}, \href{https://doi.org/10.1016/0370-2693(95)01433-0}{\emph{Phys.
  Lett. B} {\bfseries 365} (1996) 437}.

\bibitem{OPAL:1998wmk}
{\scshape OPAL} collaboration, \emph{{Measurement of the average polarization
  of b baryons in hadronic Z0 decays}},
  \href{https://doi.org/10.1016/S0370-2693(98)01387-2}{\emph{Phys. Lett. B}
  {\bfseries 444} (1998) 539}
  [\href{https://arxiv.org/abs/hep-ex/9808006}{{\ttfamily hep-ex/9808006}}].

\bibitem{DELPHI:1999hkl}
{\scshape DELPHI} collaboration, \emph{{Lambda(b) polarization in Z0 decays at
  LEP}}, \href{https://doi.org/10.1016/S0370-2693(99)01431-8}{\emph{Phys. Lett.
  B} {\bfseries 474} (2000) 205}.

\bibitem{Grunhaus}
J.~Grunhaus, S.~Popescu and D.~Rohrlich, \emph{Jamming nonlocal quantum
  correlations}, \href{https://doi.org/10.1103/PhysRevA.53.3781}{\emph{Phys.
  Rev. A} {\bfseries 53} (1996) 3781}.

\bibitem{Horodecki-Ramanathan-2019}
P.~Horodecki and R.~Ramanathan, \emph{The relativistic causality versus
  no-signaling paradigm for multi-party correlations},
  \href{https://doi.org/10.1038/s41467-019-09505-2}{\emph{Nature
  Communications} {\bfseries 10} (2019) 1701}.

\bibitem{Bell-nonlocality}
N.~Brunner, D.~Cavalcanti, S.~Pironio, V.~Scarani and S.~Wehner, \emph{Bell
  nonlocality}, \href{https://doi.org/10.1103/RevModPhys.86.419}{\emph{Rev.
  Mod. Phys.} {\bfseries 86} (2014) 419}.

\bibitem{HORODECKI1995340}
R.~Horodecki, P.~Horodecki and M.~Horodecki, \emph{Violating bell inequality by
  mixed spin-12 states: necessary and sufficient condition},
  \href{https://doi.org/https://doi.org/10.1016/0375-9601(95)00214-N}{\emph{Physics
  Letters A} {\bfseries 200} (1995) 340}.

\end{thebibliography}\endgroup
 
\end{document}